\documentclass[10pt,twocolumn,english,aps,prd,nofootinbib,preprintnumbers]{revtex4}
\usepackage[latin9]{inputenc}
\usepackage{graphicx}
\usepackage{amsmath}
\usepackage{amssymb}
\usepackage{amsfonts}
\usepackage{lscape}
\usepackage{hyperref}
\usepackage{url}
\usepackage{color}
\usepackage{bm}
\usepackage{tabularx}
\usepackage{mathtools}
\newcommand{\be}{\begin{equation}}
\newcommand{\ee}{\end{equation}}
\newcommand{\ba}{\begin{eqnarray}}
\newcommand{\ea}{\end{eqnarray}}
\newcommand{\nn}{\nonumber}

\renewcommand{\[}{\begin{equation}}
\renewcommand{\]}{\end{equation}}
\def\be{\begin{equation}}
\def\ee{\end{equation}}
\def\bea{\begin{eqnarray}}
\def\eea{\end{eqnarray}}
\def\eqi{\begin{equation}}
\def\eqf{\end{equation}}
\def\eqia{\begin{eqnarray}}
\def\eqfa{\end{eqnarray}}
\def\lcdm{$\Lambda$CDM }

\makeatother

\begin{document}

\preprint{IFT-UAM/CSIC-19-051}

\title{Designing Horndeski and the effective fluid approach}

\author{Rub\'{e}n Arjona}
\email{ruben.arjona@uam.es}

\author{Wilmar Cardona}
\email{wilmar.cardona@uam.es}

\author{Savvas Nesseris}
\email{savvas.nesseris@csic.es}

\affiliation{Instituto de F\'isica Te\'orica UAM-CSIC, Universidad Auton\'oma de Madrid,
Cantoblanco, 28049 Madrid, Spain}

\date{\today}

\begin{abstract}
We present a family of designer Horndeski models, i.e. models that have a background exactly equal to that of the \lcdm model but perturbations given by the Horndeski theory. Then, we extend the effective fluid approach to Horndeski theories, providing simple analytic formulae for the equivalent dark energy effective fluid pressure, density and velocity. We implement the dark energy effective fluid formulae in our code EFCLASS, a modified version of the widely used Boltzmann solver CLASS, and compare the solution of the perturbation equations with those of the code hi\_CLASS which already includes Horndeski models. We find that our simple modifications to the vanilla code are accurate to the level of $\sim 0.1\%$ with respect to the more complicated hi\_CLASS code. Furthermore, we study the kinetic braiding model both on and off the attractor and we find that even though the full case has a proper \lcdm limit for large $n$, it is not appropriately smooth, thus causing the quasistatic approximation to break down. Finally, we focus on our designer model (HDES), which has both a smooth $\Lambda$CDM limit and well-behaved perturbations, and we use it to perform Markov Chain Monte Carlo analyses to constrain its parameters with the latest cosmological data. We find that our HDES model can also alleviate the soft $2\sigma$ tension between the growth data and Planck 18 due to a degeneracy between $\sigma_8$ and one of its model parameters that indicates the deviation from the \lcdm model.
\end{abstract}
\maketitle

\section{Introduction \label{Section:Introduction}}

Not long ago, measurements of the distance-redshift relation from distant Supernovae type Ia (SNIa) revealed that the Universe is not only expanding as time goes by, it is actually accelerating \cite{Riess:1998cb,Perlmutter:1998np}. The consequences of these observations are far-reaching and several analyses of the data sets have been carefully carried out ever since (see Ref. \cite{Nielsen:2015pga} and references therein). Although there were some concerns about possible systematic errors, analyses of new and improved data sets have shown that results in previous works are robust  \cite{Riess:2006fw,Betoule:2014frx}. Moreover, most recent astrophysical measurements of the Cosmic Microwave Background (CMB) anisotropies and the distribution of galaxies in the Universe, when interpreted in the context of the cosmological constant cold dark matter model ($\Lambda$CDM), are in very good agreement with a late-time accelerating phase \cite{Aghanim:2018eyx,Abbott:2017wau}.

Current Bayesian analyses of astrophysical measurements indicate that \lcdm beats alternative models \cite{Heavens:2017hkr}. In spite of being successful at fitting most data sets, \lcdm is just a very good phenomenological model as its main constituents are either unknown or misunderstood. First, Cold Dark Matter (CDM) has not been directly detected thus far despite the huge effort this research field has attracted over the past years \cite{Bertone:2016nfn}. Second, there exists an important disagreement between both predicted and inferred values of the cosmological constant $\Lambda$ whose solution will possibly lead to new physics \cite{Weinberg:1988cp,Carroll:2000fy}.

Even though reconciling the quantum field theory prediction with the observed value of the cosmological constant seems unlikely, it has become clear that a Dark Energy (DE) component resembling a cosmological constant not only can alleviate several problems present in a CDM model but also can drive the current accelerating expansion of the Universe \cite{Kofman:1985fp}. Although several mechanisms have been proposed in the literature which could be responsible for speeding up the Universe, nowadays there are two main approaches. On the one hand, one finds Modified Gravity (MG) models \cite{Clifton:2011jh}. Einstein's Theory of General Relativity (GR), the theory of gravity that is assumed in the \lcdm model, seems to break down on tiny scales and possibly will require modifications on large scales to account for current observations \cite{Bertschinger:2011kk}. However, modifying GR can be laborious as several tests carried out up to cosmological scales agree very well with GR \cite{Bertotti:2003rm,Reyes:2010tr,Collett:2018gpf,PhysRevLett.116.221101,He:2018oai,PhysRevLett.121.231101,PhysRevLett.121.231102,Ishak:2018his,Luna:2018tot,Basilakos:2018arq,Perez-Romero:2017njc, Basilakos:2017rgc,Nesseris:2017vor,Basilakos:2016nyg}. On the other hand, there are DE models \cite{Copeland:2006wr} which rely on yet unobserved scalar fields that would dominate the energy content of the Universe at late times and also avoid fine-tuning issues \cite{Ratra:1987rm,ArmendarizPicon:2000dh}.

Although DE and MG models are clearly motivated by different underlying physics, it is possible to study both kinds of models on the same footing. In an effective fluid approach departures from GR can be interpreted as an effective fluid contribution in such a way that comparison with DE models might become relatively simple \cite{Kunz:2006ca,Pogosian:2010tj,Capozziello:2005mj,Capozziello:2006dj,Capozziello:2018ddp}. When interpreted as fluids, MG models can be described by an equation of state $w(a)$, a sound speed $c_s^2(a,k)$, and an anisotropic stress $\pi(a,k)$: background is affected by the behavior of $w(a)$ while perturbations are mainly governed by $c_s^2(a,k)$ and $\pi(a,k)$. Since both DE and MG models predict different behavior for these three functions, in an effective fluid approach different models can be, to a certain degree, distinguished.

It is well known that both DE and MG models can accommodate background astrophysical observations as well as the standard cosmological model \lcdm (e.g., the so-called designer $f(R)$ models \cite{Multamaki:2005zs,delaCruzDombriz:2006fj,Pogosian:2007sw,Nesseris:2013fca}). As a consequence, these models are degenerated at the background level even though there have been various attempts to disentangle them by using model independent approaches \cite{Nesseris:2010ep,Nesseris:2012tt}. Fortunately, the study of linear order perturbations might break this degeneracy because DE and MG models predict different growths of structures and could in principle be distinguishable from \lcdm \cite{Tsujikawa:2007gd,Pogosian:2007sw}.

Given the wide range of both DE and MG models it is useful to have a unified framework which encompasses several of them. It turns out that such a theory exists since 1974 when Horndeski found the most general Lorentz-invariant extension of GR in four dimensions \cite{Horndeski:1974wa}. This theory can be obtained by using a single scalar field and restricting the equations of motion to being second order in time derivatives. The Horndeski Lagrangian comprehends theories such as Kinetic Gravity Braiding, Brans-Dicke and scalar tensor gravity, single field quintessence and K-essence theories, as well as  $f(R)$ theories in their scalar-tensor formulation \cite{Baker:2012zs}. Although the range of models encompassed by the Horndeski Lagrangian was severely reduced (see, for instance, \cite{Creminelli:2017sry,Sakstein:2017xjx,Ezquiaga:2017ekz,Baker:2017hug,Amendola:2017orw,Crisostomi:2017pjs,
Frusciante:2018,Kase:2018aps,McManus:2016kxu,Lombriser:2015sxa,Copeland:2018yuh,Noller:2018eht,deRham:2018red}) with the recent discovery of gravitational waves by the LIGO Collaboration \cite{Abbott:2017oio}, an interesting remaining subclass of models (including $f(R)$ theories \cite{Sotiriou:2008rp,DeFelice:2010aj,Nojiri:2017ncd,Nojiri:2010wj} and Kinetic Gravity Braiding \cite{Deffayet:2010qz}) is well worth an investigation.

Recently we employed an effective fluid approach to study $f(R)$ theories \cite{Arjona:2018jhh}. Even though it is not easy to obtain expressions for quantities describing perturbations (e.g., pressure perturbation $\delta P$) in MG models \cite{Kunz:2006wc}, by using the quasistatic and subhorizon approximations we found analytical expressions for the effective DE perturbations as well as the quantities describing the effective DE fluid, namely, $w(a)$, $c_s^2(a,k)$, and $\pi(a,k)$. We implemented our approach in the code CLASS\footnote{\url{http://class-code.net/}} \cite{Blas:2011rf} and found excellent agreement with the so-called Equation of State (EOS) approach \cite{Battye:2015hza,Battye:2017ysh}, which does not use any approximation. In this paper we extend our work \cite{Arjona:2018jhh} to the remaining part of the Horndeski Lagrangian which contains $f(R)$ theories as a special case. Horndeski theories have been implemented in the code hi\_CLASS \cite{Zumalacarregui:2016pph} which solves the full set of dynamical equations without using the quasistatic approximation. In our approach we find analytical expressions for the effective DE perturbations that give us a better understanding of the underlying physics and also allow us to compare with our numerical implementation. Moreover, we show that it is possible to find `designer Horndeski theories' matching a given background evolution. We implement one such a model in the hi\_CLASS code and show  there is good agreement with our approach, namely, our effective fluid approach assuming both quasistatic and subhorizon approximations performs quite well.

The paper is organized as follows. In Sec. \ref{Section:theoretical-framework} we discuss the equations for perturbations in a Friedmann-Lemaitre-Robertson-Walker (FLRW) metric and set our notation. Then, we introduce the Horndeski Lagrangian and discuss both background and perturbation equations in Sec. \ref{Section:Horndeski}. In Sec. \ref{Section:EFA} we study the remaining subclass of Horndeski theories by utilizing the effective fluid approach, we discuss the subhorizon and quasistatic approximations and present analytical results for two classes of models, those in which we have dark energy anisotropic stress and those in which we do not. In Sec. \ref{Section:DES} we show analytical results for a family of models named `designer Horndeski' which mimic the \lcdm background and in Sec. \ref{Section:Numerical-Solution} we compare our analytical solutions for DE perturbations with a fully numerical solution of the system of differential equations and show they are in very good agreement. We then constrain the parameter space for a viable designer Horndeski model in Sec. \ref{Section:mcmc} and in Sec. \ref{Section:conclusions} we present our conclusions. In Appendices \ref{Section:appendix-A} and \ref{Section:appendix-B} we give details about our analytical computations.

\section{Theoretical framework \label{Section:theoretical-framework}}

In the standard cosmological model one assumes the Einstein-Hilbert action
\be
S=\int d^{4}x\sqrt{-g}\left[  \frac{1}{2\kappa} R
+\mathcal{L}_{m}\right],  \label{eq:action-GR}%
\ee
where $g$ is the determinant of the metric $g_{\mu\nu}$, $R$ is the Ricci scalar, $\kappa\equiv \frac{8\pi G_N}{c^4}$ and $\mathcal{L}_{m}$ is the Lagrangian for matter fields.\footnote{Throughout this paper we set the speed of light $c=1$ and $\kappa=8\pi G_N$ with $G_N$ being the bare Newton's constant. Our conventions are: $(-+++)$ for the metric signature, the Riemann and Ricci tensors are given respectively by $V_{b;cd}-V_{b;dc}=V_a R^a_{bcd}$ and $R_{ab}=R^s_{asb}$.} Applying the principle of least action to Eq. \eqref{eq:action-GR} one obtains the field equations
\be
G_{\mu\nu} = \kappa\,T_{\mu\nu}^{(m)},
\label{eq:EE}
\ee
where $G_{\mu\nu} \equiv R_{\mu\nu} - \frac{1}{2} g_{\mu\nu} R $ is the Einstein tensor and $T_{\mu\nu}^{(m)}$ is the energy-momentum tensor for matter fields. At this point one needs to make more assumptions about the geometrical properties and the matter content in the Universe. First, since observations indicate the Universe on large scales is statistically homogeneous and isotropic \cite{Hogg:2004vw,Ade:2015hxq} (also having tiny inhomogeneities which can be treated within linear perturbation theory), one further assumes a perturbed FLRW metric
\be
ds^2=-\left(1+2\Psi(\vec{x},t)\right)dt^2+a(t)^2(1+2\Phi(\vec{x},t))d\vec{x}^2,
\label{eq:FRWpert}
\ee
where $a$ is the scale factor, $\vec{x}$ represents spatial coordinates, 
$t$ is the cosmic time and $\Psi$ and $\Phi$ are the gravitational potentials in the Newtonian gauge. Second, one can suppose the matter fields are ideal fluids (with small perturbations) having an energy-momentum tensor given by
\be
T^\mu_{\nu}=P\delta^\mu_{\nu}+(\rho+P)U^\mu U_\nu,\label{eq:enten}
\ee
where $P$ is the pressure, $\rho$ is the energy density, and $U^\mu=\left(1-\Psi,\frac{\vec{u}}{a(t)}\right)$ is the velocity four-vector. As a result, the elements of the energy-momentum tensor up to first order are given by :
\bea
T^0_0&=&-(\bar{\rho}+\delta \rho), \label{eq:effectTmn1}\\
T^0_i&=&(\bar{\rho}+\bar{P})a(t)u_i,\label{eq:effectTmnvde}\\
T^i_j&=& (\bar{P}+\delta P)\delta^i_j+\Sigma^i_j, \label{eq:effectTmn}
\eea
where $\bar{\rho}$ is the background energy density, $\bar{P}$ is the background pressure, $u_i=a(t)\dot{x_i}$, $\Sigma^i_j(\vec{x},\tau) \equiv T^i_j-\delta^i_j T^k_k/3$ is an anisotropic stress tensor, and $\delta \rho(\vec{x},\tau)$ and $\delta P(\vec{x},\tau)$ are the density and pressure perturbations, respectively.\footnote{In our notation,  a dot over a function $f$ denotes the derivative with respect to the cosmic time :  $\dot{f}\equiv\frac{df}{dt}$. In addition, Greek indices run from $0$ to $3$ whereas Latin indices take on values from $1$ to $3$.}

\subsection{Background}

If one only considers zero order quantities in the Einstein field equations \eqref{eq:EE}, then there are two independent Friedmann equations describing the background evolution of the Universe:
\bea
H^2&=&\frac{\kappa}{3}\bar{\rho}, \label{eq:Friedmann-zero-order-1} \\
H^2+\dot{H}&=&-\frac{\kappa}{6} \left(\bar{\rho} + 3\bar{P}\right),
\label{eq:Friedmann-zero-order-2}
\eea
where $H\equiv\frac{\dot{a}}{a}$ is the cosmic Hubble parameter.\footnote{The conformal Hubble parameter $\mathcal{H}$ and the Hubble parameter $H$ are related via $\mathcal{H} = a H$.}

\subsection{Linear perturbations}

Considering just the first order perturbations in the Einstein field equations \eqref{eq:EE} we obtain
\be
-\frac{k^2}{a^2}\Phi+3\frac{\dot{a}}{a}\left(\frac{\dot{a}}{a}\Psi-\dot{\Phi}\right) = \frac{\kappa}{2} \delta T^0_0, \label{eq:phiprimeeq}
\ee
\be
k^2\left(\frac{\dot{a}}{a}\Psi-\dot{\Phi}\right) = \frac{\kappa}{2} a (\bar{\rho}+\bar{P})\theta,\label{eq:phiprimeeq1}
\ee
\be
-\frac{k^2}{3a^2}(\Phi+\Psi)+\left(2\frac{\ddot{a}}{a}+\frac{\dot{a}^2}{a^2}\right)\Psi+\frac{\dot{a}}{a}\left(\dot{\Psi}-3\dot{\Phi}\right)-\ddot{\Phi}
=\frac{\kappa}{6} \delta T^i_i,
\ee
\be
-k^2(\Phi+\Psi) = \frac{3\kappa}{2} a^2 (\bar{\rho}+\bar{P})\sigma \label{eq:anisoeq},
\ee
where we defined the velocity $\theta\equiv ik^ju_j$ and wrote the anisotropic stress as  $(\bar{\rho}+\bar{P})\sigma\equiv-(\hat{k}_i\hat{k}_j-\frac13 \delta_{ij})\Sigma^{ij}$.

From the conservation of the energy-momentum tensor $T^{\mu\nu}_{;\nu}=0$ one obtains the equations for the evolution of perturbations. Defining the equation of state parameter as $w\equiv\frac{\bar{P}}{\bar{\rho}}$ and the sound speed  $c_s^2\equiv\frac{\delta P}{\delta \rho}$ we find the equations governing the evolution of density and pressure perturbations are given by
\be
\dot{\delta} = -(1+w)(\frac{\theta}{a}+3\dot{\Phi})-3\frac{\dot{a}}{a}\left(c_s^2-w\right)\delta,
\label{eq:cons1}
\ee
\be
\dot{\theta} = -\frac{\dot{a}}{a}(1-3w)\theta-\frac{\dot{w}}{1+w}\theta+\frac{c_s^2}{1+w}\frac{k^2}{a}\delta-\frac{k^2}{a}\sigma+\frac{k^2}{a}\Psi,
\label{eq:cons2}
\ee

The system of differential equations \eqref{eq:cons1}-\eqref{eq:cons2} presents problems when the equation of state crosses $-1$ because there is a singularity. However, a simple change of variable turns out to be helpful in solving this inconvenience. We will use the scalar velocity perturbation $V\equiv i k_jT^j_0/\rho=(1+w)\theta$ instead of the velocity $\theta$. In terms of this new variable the evolution equations \eqref{eq:cons1}-\eqref{eq:cons2} become
\bea
\delta' &=&- 3(1+w) \Phi'-\frac{V}{a^2 H}-\frac{3}{a}\left(\frac{\delta P}{\bar{\rho}}-w\delta\right),
\label{Eq:evolution-delta}
\eea
\bea
V' &=& -(1-3w)\frac{V}{a}+\frac{k^2}{a^2 H}\frac{\delta P}{\bar{\rho}} +(1+w)\frac{k^2}{a^2 H} \Psi \nn \\
 &-&\frac23 \frac{k^2}{a^2 H} \pi,
\label{Eq:evolution-V}
\eea
where a prime $'$ denotes a derivative with respect to the scale factor and we defined the anisotropic stress parameter $\pi\equiv\frac32(1+w)\sigma$.

\section{Horndeski}
\label{Section:Horndeski}

Horndeski theory constitutes the most general Lorentz-invariant extension of GR in four dimensions and encompasses several DE and MG models. Although in its most general form the Horndeski Lagrangian has several free functions, the recent discovery of gravitational waves by the LIGO Collaboration significantly constrained the allowed models. In particular, it has been shown that the constraint on the speed of Gravitational Waves (GWs) must satisfy \cite{Ezquiaga:2017ekz}
\bea
-3 \cdot 10^{-15} \le c_g/c-1 \le 7 \cdot 10^{-16},
\eea
which for Horndeski theories implies that
\bea
G_{4X}\approx 0, \hspace{2mm} G_5 \approx \text{const.},
\eea
as can be seen from the sound speed formula for tensor perturbations \cite{Kobayashi:2011nu}
\bea
c^2_T=\frac{G_4-XG_{5\phi}-XG_{5X}\ddot{\phi}}{G_4-2XG_{4X}-X\left(G_{5X}\dot{\phi}H-G_{5\phi}\right)}.
\eea

In this section we will derive evolution equations for the remaining parts of the Horndeski Lagrangian, namely,
\be
S[g_{\mu \nu}, \phi] = \int d^{4}x\sqrt{-g}\left[\sum^{4}_{i=2} \mathcal{L}_i\left[g_{\mu \nu},\phi\right] + \mathcal{L}_m \right],
\label{eq:action1}
\ee
where
\bea
\mathcal{L}_2&=& G_2\left(\phi,X\right) \equiv K\left(\phi,X\right),\\
\mathcal{L}_3&=&-G_3\left(\phi,X\right)\Box \phi,\\
\mathcal{L}_4&=&G_4\left(\phi\right) R.
\eea
Here $\phi$ is a scalar field, $X \equiv -\frac{1}{2}\partial_{\mu}\phi\partial^{\mu}\phi$ is a kinetic term, and $\Box \phi \equiv g^{\mu \nu}\nabla_{\mu}\nabla_{\nu}\phi$; $K$, $G_3$ and $G_4$ are free functions of $\phi$ and $X$.\footnote{From now on we define $G_i \equiv G_i\left(\phi,X\right)$, $G_{i,X} \equiv G_{iX} \equiv \frac{\partial G_i}{\partial X}$ and $G_{i,\phi} \equiv G_{i\phi} \equiv \frac{\partial G_i}{\partial \phi}$ where $i=2,3,4$.} Since we are mainly interested in the late-time dynamics of the Universe, hereafter we will further assume $\mathcal{L}_m$ is the Lagrangian of a CDM component. As has been mentioned in \cite{Frusciante:2018}, although the functions $K$, $G_3$ and $G_4$  are able to modify the background with a general dependence on $X$ and $\phi$, this does not hold at the perturbations level. For instance, $K(\phi,X)$ encloses the k-essence and quintessence theory and is partly responsible for the background and the perturbations, however $K(\phi)$ does not contribute to the perturbations.

The term $G_3(\phi,X)$ includes the kinetic gravity braiding with $G_{3X}\neq 0$ being in charge of combining the kinetic term of the scalar and the metric, but the term $G_3(\phi)$ only modifies the background as a dynamical dark energy. Finally, $G_4$ is the only function that is able to modify the non-minimal coupling of the scalar to the Ricci curvature.

Among the theories embedded in the action \eqref{eq:action1} one finds, for example:
\begin{itemize}
\item  \textbf{f(R) theories.} When interpreted as a non-minimal coupled scalar field, these theories can be written using \cite{Chiba:2003ir}
\bea
K &=& -\frac{Rf_{,R}-f}{2\kappa}, \label{eq:fR-horndeski-1}\\
G_4 &=&\frac{\phi}{2\sqrt{\kappa}}, \label{eq:fR-horndeski}
\eea
where $\phi \equiv \frac{f_{,R}}{\sqrt{\kappa} }$ has units of mass and $f_{,R} \equiv \dfrac{df}{dR}$.
\item \textbf{Brans-Dicke theories.} In our notation we have
\bea
K &=& \frac{\omega_{BD} X}{\phi \sqrt{\kappa}}-V(\phi), \\
G_4 &=& \frac{\phi}{2 \sqrt{\kappa}},
\eea
where $V(\phi)$ is the field potential and $\omega_{BD}$ is the Brans-Dicke parameter \cite{Brans:1961sx}.
\item \textbf{Kinetic gravity braiding.} This kind of scalar-tensor models exhibit mixing of scalar and tensor kinetic terms \cite{Deffayet:2010qz} and can be written as
\bea
K &=&K(X), \\
G_3 &=& G_3(X), \\
G_4 &=& \frac{1}{2\kappa}.
\eea
\end{itemize}

\begin{itemize}
\item \textbf{Non-minimal coupling (NMC) model \cite{Quiros:2019ktw}.} In our notation and for a coupling constant $\zeta$
\bea
K &=& \omega(\phi)X-V(\phi),\\
G_4 &=& \left(\frac{1}{2\kappa}-\frac{\zeta \phi^2}{2}\right), \\
G_3 &=& 0. 
\eea
In the context of inflation, a Higgs-like inflation model corresponds to $\omega(\phi)=1$, $V(\phi)=\lambda\left(\phi^2-\nu^2\right)^2/4$.
\item \textbf{Cubic Galileon \cite{Quiros:2019ktw}.} The simplest case is when
\bea
K &=& -X,\\
G_3 &\propto&  X,\\ 
G_4 &=& \frac{1}{2\kappa},
\eea

\item \textbf{4-dimensional static and spherical symmetric solution of Black Hole with scalar hair \cite{Fang:2018vog}.}
\bea
K &=& X,\\
G_3 &=& -\frac{\alpha \log(-X)}{\sqrt{\kappa}}, \\
G_4 &=& \frac{1}{2\kappa}.
\eea

\end{itemize}
As previously done for the Einstein-Hilbert action \eqref{eq:action-GR}, here we apply the principle of least action to \eqref{eq:action1} in order to find evolution equations for both the gravitational field and the scalar field. Varying Eq. \eqref{eq:action1} with respect to the metric and the scalar field one finds\footnote{See Appendix \ref{Section:appendix-A} for a derivation of the field equations.} \cite{Kobayashi:2011nu}
\bea
& & \delta \left(\sqrt{-g}\sum^{4}_{i=2} \mathcal{L}_i\right)=\sqrt{-g}\left[\sum^{4}_{i=2}\mathcal{G}^{i}_{\mu \nu}\delta g^{\mu \nu} \right. \nn \\
& & \left. +\sum^{4}_{i=2}\left(P_{\phi}^i-\nabla^{\mu}J_{\mu}^i\right)\delta \phi\right]+ \text{total derivative},
\eea
which allows us to find the field equations. First, the gravitational field equation is given by
\be
\label{eq:aa}
\sum^{4}_{i=2}\mathcal{G}^{i}_{\mu \nu}=\frac{1}{2} T_{\mu \nu}^{(m)},
\ee
where we have defined
\bea
\mathcal{G}^2_{\mu \nu}&=&-\frac{1}{2}K_{X}\nabla_{\mu}\phi\nabla_{\nu}\phi-\frac{1}{2}Kg_{\mu \nu} \label{eq:def2}\\
\mathcal{G}^3_{\mu \nu}&=&\frac{1}{2}G_{3X}\Box\phi\nabla_{\mu}\phi\nabla_{\nu}\phi+\nabla_{(\mu}G_3\nabla_{\nu)}\phi \nn \\
&-&\frac{1}{2}g_{\mu \nu}\nabla_{\lambda}G_3\nabla^{\lambda}\phi\\
\mathcal{G}^4_{\mu \nu}&=&G_4G_{\mu \nu}+g_{\mu \nu}\left(G_{4\phi}\Box\phi-2XG_{4\phi\phi}\right)-G_{4\phi}\nabla_{\mu}\nabla_{\nu}\phi  \nn \\
&-& G_{4\phi\phi}\nabla_{\mu}\phi\nabla_{\nu}\phi, \label{eq:def4}
\eea
and $T_{\mu \nu}^{(m)}$ is the energy-momentum tensor of a CDM component. Note that from Eq. \eqref{eq:aa} we retrieve the GR field equations \eqref{eq:EE} if we set $K=G_3=0$ and $G_4=\frac{1}{2\kappa}$. Second, the scalar field equation reads
\begin{equation}
\label{eq:abb}
\nabla^{\mu}\left(\sum^4_{i=2}J^i_{\mu}\right)=\sum^{4}_{i=2}P^i_{\phi},
\end{equation}
where
\bea
P^2_{\phi}&=&K_{\phi}\\
P^3_{\phi}&=&\nabla_{\mu}G_{3\phi}\nabla^{\mu}\phi,\\
P^4_{\phi}&=&G_{4\phi}R, \\
J^2_{\mu}&=&-\mathcal{L}_{2X}\nabla_{\mu}\phi\\
J^3_{\mu}&=&-\mathcal{L}_{3X}\nabla_{\mu}\phi+G_{3X}\nabla_{\mu}X+2G_{3\phi}\nabla_{\mu}\phi,\\
J^4_{\mu}&=&0.
\eea

As it is mentioned in Ref. \cite{Kobayashi:2011nu}, one could think $\nabla^{\mu}J^i_{\mu}$ leads to higher than  second-order derivatives. However, this is not the case since commutations of higher derivatives can be substituted by the curvature tensor and are hence canceled. In particular, one can prove that
\bea
\label{eq:comm}
& & \nabla_{\mu}\left(\Box \phi \nabla^{\mu}\phi+\nabla^{\mu}X\right)=\left(\Box \phi\right)^2-\left(\nabla_{\alpha}\nabla_{\beta}\phi\right)^2 \nn \\
& & -R_{\mu \nu}\nabla^{\mu}\phi \nabla^{\nu}\phi,
\eea
which will be of paramount importance when we will discuss perturbation equations.

It is possible to find a relatively simple expression for the scalar field equation \eqref{eq:abb} if we consider the case $i=3$, namely,
\bea
&0& = 2G_{3\phi}\Box \phi+\nabla^{\mu}G_{3\phi}\nabla_{\mu}\phi+\nabla_{\mu}\phi\nabla^{\mu}G_{3X}\Box \phi  \label{eq:i3} \\
&+& \underbrace{\nabla^{\mu}\left(G_{3X}\nabla_{\mu}X\right)+G_{3X}\left(\Box \phi\right)^2+G_{3X}\nabla_{\mu}\phi\nabla^{\mu}\Box \phi}. \nn
\eea
The terms on top of the brace in Eq. \eqref{eq:i3} can be expanded as
\bea
\label{eq:plt}
& & \nabla^{\mu}G_{3X}\nabla_{\mu}X \\
& & +\underbrace{G_{3X}\Box X+G_{3X}\left(\Box \phi\right)^2+G_{3X}\nabla^{\mu}\phi\nabla_{\mu}\Box \phi}=0, \nn
\eea
and the terms on top of the brace in Eq. \eqref{eq:plt} can in turn be written as
\bea
\label{eq:cvb}
& & G_{3X}\left[\nabla^{\mu}\phi\nabla_{\mu}\Box \phi+\left(\Box \phi\right)^2+\Box X\right]= \nn \\
& & G_{3X}\left[\nabla_{\mu}\left(\Box \phi \nabla^{\mu}\phi+\nabla^{\mu}X\right)\right].
\eea
Using Eq. \eqref{eq:comm} in Eq. \eqref{eq:cvb} we find
\bea
& & G_{3X}\left[\nabla^{\mu}\phi\nabla_{\mu}\Box \phi+\left(\Box \phi\right)^2+\Box X\right]= \nn \\
& & G_{3X}\left[\left(\Box \phi\right)^2-\left(\nabla_{\alpha}\nabla_{\beta}\phi\right)^2-R_{\mu \nu}\nabla^{\mu}\phi \nabla^{\nu}\phi\right],
\eea
and the scalar field equation \eqref{eq:abb} can be written as
\bea
& & -\nabla_{\mu}K_{X}\nabla^{\mu}\phi - K_{X}\Box \phi - K_{\phi} + 2G_{3\phi}\Box \phi + \nabla_{\mu}G_{3\phi}\nabla^{\mu}\phi \nn \\
& & +\nabla_{\mu}G_{3X}\Box \phi \nabla^{\mu}\phi + \nabla_{\mu}G_{3X}\nabla^{\mu}X + G_{3X}\left[\left(\Box \phi\right)^2 - \right. \nn \\
& & \left. \left(\nabla_{\alpha}\nabla_{\beta}\phi\right)^2 - R_{\mu \nu}\nabla^{\mu}\phi \nabla^{\nu}\phi\right]-G_{4\phi}R=0.
\label{eq:scalar-field-equation-horndeski}
\eea

In what follows, in order to simplify the notation we will denote the kinetic term of the scalar field evaluated at the background simply by $X$ and its linear order perturbation as $\delta X$.

\subsection{Background}

Thus far the discussion of the field equations has been quite general. Now, as previously done in Sec.~\ref{Section:theoretical-framework}, we assume a perturbed FLRW as given in Eq.~\eqref{eq:FRWpert}. If we consider only zero order quantities in the gravitational field equation \eqref{eq:aa}, we obtain
\bea
\mathcal{E}\equiv \sum^{4}_{i=2}\mathcal{E}_i&=&-\rho_m,\label{eq:epsi}\\
\mathcal{P}\equiv \sum^{4}_{i=2}\mathcal{P}_i&=&0,\label{eq:pp}
\eea
where
\begin{align}
\mathcal{E}_2 &\equiv 2XK_{X}-K, \\
\mathcal{E}_3 &\equiv 6X\dot{\phi}HG_{3X}-2XG_{3\phi}, \\
\mathcal{E}_4 &\equiv -6H^2G_4-6H\dot{\phi}G_{4\phi}, \\
\mathcal{P}_2 &\equiv K, \\
\mathcal{P}_3 &\equiv -2X\left(G_{3\phi}+\ddot{\phi}G_{3X}\right), \\
\mathcal{P}_4 &\equiv 2\left(3H^2+2\dot{H}\right)G_4+2\left(\ddot{\phi}+2H\dot{\phi}\right)G_{4\phi}+2\dot{\phi}^2G_{4\phi\phi}.
\end{align}
Eqs.~\eqref{eq:epsi}-\eqref{eq:pp} are the modified Friedmann equations describing the background evolution of the Universe. Collecting terms they respectively read
\bea
& & 2XK_{X}-K+6X\dot{\phi}HG_{3X}-2XG_{3\phi}-6H^2G_4 \nn \\
& & -6H\dot{\phi}G_{4\phi}+\rho_m=0,
\label{eq:friedmann1-horndeski}
\eea
\bea
& & K - 2X\left(G_{3\phi} + \ddot{\phi}G_{3X}\right) + 2\left(3H^2+2\dot{H}\right)G_4 \nn \\
& & +2\left(\ddot{\phi}+2H\dot{\phi}\right)G_{4\phi}+2\dot{\phi}^2G_{4\phi\phi}=0.
\label{eq:friedmann2-horndeski}
\eea
Note that from Eqs.~\eqref{eq:friedmann1-horndeski}-\eqref{eq:friedmann2-horndeski} we respectively retrieve the Friedmann equations \eqref{eq:Friedmann-zero-order-1}-\eqref{eq:Friedmann-zero-order-2} if we set $K=G_3=0$ and $G_4=\frac{1}{2\kappa}$. Rearranging terms in Eqs. \eqref{eq:friedmann1-horndeski}-\eqref{eq:friedmann2-horndeski} we can define an effective DE density
\bea
\label{eq:dde}
& & \bar{\rho}_{DE} = \dot{\phi}^2K_{X}- K + 3\dot{\phi}^3HG_{3X} -\dot{\phi}^2G_{3\phi} \nn \\
& & + 3H^2\left(\frac{1}{\kappa}-2G_{4}\right)-6H\dot{\phi}G_{4\phi},
\eea
and an effective DE pressure
\bea
\label{eq:ppe}
& & \bar{P}_{DE}= K - \dot{\phi}^2\left(G_{3\phi} + \ddot{\phi}G_{3X}\right)+ 2\dot{\phi}^2G_{4\phi\phi}  \nn \\
& &  + 2\left(\ddot{\phi} +  2H\dot{\phi}\right)G_{4\phi} - \left(3H^2 + 2\dot{H}\right)\left(\frac{1}{\kappa}-2G_{4}\right),~~~~~~
\eea
in such a way that we can write the modified Friedmann equations Eqs. \eqref{eq:friedmann1-horndeski}-\eqref{eq:friedmann2-horndeski} as
\begin{align}
3H^2=\kappa\left(\bar{\rho}_{DE}+\rho_m\right) \label{eq:327} \\
 -\left(2\dot{H}+3H^2\right)=\kappa\bar{P}_{DE} ,
\end{align}
where we are assuming that matter is pressureless $\bar{P}_m=0$ as indicated by current constraints \cite{PhysRevLett.120.221102}. The effective DE density and pressure in Eqs. \eqref{eq:dde}-\eqref{eq:ppe} allow us to define an effective DE equation of state as
\begin{widetext}
\begin{equation}
\label{eq:ww}
    w_{DE}=\frac{K-\dot{\phi}^2\left(G_{3\phi}+\ddot{\phi}G_{3X}\right)-\left(3H^2+2\dot{H}\right)\left(\frac{1}{\kappa}-2G_{4}\right)+2\left(\ddot{\phi}+2H\dot{\phi}\right)G_{4\phi}+2\dot{\phi}^2G_{4\phi\phi}}{\dot{\phi}^2K_{X}-K+3\dot{\phi}^3HG_{3X}-\dot{\phi}^2G_{3\phi}+3H^2\left(\frac{1}{\kappa}-2G_{4}\right)-6H\dot{\phi}G_{4\phi}}.
\end{equation}
\end{widetext}

Let us now consider the scalar field equation \eqref{eq:scalar-field-equation-horndeski} and only keep
zero order quantities, that is to say,
\bea
\label{eq:scalar-field-equation-zeroorder}
& & K_{\phi}-\left(K_{X}-2G_{3\phi}\right)\left(\ddot{\phi}+3H\dot{\phi}\right)-K_{\phi X}\dot{\phi}^2 -\nn \\
& & K_{XX}\ddot{\phi}\dot{\phi}^2 + G_{3\phi\phi}\dot{\phi}^2 + G_{3\phi X}\dot{\phi}^2\left(\ddot{\phi} - 3H\dot{\phi}\right) - \nn \\
& & 3G_{3X}\left(2H\dot{\phi}\ddot{\phi} + 3H^2\dot{\phi}^2+\dot{H}\dot{\phi}^2\right) - 3G_{3XX}H\dot{\phi}^3\ddot{\phi} \nn \\
& & + 6G_{4\phi}\left(2H^2+\dot{H}\right)=0,
\eea
which fully agrees with \cite{Kimura:2010di}. Note that defining
\bea
J_{\mu} &\equiv & \sum_{i=2}^4 J^i_{\mu}, \\
P_{\phi} & \equiv & \sum^{4}_{i=2}P^i_{\phi}
\eea
we can write the scalar field equation \eqref{eq:abb} as
\begin{equation}
\nabla_{\mu}J^{\mu}=P_{\phi}
\end{equation}
and it becomes clear that there exists a Noether current for Lagrangians invariant under constant shifts of the field $\phi \rightarrow \phi + c$ \cite{Deffayet:2010qz}, namely,
\begin{equation}
J_{\mu}=\left(\mathcal{L}_{2X}+\mathcal{L}_{3X}-2G_{3\phi}\right)\nabla_{\mu}\phi-G_{3X}\nabla_{\mu}X.
\end{equation}
Taking into consideration that $X=\frac{1}{2}\dot{\phi}^2$, the charge density of the Noether current can be written as
\begin{equation}
\label{eq:jj}
    J \equiv J_0=\dot{\phi}\left(K_X-2G_{3\phi}+3H\dot{\phi}G_{3X}\right),
\end{equation}
so that the scalar field equation is given by the simple expression
\begin{equation}
    \dot{J}+3HJ=P_{\phi}.
\end{equation}
When $P_{\phi}=0$ then it is easy to see that the solution to the previous equation is
\be
J=\frac{J_c}{a^3},
\ee
where $J_c$ is a constant. When $J_c=0$, then the system is on the attractor solution, but when $J_c\neq0$ then the system is not on the attractor and as we will see in Sec.~\ref{kgb-attractor} interesting dynamics may arise.

\subsection{Linear perturbations}

Considering only first order quantities in the gravitational field equations \eqref{eq:aa} one obtains \cite{DeFelice:2011hq,Matsumoto:2018dim}
\bea
& & A_1\dot{\Phi}+A_2\dot{\delta \phi}+A_3\frac{k^2}{a^2}\Phi+A_4\Psi +\left(A_6\frac{k^2}{a^2} -\mu \right)\delta \phi \nn \\
& & -\rho_m\delta_m=0, \label{eq:1}\\
& & C_1\dot{\Phi}+C_2\dot{\delta \phi}+C_3\Psi+C_4\delta \phi-\frac{a \rho_m V_m}{k^2}=0, \label{eq:field-equation-horndeski-3}\\
& & B_1\ddot{\Phi}+B_2\ddot{\delta \phi}+B_3\dot{\Phi}+B_4\dot{\delta \phi}+B_5\dot{\Psi}+B_6\frac{k^2}{a^2}\Phi \nn \\
& & +\left(B_7\frac{k^2}{a^2}+3\nu\right)\delta \phi+\left(B_8\frac{k^2}{a^2}+B_9\right)\Psi=0, \label{eq:2}\\
& & G_4\left(\Psi+\Phi\right)+G_{4\phi}\delta \phi=0. \label{eq:field-equation-horndeski-4}
\eea
Note that when $K = G_3 = 0$ and $G_4 = \frac{1}{2\kappa}$, Eqs.~\eqref{eq:1}-\eqref{eq:field-equation-horndeski-4} respectively correspond to the GR limit given by Eqs.~\eqref{eq:phiprimeeq}-\eqref{eq:anisoeq} with $\sigma_m=0$.

If we now consider the scalar field equation \eqref{eq:scalar-field-equation-horndeski} and take into account only first order quantities we find
\bea
\label{eq:3}
& & D_1\ddot{\Phi}+D_2\ddot{\delta \phi}+D_3\dot{\Phi}+D_4\dot{\delta \phi}+D_5\dot{\Psi}+\left(D_7\frac{k^2}{a^2}+D_8\right)\Phi \nn \\
& & +\left(D_9\frac{k^2}{a^2}-M^2\right)\delta \phi +\left(D_{10}\frac{k^2}{a^2}+D_{11}\right)\Psi=0.
\eea
Expressions for the coefficients $A_i$, $\mu$, $\nu$, $B_i$, $C_i$ and $D_i$ can be found in Appendix \ref{Section:appendix-B} and are in agreement with those found in \cite{DeFelice:2011hq,Matsumoto:2018dim}, except for $D_8$ which is actually equal to zero  as can be seen by using the expression found in \cite{Matsumoto:2018dim} and using the background equations of motion for the scalar field.


\section{The effective fluid approach \label{Section:EFA}}

We have seen in the previous section that the gravitational field equations for the Horndeski Lagrangian can be written in such a way that they resemble those found in Sec. \ref{Section:theoretical-framework} where we assumed GR and a perfect fluid. Indeed, defining an effective DE density and pressure given by Eqs.~\eqref{eq:dde}-\eqref{eq:ppe} makes it possible to obtain an effective DE equation of state (see Eq. \eqref{eq:ww}). As mentioned in Sec. \ref{Section:Introduction}, a fluid can be described by its equation of state, sound speed, and anisotropic stress, so in what follows we will explicitly derive those quantities.

In this section we will present relatively simple expressions for the effective DE sound speed and anisotropic stress under the subhorizon and quasistatic approximations. Actually, by defining an effective DE fluid we are considering a DE effective energy-momentum tensor $T_{\mu\nu}^{DE}$ obtained via the gravitational field equations \eqref{eq:aa} and defined explicitly as follows:
\bea
G_{\mu\nu}&=&\kappa\left(T_{\mu\nu}^{(m)}+T_{\mu\nu}^{(DE)}\right),\nn \\
\kappa T_{\mu\nu}^{(DE)}&=&G_{\mu\nu}-2\kappa \sum^{4}_{i=2}\mathcal{G}^{i}_{\mu \nu}.
\eea
Since we are taking into consideration expressions up to linear order, $T_{\mu\nu}^{DE}$ also contains small perturbations which allow us to define quantities such as DE effective perturbations in the pressure, density, and velocity. These can be extracted from the DE effective energy-momentum tensor $T_{\mu\nu}^{DE}$ by considering the decomposition of the tensor into its components, given by Eqs.~\eqref{eq:effectTmn1}-\eqref{eq:effectTmn}. Qualitatively, these expressions have the following structure:
\bea
\frac{\delta P_{DE}}{\bar{\rho}_{DE}}&=&(...)\delta \phi+(...)\dot{\delta \phi}+(...)\ddot{\delta \phi}+(...)\Psi \nn \\
&+&(...)\dot{\Psi} +(...)\Phi +(...)\dot{\Phi}+(...)\ddot{\Phi},\label{eq:effdenp0}\\
\delta_{DE}&=&(...)\delta \phi+(...)\dot{\delta \phi}+(...)\Psi \nn \\
&+& (...)\Phi
+(...)\dot{\Phi},\label{eq:effprp0} \\
V_{DE}&=&(...)\delta \phi+(...)\dot{\delta \phi}+(...)\Psi \nn \\
&+&(...)\Phi
+(...)\dot{\Phi}.\label{eq:efftheta0}
\eea
where $(...)$ indicates expressions which might be cumbersome. It is therefore very helpful to work out these expressions under the subhorizon and quasistatic approximations in order to gain a better understanding.

We have explained in great detail the way we carry out the subhorizon and quasistatic approximations in our previous paper (see Sec. II.A.1 in Ref. \cite{Arjona:2018jhh}), but in a nutshell, the former refers to only considering modes deep in the Hubble radius, i.e. those for which $k^2 \gg a^2 H^2$, while the latter refers to neglecting derivatives of the potentials during matter domination as they are roughly constant but also terms of similar order as $\partial_\eta\sim 1/\eta\sim aH(a)$. For example, the perturbation in the Ricci scalar is
\bea
\delta R&=&-\frac{12  (\mathcal{H}^2+\dot{\mathcal{H}})}{a^2}\Psi-\frac{4 k^2}{a^2}\Phi+\frac{2 k^2 }{a^2}\Psi \nn \\
&-&\frac{18 \mathcal{H} }{a^2}\dot{\Phi}-\frac{6 \mathcal{H} }{a^2}\dot{\Psi}-\frac{6 \ddot{\Phi}}{a^2}, \nn\label{eq:ricciexact}\\
&\simeq & -\frac{4 k^2}{a^2}\Phi+\frac{2 k^2}{a^2}\Psi.\nn
\eea
Following the same procedure and applying the subhorizon approximation to the linearized gravitational field equations \eqref{eq:1},\eqref{eq:2}, and to the linearized scalar field equation \eqref{eq:3}, one finds, respectively,
\bea
\label{eq:bla1}
& & A_3\frac{k^2}{a^2}\Phi+A_6\frac{k^2}{a^2}\delta \phi-\kappa \rho_m\delta_m \simeq 0,\\
\label{eq:trace}
& & B_6\frac{k^2}{a^2}\Phi +  B_8\frac{k^2}{a^2}\Psi + B_7\frac{k^2}{a^2}\delta \phi  \simeq 0,\\
\label{eq:bla}
& & D_7\frac{k^2}{a^2}\Phi+\left(D_9\frac{k^2}{a^2}-M^2\right)\delta \phi+D_{10}\frac{k^2}{a^2}\Psi \simeq 0.
\eea
 Note that since $B_7=4G_{4\phi}$ and $B_6=B_8$ (see Appendix \ref{Section:appendix-B}), Eq. (\ref{eq:trace}) leads to no anisotropic stress $\Phi=-\Psi$ when $G_4$ is a constant.

Solving Eqs.~\eqref{eq:bla1}-\eqref{eq:bla} for $\Phi$, $\Psi$ and $\delta \phi$ one finds
\bea
& & \frac{k^2}{a^2}\Psi=-\frac{\kappa}{2}\frac{G_{\textrm{eff}}}{G_N}\bar{\rho}_m\delta,\\
& & \frac{k^2}{a^2}\Phi=\frac{\kappa}{2} Q_{\textrm{eff}}\bar{\rho}_m\delta, \\
& & \delta \phi=\frac{\left(A_6B_6-B_6B_7\right)\rho_m\delta_m}{\left(A^2_6B_6-2A_6B_6B_7+B^2_6D_9\right)\frac{k^2}{a^2}-B^2_6M^2},~~~
\eea
where $G_{\textrm{eff}}$ and $Q_{\textrm{eff}}$ are Newton's effective constant
\bea
& & \frac{G_{\textrm{eff}}}{G_N} =\frac{2\left[\left(B_6D_9-B^2_7\right)\frac{k^2}{a^2}-B_6M^2\right]}{\left(A^2_6B_6+B^2_6D_9-2A_6B_7B_6\right)\frac{k^2}{a^2}-B^2_6M^2},~~~~~~\label{eq:ODEGeff}\\
& & Q_{\textrm{eff}} =\frac{2\left[\left(A_6B_7-B_6D_9\right)\frac{k^2}{a^2}+B_6M^2\right]}{\left(A^2_6B_6+B^2_6D_9-2A_6B_7B_6\right)\frac{k^2}{a^2}-B^2_6M^2},~~~~~~
\eea
and we make use of the following correspondence $A_3=B_6=B_8$, $D_7=B_7$ and $D_{10}=A_6$ (see Appendix \ref{Section:appendix-B}). One can also define the following anisotropic stress parameters
\begin{align}
    \eta &\equiv \frac{\Psi+\Phi}{\Phi}=\frac{\left(A_6-B_7\right)B_7\frac{k^2}{a^2}}{\left(A_6B_7-B_6D_9\right)\frac{k^2}{a^2}+B_6M^2},\\
    \gamma &\equiv -\frac{\Phi}{\Psi}=\frac{\left(A_6B_7-B_6D_9\right)\frac{k^2}{a^2}+B_6M^2}{\left(B^2_7-B_6D_9\right)\frac{k^2}{a^2}+B_6M^2}.
\end{align}
The aforementioned expressions for Newton's effective constant and the anisotropic stress parameters are in agreement with the ones in Ref.~\cite{DeFelice:2011hq}.

The subhorizon approximation is also useful as the evolution equations for the growth of matter perturbations $\delta_m$ given by Eqs.~\eqref{Eq:evolution-delta}-\eqref{Eq:evolution-V} can be reduced to a single differential equation, where the variable $G_{\textrm{eff}}$ plays a primary role:
\be
\delta_m''(a)+\left(\frac3{a}+\frac{H'(a)}{H(a)}\right)\delta_m'(a)-\frac32 \frac{\Omega_{m,0}G_{\textrm{eff}}/G_N}{a^5 H(a)^2/H_0^2} \delta_m(a)=0, \label{eq:ODE-growth}
\ee
with $G_{\textrm{eff}}$ given by Eq.~\eqref{eq:ODEGeff} and initial conditions $\delta_m(a_i)=a_i$ and $\delta_m'(a_i)=1$ for an initial value for the scale factor $a_i$ deep in the matter era.

In what follows, we will present the effective DE perturbations under the subhorizon and quasistatic approximations for two classes of models: those in which there is DE anisotropic stress and those where DE anisotropic stress vanishes.

\subsection{Horndeski models with DE anisotropic stress}
\label{Section:Horndeski_anisotropy}

We now apply the subhorizon and quasistatic approximations in Eqs.~\eqref{eq:effdenp0}-\eqref{eq:efftheta0} using the same prescription as in Ref.~\cite{Arjona:2018jhh}. We also found, in agreement with Ref.~\cite{Kimura:2010di}, that the quasistatic approximation breaks down for this model due to the rapid oscillations of the scalar field, so if we eliminate the scalar field, then this can slightly increase the accuracy of the numerical solutions of the effective fluid equations. To eliminate $\delta \phi$ and its derivatives, we use Eq.~\eqref{eq:field-equation-horndeski-4} and insert the resulting equations in Eqs.~\eqref{eq:effdenp0}-\eqref{eq:efftheta0}.

Then, by keeping the dominant $k^2$ terms (the subhorizon approximation) and dropping time derivatives of the potentials (the quasistatic approximation) in Eqs.~\eqref{eq:effdenp0}-\eqref{eq:efftheta0} we find
\bea
\frac{\delta P_{DE}}{\bar{\rho}_{DE}} & \simeq & \frac{1}{3\mathcal{F}_4}\frac{\frac{k^4}{a^4}\mathcal{F}_1+\frac{k^2}{a^2}\mathcal{F}_2+\mathcal{F}_3}{\frac{k^4}{a^4}\mathcal{F}_5+\frac{k^2}{a^2}\mathcal{F}_6}\frac{\bar{\rho}_m}{\bar{\rho}_{DE}} \delta_m,\label{eq:effpres} \\
\delta_{DE} & \simeq & \frac{\frac{k^4}{a^4}\mathcal{F}_7+\frac{k^2}{a^2}\mathcal{F}_8+\mathcal{F}_9}{\frac{k^4}{a^4}\mathcal{F}_5+\frac{k^2}{a^2}\mathcal{F}_{6}}\frac{\bar{\rho}_m}{\bar{\rho}_{DE}} \delta_m,\label{eq:effder} \\
V_{DE} & \simeq & a\frac{\frac{k^2}{a^2}\mathcal{F}_{10}+\mathcal{F}_{11}}{\frac{k^2}{a^2}\mathcal{F}_5+\mathcal{F}_6}\frac{\bar{\rho}_m}{\bar{\rho}_{DE}} \delta_m,\label{eq:efftheta}
\eea
for the effective DE pressure perturbation, effective DE density perturbation, and effective DE velocity perturbation, respectively (the interested reader can find the expressions for $\mathcal{F}_i$ in Appendix \ref{Section:appendix-B}). It is now also possible to obtain an expression for the effective DE anisotropic stress under the subhorizon approximation
\bea
\pi_{DE}&=& \frac{\frac{k^2}{a^2} (\Phi+\Psi)}{\kappa~ \bar{\rho}_{DE}}
\simeq \frac{\frac{k^2}{a^2}\mathcal{F}^2_4B_7\left(B_7-A_6\right)}{\frac{k^2}{a^2}\mathcal{F}_5+\mathcal{F}_6}\frac{\bar{\rho}_m}{\bar{\rho}_{DE}} \delta_m\nn\\
&\simeq &\frac{\frac{k^4}{a^4}\mathcal{F}^2_4B_7\left(B_7-A_6\right)}{\frac{k^4}{a^4}\mathcal{F}_7+\frac{k^2}{a^2}\mathcal{F}_8+\mathcal{F}_9}\delta_{DE}.\label{eq:effpi}
\eea

Having found expressions for the effective DE equation of state (see Eq.~\eqref{eq:ww}) and the effective DE anisotropic stress (Eq. \eqref{eq:effpi}), the only missing ingredient for an effective fluid description of the Horndeski Lagrangian is the sound speed. This quantity can easily be found using our equations for the effective DE pressure perturbation \eqref{eq:effpres} and the effective DE density perturbation \eqref{eq:effder}. The DE sound speed reads
\bea
c_{s,DE}^2 & \equiv & \frac{\delta P_{DE}}{\delta \rho_{DE}} \nn \\
& = & \frac{1}{3}\frac{\frac{k^4}{a^4}\mathcal{F}_1+\frac{k^2}{a^2}\mathcal{F}_2+\mathcal{F}_3}{\frac{k^4}{a^4}\mathcal{F}_7+\frac{k^2}{a^2}\mathcal{F}_8+\mathcal{F}_9}.
\label{eq:DE-sound-speed}
\eea
Due to the presence of anisotropic stress, perturbations on subhorizon scales in the effective DE fluid are not driven by the sound speed \eqref{eq:DE-sound-speed}, but by an effective DE sound speed defined as \cite{Cardona:2014iba,Arjona:2018jhh}
\bea
c_{s,eff}^2 & \equiv & c_{s,DE}^2-\frac{2}{3}\pi_{DE}/\delta_{DE} \label{eq:eff-DE-sound-speed} \\
& = &\frac{1}{3}\frac{\frac{k^4}{a^4}\left(\mathcal{F}_1-2\mathcal{F}^2_4B_7\left(B_7-A_6\right)\right)+\frac{k^2}{a^2}\mathcal{F}_2+\mathcal{F}_3}{\frac{k^4}{a^4}\mathcal{F}_7+\frac{k^2}{a^2}\mathcal{F}_8+\mathcal{F}_9}. \nn
\eea

Finally, it is clear that for the cosmological constant model, i.e. $\mathcal{L}_2=-\frac{\Lambda}{\kappa}$, $\mathcal{L}_3=0$, $\mathcal{L}_4=\frac{1}{2\kappa}R$, $\mathcal{L}_5=0$, we have $K=-\frac{\Lambda}{\kappa}$, $G_3=0$, $G_4=\frac{1}{2\kappa}$ and $G_5=0$, which implies that $w_{DE}=-1$ and $\left(\delta P_{DE},\delta \rho_{DE},\pi_{DE}\right)=\left(0,0,0\right)$ as expected.

\subsubsection{f(R) models}
Thus far we have kept the discussion quite general, that is to say, we did not specify any function in the Horndeski Lagrangian \eqref{eq:action1}. To mention an example, we will present the results for $f(R)$ models. With the definitions in Eqs. \eqref{eq:fR-horndeski-1}-\eqref{eq:fR-horndeski} and using units where $\kappa=1$, one obtains

\bea
B_7 & = & 2A_6 = 2, \quad B_6  =  B_8 = 2\phi, \quad D_9  =  0, \nn \\
\mathcal{F}_1 & = & \mathcal{F}_4 = -1/2, \quad \mathcal{F}_2  =  -\frac{15\ddot{F}}{4}, \quad \mathcal{F}_3  =  -\frac{2F\ddot{F}}{4F_{,R}}, \nn \\
\mathcal{F}_5 & = & -\frac{3F}{2}, \quad \mathcal{F}_6  =  -\frac{F^2}{2F_{,R}}, \quad \mathcal{F}_7  = -1+\frac{3F}{2}, \nn \\
\mathcal{F}_8 & = & \frac{\left(F-1\right)F}{2F_{,R}}, \quad \mathcal{F}_9 = 0, \quad \mathcal{F}_{10} =  -\frac{3\dot{F}}{2}, \nn \\
\mathcal{F}_{11} & = & -\frac{F\dot{F}}{4F_{,R}}, \quad M^2=-K_{\phi\phi}=\frac{1}{2f_{RR}},
\eea
where
\bea
K_{\phi}&=&\frac{dK}{d\phi}=\frac{dK}{dR}\frac{dR}{d\phi}\nn\\
&=&\frac{1}{2f_{,RR}}\left(Rf_{,RR}\right)=-\frac{R}{2},\\
K_{\phi\phi}&=&\frac{d}{d\phi}\left(\frac{dK}{d\phi}=-\frac{R}{2}\right)\nn \\ &=&\frac{1}{f_{,RR}}\frac{d}{dR}\left(-\frac{R}{2}\right)=-\frac{1}{2f_{RR}},~~~
\eea
and $F=f_{,R}$, $F_{,R}=f_{,RR}$. Then, the effective DE fluid quantities read
\bea
\frac{\delta P_{DE}}{\bar{\rho}_{DE}}&\simeq&\frac{1}{3F}\frac{2\frac{k^2}{a^2}\frac{F_{,R}}{F}+3(1+5\frac{k^2}{a^2}\frac{F_{,R}}{F})\ddot{F}k^{-2}}{1+3\frac{k^2}{a^2}\frac{F_{,R}}{F}}\frac{\bar{\rho}_m}{\bar{\rho}_{DE}} \delta_m \label{eq:effpres1}, \nn \\
& & \\
\delta_{DE}&\simeq&\frac{1}{F}\frac{1-F+\frac{k^2}{a^2}(2-3F)\frac{F_{,R}}{F}}{1+3\frac{k^2}{a^2}\frac{F_{,R}}{F}}\frac{\bar{\rho}_m}{\bar{\rho}_{DE}} \delta_m\label{eq:effder1}, \\
V_{DE} & \simeq & \frac{a\dot{F}}{2F}\frac{1+6\frac{k^2}{a^2}\frac{F_{,R}}{F}}{1+3\frac{k^2}{a^2}\frac{F_{,R}}{F}}\frac{\bar{\rho}_m}{\bar{\rho}_{DE}} \delta_m.\label{eq:vdefR}
\eea
\bea
\pi_{DE} & \simeq & \frac{1}{F}\frac{\frac{k^2}{a^2}\frac{F_{,R}}{F}}{1+3\frac{k^2}{a^2}\frac{F_{,R}}{F}}\frac{\bar{\rho}_m}{\bar{\rho}_{DE}} \delta_m \nn\\
& \simeq & \frac{\frac{k^2}{a^2}\frac{F_{,R}}{F}}{1-F+\frac{k^2}{a^2}(2-3F)\frac{F_{,R}}{F}}\delta_{DE}.
\eea
\be
c_{s,DE}^2\simeq\frac13 \frac{2\frac{k^2}{a^2}\frac{F_{,R}}{F}+3(1+5\frac{k^2}{a^2}\frac{F_{,R}}{F})\ddot{F}k^{-2}}{1-F+\frac{k^2}{a^2}(2-3F)\frac{F_{,R}}{F}},
\ee
\bea
c_{s,eff}^2 & \simeq & \frac{(1+5\frac{k^2}{a^2}\frac{F_{,R}}{F})\ddot{F}k^{-2}}{1-F+\frac{k^2}{a^2}(2-3F)\frac{F_{,R}}{F}}.
\eea
These results are in perfect agreement with our previous work \cite{Arjona:2018jhh}.

\subsection{Horndeski models with no dark energy anisotropic stress}
With the same approach that we followed in (\ref{Section:Horndeski_anisotropy}) we compute the DE perturbations for models where the is no DE anisotropic stress, i.e $\Phi=-\Psi$. With this restriction it is easy to see from Eq. (\ref{eq:field-equation-horndeski-4}) that $G_{4\phi}=0$. Then applying this condition under the subhorizon approximation in Eqs.~\eqref{eq:effdenp0}-\eqref{eq:efftheta0} leads to
\bea
\frac{\delta P_{DE}}{\bar{\rho}_{DE}}&\simeq&\frac{1}{3}\frac{\frac{k^2}{a^2}\hat{\mathcal{F}}_2+\hat{\mathcal{F}}_3}{\frac{k^4}{a^4}\hat{\mathcal{F}}_5+\frac{k^2}{a^2}\hat{\mathcal{F}}_6}\frac{\bar{\rho}_m}{\bar{\rho}_{DE}} \delta_m,\label{eq:effpreskgb}
\\
\delta_{DE}&\simeq&\frac{\frac{k^4}{a^4}\hat{\mathcal{F}}_7+\frac{k^2}{a^2}\hat{\mathcal{F}}_8+\hat{\mathcal{F}}_9}{\frac{k^4}{a^4}\hat{\mathcal{F}}_5+\frac{k^2}{a^2}\hat{\mathcal{F}}_{6}}\frac{\bar{\rho}_m}{\bar{\rho}_{DE}} \delta_m,\label{eq:effderkgb} \\
V_{DE} & \simeq & a\frac{\frac{k^2}{a^2}\hat{\mathcal{F}}_{10}+\hat{\mathcal{F}}_{11}}{\frac{k^2}{a^2}\hat{\mathcal{F}}_5+\hat{\mathcal{F}}_6}\frac{\bar{\rho}_m}{\bar{\rho}_{DE}} \delta_m,\label{eq:effthetakgb}
\eea
and since $\Phi=-\Psi$ the anisotropic parameters read
\begin{align}
    \eta &\equiv \frac{\Psi+\Phi}{\Phi}=0,\\
    \gamma &\equiv -\frac{\Phi}{\Psi}=1,
\end{align}
as expected, while the DE anisotropic stress parameter is zero $\pi_{DE}=0$. Our general expression for the DE sound speed \eqref{eq:DE-sound-speed} reduces in this case to
\begin{equation}
\label{eq:csde}
 c_{s,DE}^2=\frac{\frac{k^2}{a^2}\hat{\mathcal{F}}_2+\hat{\mathcal{F}}_3}{\frac{k^4}{a^4}\hat{\mathcal{F}}_7+\frac{k^2}{a^2}\hat{\mathcal{F}}_8+\hat{\mathcal{F}}_9},
\end{equation}
which is equal to the DE effective sound speed since $\pi_{DE}=0$. Here we will show results for a few specific models embedded in the Horndeski Lagrangian.

\subsubsection{Quintessence}

We can recover the Lagrangian of Quintessence by choosing the following functions
\begin{equation}
K = X-V(\phi), \hspace{5mm} G_4=\frac{1}{2\kappa}
\end{equation}
where $\phi$ is the scalar field, $X$ is the kinetic term defined as $X=-\frac{1}{2}g^{\mu \nu}\partial_{\mu}\phi\partial_{\nu}\phi$ and $V(\phi)$ is the potential. Using a variational approach one finds that the effective pressure, density and velocity perturbations for Quintessence theories are given by
 \bea
\delta P_{DE}&=&\left(\dot{\phi}\dot{\delta \phi}-\Psi\dot{\phi}^2\right)-V_{\phi}\delta\phi, \nn \\
\rho_{DE}\delta_{DE}&=&\left(\dot{\phi}\dot{\delta \phi}-\Psi\dot{\phi}^2\right)+V_{\phi}\delta\phi, \\
V_{DE}& = &\frac{k^2}{a}\dot{\phi}\delta\phi,\label{eq:vdequint}
\eea
and these expressions are in agreement with \cite{Amendola:2015ksp}. Also, the DE anisotropic stress parameter $\pi_{DE}$ is zero since for Quintessence $\Psi=-\Phi$. We find that under the subhorizon approximation
\bea
A_6 = 0, \quad B_6 = -2, \quad D_9 = -K_X, \quad M^2 = -K_{\phi\phi},~~
\eea
so that the effective pressure, density and velocity perturbations for Quintessence theories are given by
 \bea
\frac{\delta P_{DE}}{\bar{\rho}_{DE}}&\simeq&\frac{\dot{\phi}^2}{2k^2/a^2}\frac{\bar{\rho}_m}{\bar{\rho}_{DE}} \delta_m \label{eq:effpressquint},\\
\delta_{DE}&\simeq&\frac{\dot{\phi}^2}{2k^2/a^2}\frac{\bar{\rho}_m}{\bar{\rho}_{DE}} \delta_m,\label{eq:effdequint} \\
V_{DE}& \simeq & 0.
\eea
It is thus straightforward, using Eqs.~\eqref{eq:effpressquint} and \eqref{eq:effdequint}, to see that the DE sound speed is given by
\be
c_{s,DE}^2 = 1.
\ee
Moreover, we also find that in the subhorizon approximation
\bea
\delta\phi&\simeq&0,\nn \\
\Psi &\simeq&-\frac{\bar{\rho}_m\delta_m a^2}{2k^2},
\eea

\subsubsection{K-essence}
In our notation the Lagrangian of K-essence theories is specified by the functions \cite{Scherrer:2004au,dePutter:2007ny}
\begin{equation}
K\left(\phi,X\right) = P\left(\phi,X\right), \hspace{5mm} G_4=\frac{1}{2\kappa},
\end{equation}
 and as usual through the variation of the action it is possible to find expressions for the pressure, density, and velocity perturbations
\bea
\delta P_{DE} & = & P_{\phi}\delta\phi+P_X\left(\dot{\phi}\dot{\delta \phi}-\dot{\phi}^2\Psi\right),\\
\rho_{DE}\delta_{DE}&=&\delta\phi \left(P_{X\phi}\dot{\phi}^2 - P_{\phi}\right) \nn \\
&-& \dot{\phi}\left(P_X+P_{XX}\dot{\phi}^2\right)\left(\dot{\phi}\Psi-\dot{\delta \phi}\right),\\
V_{DE} & = & \frac{k^2}{a}P_X\dot{\phi}\delta\phi.\label{eq:vdekess}
\eea
Since for K-essence $\Psi=-\Phi$ the DE anisotropic stress parameter $\pi_{DE}$ vanishes. We find that under the subhorizon approximation
\bea
A_6 = 0, \quad B_6 = -2, \quad D_9 = -P_X, \quad M^2 = -P_{\phi\phi},~~
\eea
and therefore the DE perturbations for K-essence theories are given by
\bea
\frac{\delta P_{DE}}{\bar{\rho}_{DE}}&\simeq&\frac{P_X\dot{\phi}^2}{2k^2/a^2}\frac{\bar{\rho}_m}{\bar{\rho}_{DE}} \delta_m \label{eq:effpreskess}, \nn \\
& & \\
\delta_{DE}&\simeq&\frac{\dot{\phi}^2\left(P_X+P_{XX}\dot{\phi}^2\right)}{2k^2/a^2}\frac{\bar{\rho}_m}{\bar{\rho}_{DE}} \delta_m,\label{eq:effderkess} \\
V_{DE} & \simeq & 0,
\eea
and the DE sound speed reads
\be
c_{s,DE}^2 = \frac{P_X}{P_X+2 X P_{XX}},
\ee
in agreement with Refs.~\cite{Scherrer:2004au,dePutter:2007ny}. The perturbations of the scalar field and the gravitational potential are respectively given by
\bea
\delta\phi&\simeq&0,\nn \\
\Psi &\simeq&-\frac{\bar{\rho}_m\delta_m a^2}{2k^2}.
\eea

\subsubsection{Kinetic gravity braiding \label{kgb-attractor}}

An interesting DE model is the kinetic gravity braiding (KGB) which is characterized by the following Lagrangian
\begin{equation}
K=K(X), \hspace{5mm} G_3=G_3(X), \hspace{5mm} G_4=\frac{1}{2 \kappa}.\label{eq:KGB:definition}
\end{equation}
Since $G_4$ is constant it is easily shown from Eq. \eqref{eq:trace} that the KGB model has no DE anisotropic stress and therefore the anisotropic parameters
\begin{align}
    \eta &\equiv \frac{\Psi+\Phi}{\Phi}=0,\\
    \gamma &\equiv -\frac{\Phi}{\Psi}=1.
\end{align}
Furthermore, it follows that the effective Newton's constant $G_{\textrm{eff}}/G_N$ is given by
\begin{align}
    G_{\textrm{eff}}/G_N =\frac{M^2-D_9\frac{k^2}{a^2}}{M^2-\left(D_9+A^2_6/2\right)\frac{k^2}{a^2}}.
\end{align}

The effective DE density and pressure $\bar{\rho}_{DE}$ and $\bar{P}_{DE}$ read, respectively,
\bea
\label{eq:dekgb}
   \kappa\bar{\rho}_{DE}&=&-K+\dot{\phi}^2\left(-G_{3\phi}+K_X+3G_{3X}H\dot{\phi}\right),\\
 \label{eq:pkgb}
    \kappa\bar{P}_{DE}&=&K-\dot{\phi}^2\left(G_{3\phi}+G_{3X}\ddot{\phi}\right),
\eea
and therefore the DE equation of state is given by
\begin{equation}
\label{eq:wkgb}
    w_{DE}=\frac{K-\dot{\phi}^2\left(G_{3\phi}+G_{3X}\ddot{\phi}\right)}{-K+\dot{\phi}^2\left(-G_{3\phi}+K_X+3G_{3X}H\dot{\phi}\right)}.
\end{equation}
We also find that the scalar field equation at the background level is
\bea
& & K_{\phi}-\left(K_X-2G_{3\phi}\right)\left(\ddot{\phi}+3H\dot{\phi}\right)-K_{X\phi}\dot{\phi}^2-K_{XX}\ddot{\phi}\dot{\phi}^2 \nn \\
& & + G_{3\phi\phi}\dot{\phi}^2+G_{3X\phi}\dot{\phi}^2\left(\ddot{\phi}-3H\dot{\phi}\right) -3G_{3X}\left(2H\ddot{\phi}\dot{\phi} \right. \nn \\
& & \left. + 3H^2\dot{\phi}^2 + \dot{H}\dot{\phi}^2\right) - 3G_{3XX}H\ddot{\phi}\dot{\phi}^3=0.
\eea

As a specific example we now discuss the KGB model of Ref.~\cite{Kimura:2010di} defined by
\begin{align}
K(X)&=-X \label{eq:KGB:1}\\
G_3(X)&= \frac{1}{\sqrt{\kappa}} \left( \kappa r^2_c X \right)^n=\alpha X^n \label{eq:KGB:2},
\end{align}
where $n$ and $\alpha$ are parameters in the model. A number of reasons make the KGB an attractive model. First, it passes the recent observational constraints from gravitational waves. Second, it is known that this model connects the original Galileon model \cite{Deffayet:2010qz} and the $\Lambda \text{CDM}$ model by the parameter $n$, at least for the background and first order perturbations: linear perturbations of the KGB model reduce to those of \lcdm (original Galileon) for $n=\infty$ ($n=1$) \cite{Kimura:2010di}.

The charge density of the Noether current Eq. \eqref{eq:jj} is in this case
\begin{equation}
\label{eq:jo}
     J_0=\dot{\phi}\left(3\dot{\phi}G_{3X}H-1\right),
\end{equation}
and satisfies the differential equation
\begin{equation}
\dot{J_0} + 3 H J_0 = 0,
\end{equation}
whose solution reads
\begin{equation}
J_0 = \frac{J_c}{a^3}
\end{equation}
with $J_c$ a constant. It is therefore clear that $J_0$ approaches zero as the Universe expands. The simplest attractor solution is located at $J_0=0$ and has two branches, namely,
\begin{equation}
\dot{\phi}=0
\end{equation}
and
\begin{equation}
\label{eq:att}
\dot{\phi}=\frac{1}{3G_{3X}H}.
\end{equation}
Because the first case has ghostly perturbations, as it is shown in \cite{Kimura:2010di}, we will focus on the attractor solution Eq. \eqref{eq:att}. Using Eqs.~\eqref{eq:327} and \eqref{eq:jo} we find that the modified Friedmann equation is given by
\bea
\label{eq:hubn}
\left(\frac{H}{H_0}\right)^2&=&\left(1-\Omega_{m,0}\right)\left(\frac{H}{H_0}\right)^{-\frac{2}{2n-1}}+ \Omega_{m,0} a^{-3},~~~~
\eea
where we have neglected radiation. The background equation of the KGB model reduces to that of  $\Lambda \text{CDM}$ for $n=\infty$ as can be seen from Eq.~\eqref{eq:hubn}. Also, one can easily find an expression for the parameter $\alpha$ by using Eq.~\eqref{eq:hubn} at the present epoch
\begin{equation}
    \alpha=\left(\frac{2^{n-1}}{3n}\right)\left(\frac{1}{6\left(1-\Omega_{m,0}\right)}\right)^{\frac{2n-1}{2}}.
\end{equation}
The DE equation of state becomes
\begin{equation}
    w_{DE}=\frac{\bar{P}_{DE}}{\bar{\rho}_{DE}}=\frac{2\dot{H}}{3\left(2n-1\right)}-1,
\end{equation}
and through Eq. \eqref{eq:att} it is also possible to find an analytical expression for the kinetic term
\bea
X&=&\frac12 a^2 H^2 \phi'(a)^2 \nn\\
&=&3 H_0^2 (1-\Omega_{m,0})\left(\frac{H}{H_0}\right)^{\frac{2n}{1-2n}},
\eea
where the prime stands for the derivative with respect to the scale factor.

To derive the \lcdm limit for the perturbations in this model we rewrite Eqs.~\eqref{eq:2} and \eqref{eq:3} in terms of the kinetic term perturbation $\delta X=\dot{\phi} \dot{\delta \phi}-\dot{\phi}^2 \Psi$. Then, for $n\rightarrow \infty$ the former equation reduces to $\delta X \left(-\frac{2}{a}-\frac{H'(a)}{H(a)}+O(1/n)\right)$, while the latter equation gives
\be
\dot{\delta X}+3 H \delta X=0,
\ee
which implies that the kinetic term perturbation decays as $\delta X\sim 1/a^3$ and thus can be ignored at late time. Since DE perturbations in the KGB model are proportional to $\delta X$ for large $n$, then they reduce to zero as expected for the \lcdm model.

Finally, it should be noted that a standard hydrodynamical description of the KGB in terms of an effective fluid, has been studied in Ref.~\cite{Pujolas:2011he}. There, it was shown that the KGB model can also be described in terms of an imperfect fluid with a chemical potential, in which the equations of motion reduce to the standard diffusion equation. However, in our current analysis we will only focus on the ideal fluid approach, which is totally equivalent, as we are interested in finding simple analytic solutions and with comparing with our previous work.

\section{Designer Horndeski}
\label{Section:DES}

In this section we will address the shortcomings found in the KGB model defined by Eqs. \eqref{eq:KGB:1}-\eqref{eq:KGB:2}. We will show that it is possible, starting from the Lagrangian \eqref{eq:KGB:definition}, to find a model corresponding to a given background but yet having different perturbations. Using the modified Friedmann equation and the scalar field conservation equation, we can find specific designer models such that the background is always that of the \lcdm model, namely, having $w_{DE}=-1$. This is particularly useful in detecting deviations from \lcdm at the perturbations level and is a natural expansion of our earlier work \cite{Nesseris:2013fca,Arjona:2018jhh}. We start with the modified Friedmann equation, which can be written as \bea
\label{eq:friedeq}
& & -H(a)^2-\frac{K(X)}{3}+H^2_0\Omega_m(a)+2\sqrt{2}X^{3/2}H(a)G_{3X} \nn \\
& & +\frac{2}{3}X K_X=0.
\eea
while the scalar field conservation equation can be written as
\begin{equation}
\label{eq:scfeq}
    \frac{J_c}{a^3}-6XH(a)G_{3X}-\sqrt{2}\sqrt{X}K_X=0
\end{equation}
where $J_c$ is a constant which quantifies our deviation from the attractor, as in the case of the KGB model \cite{Kimura:2010di}. We now have two equations given by \eqref{eq:friedeq} and \eqref{eq:scfeq}, but three unknown functions $(G_{3X}(X),K(X), H(a))$ thus the system is undetermined. Therefore, we need to specify one of the three unknown functions $(G_{3X}(X),K(X), H(a))$ and determine the other two using Eqs.~\eqref{eq:friedeq} and \eqref{eq:scfeq}. To facilitate this, we express the Hubble parameter as a function of the kinetic term $X$, ie $H=H(X)$ and then solve the previous equations to find $(G_{3X}(X),K(X))$. Doing so yields:
\bea
\label{eq:systemdes}
K(X) &=& -3 H_0^2 \Omega_{\Lambda,0}+\frac{J_c \sqrt{2X} H(X)^2}{H_0^2 \Omega_{m,0}}-\frac{J_c \sqrt{2X} \Omega_{\Lambda,0}}{\Omega_{m,0}} \nn\\
G_{3X}(X) &=& -\frac{2 J_c H'(X)}{3 H_0^2 \Omega_{m,0}}.
\eea
With Eqs.~\eqref{eq:systemdes} we can make a whole family of designer models that behave as \lcdm at the background level but have different perturbations. We now proceed to specify some examples using our formalism.

\subsection{Example 1}

Choosing $K(X)=0$ and solving Eqs.~\eqref{eq:systemdes} we find
\begin{align}
    K(X)&=0,\nn\\
    G_{3}(X)&=-\frac{\sqrt{2J_c} \sqrt{\Omega_{\Lambda,0}\left(2J_c\sqrt{X}+3\sqrt{2}H^2_0\Omega_{m,0}\right)}}{3H_0X^{1/4}\Omega_{m,0}},
 \end{align}
and the derivative of the scalar field $\phi'(a)$ is
 \begin{equation}
     \phi'(a)=\frac{3a^2H^2_0\Omega_{\Lambda,0}}{J_cH(a)}
 \end{equation}
where the prime is the derivative with respect to the scale factor. However, this model has the problem that it does not have a smooth limit to \lcdm when $J_c=0$.

\subsection{Example 2}

On the other hand, specifying $G_3(X)$ leads to another interesting designer model, defined as
\bea
G_{3}(X)&=&G_{30}X,\nn\\
K(X)&=& -3 H_0^2 \Omega_{\Lambda,0}+\frac{9 H_0^2(X-X_0)^2 G_{30}^2 \sqrt{X} \Omega_{m,0}}{2 \sqrt{2} J_c}\nn\\
&-&\frac{\sqrt{2} J_c \sqrt{X}  \Omega_{\Lambda,0}}{\Omega_{m,0}}
\eea
where the kinetic term is defined as
\begin{equation}
X=\frac{3G_{30}H_0X_0\Omega_{m,0}-2J_cH(a)}{3G_{30}H^2_0\Omega_{m,0}}
\end{equation}
and $X_0$ is an integration constant. However, this model has the problem that at early times the perturbations do not go to zero and we do not recover GR, since the kinetic term goes to infinity as it grows as $X\sim H(a)$.

\subsection{Example 3 (HDES)}

To solve the previous shortcomings we follow a different approach. First, we demand that the kinetic term behaves as $X= \frac{c_0}{H(a)^n}$, where $c_0>0$ and $n>0$. Then, from Eqs.~\eqref{eq:scfeq} and \eqref{eq:friedeq} we find:
\bea
\label{eq:bestdes}
G_{3}(X)&=&-\frac{2 J_c c_0^{1/n} X^{-1/n}}{3 H_0^2 \Omega_{m,0}},\\
K(X)&=&\frac{\sqrt{2} J_c c_0^{2/n} X^{\frac{1}{2}-\frac{2}{n}}}{H_0^2 \Omega_{m,0}}-3 H_0^2 \Omega_{\Lambda,0}-\frac{\sqrt{2} J_c \sqrt{X} \Omega_{\Lambda,0}}{\Omega_{m,0}}.\nn
\eea
This specific model solves both previous problems, i.e., it has a smooth limit to \lcdm and it also recovers GR when $J_c\sim0$, thus we will designate this model as \textit{HDES} and focus on it in what follows.

\subsection{Comparison with the $\alpha$ parameters}
To facilitate comparisons with the literature we also provide the expressions for our designer HDES model in terms of the $\alpha_i$ functions, where $i=M,K,B,T$. The functions $G_i(\phi,X)$ and $\alpha_i$ are connected in the following manner \cite{Zumalacarregui:2016pph}:
\bea
M_{\ast}^2& \equiv & 2\left(G_4-2XG_{4X}-\dot{\phi}HXG_{5X}+XG_{5\phi}\right),\nn\\
\alpha_M & \equiv & \frac{d\ln  M_{\ast}^2}{d\ln a},\nn\\
H^2 M_{\ast}^2\alpha_K & \equiv & 2X\left(G_{2X}+2XG_{2XX}-2G_{3\phi}-2XG_{3\phi X}\right)\nn\\
& + & 12H\dot{\phi}X\big[G_{3X}+XG_{3XX}-3G_{4\phi X} \nn\\
& - & 2XG_{4 \phi XX}\big] \nn\\
& + & 12H^2X\big[G_{4X}-G_{5\phi}+X\left(8G_{4XX}-5G_{5\phi X}\right)\nn\\
& + & 2X^2\left(2G_{4XXX}-G_{5\phi XX}\right)\big] \nn\\
& + & 4H^3\dot{\phi}X\left(3G_{5X}+7XG_{5XX}+2X^2G_{5XXX}\right), \nn\\
H^2 M_{\ast}^2\alpha_B & \equiv & 2\dot{\phi}\left(XG_{3X}-G_{4\phi}-2XG_{4\phi X}\right)\nn\\
&+& 8HX\big(G_{4X}+2XG_{4XX}-G_{5\phi}-XG_{5\phi X}\big)\nn\\
& + & \frac{2H^2\phi'X}{a}\left(3G_{5X}+2XG_{5XX}\right),\nn\\
M_{\ast}^2\alpha_T & \equiv & 4X\left(G_{4X}-G_{5\phi}\right)-2X\left(\ddot{\phi}-2H\dot{\phi}\right)G_{5X},~~~~~~~\label{eq:alphas}
\eea
where the dot is the derivative with respect to the cosmic time, $M_{\ast}^2(\tau)$ is the cosmological strength of gravity, $\alpha_T$ is the tensor speed excess, $\alpha_B$ is called the braiding and $\alpha_K$ is referred to as the kineticity. For more information on these $\alpha_i$ functions see \cite{Ezquiaga:2018btd}. At all times we require $D=\alpha_K+\frac32 \alpha_B^2 > 0$ so that there are no ghostly instabilities and that $\alpha_{M,K,B,T}\simeq0$ at early times, so as to recover GR.

For our HDES designer model given by Eqs.~\eqref{eq:bestdes}, we have that the $\alpha_i$ functions of Eq.~\eqref{eq:alphas} are given by
\bea
\label{eq:bestdesalpha}
M_{\ast}^2&\equiv & 1,\\
\alpha_M &\equiv & \frac{d\ln  M_{\ast}^2}{d\ln a}=0,\\
\label{eq:alphak}
\alpha_K &\equiv & -\frac{4 \sqrt{2} \sqrt{c_0} J_c (n-2) H(a)^{-\frac{n}{2}}}{H_0^2 n^2 \Omega_{m,0}},\\
\label{eq:alphab}
\alpha_B &\equiv & \frac{4 \sqrt{2} \sqrt{c_0} J_c H(a)^{-\frac{n}{2}}}{3 H_0^2 n \Omega_{m,0}},\\
\alpha_T &\equiv & 0. \label{eq:bestdesalphaT}
\eea
Since in Eqs.~\eqref{eq:alphak}-\eqref{eq:alphab} we have a degeneracy with the coefficients $c_0$ and $J_c$, they appear together as $\sqrt{c_0}J_c$, we can choose to absorb $c_0$ in the definition of $J_c$. Finally, it is straightforward to see that our $\alpha_i$ functions are dimensionless since through dimensional analysis we found that $[c_0]=H^{n+2}_0$, $[J_c]=H_0$, the kinetic term $[X]=H^2_0$, $[K]=H^2_0$ and $[G_{3X}]=H^{-2}_0$.

Notice that not all designer models satisfy the above conditions, so in what follows we consider only HDES, given by Eq.~\eqref{eq:bestdes}. Then, the stability condition $D=\alpha_K+\frac32 \alpha_B^2 > 0$ for our model Eq.~\eqref{eq:bestdes} gives  \be
\tilde{J}_c\left(4 \tilde{J}_c-3 \sqrt{2} (n-2) \Omega_{m,0} \left(\frac{H(a)}{H_0}\right)^{n/2}\right)>0,\label{eq:ineqD}
\ee
where we have set $\tilde{J}_c=J_c/H_0$ and $\tilde{c}_0=c_0/H^{n+2}_0=1$.

\begin{figure*}[!t]
	\centering
	\includegraphics[width=0.32\textwidth]{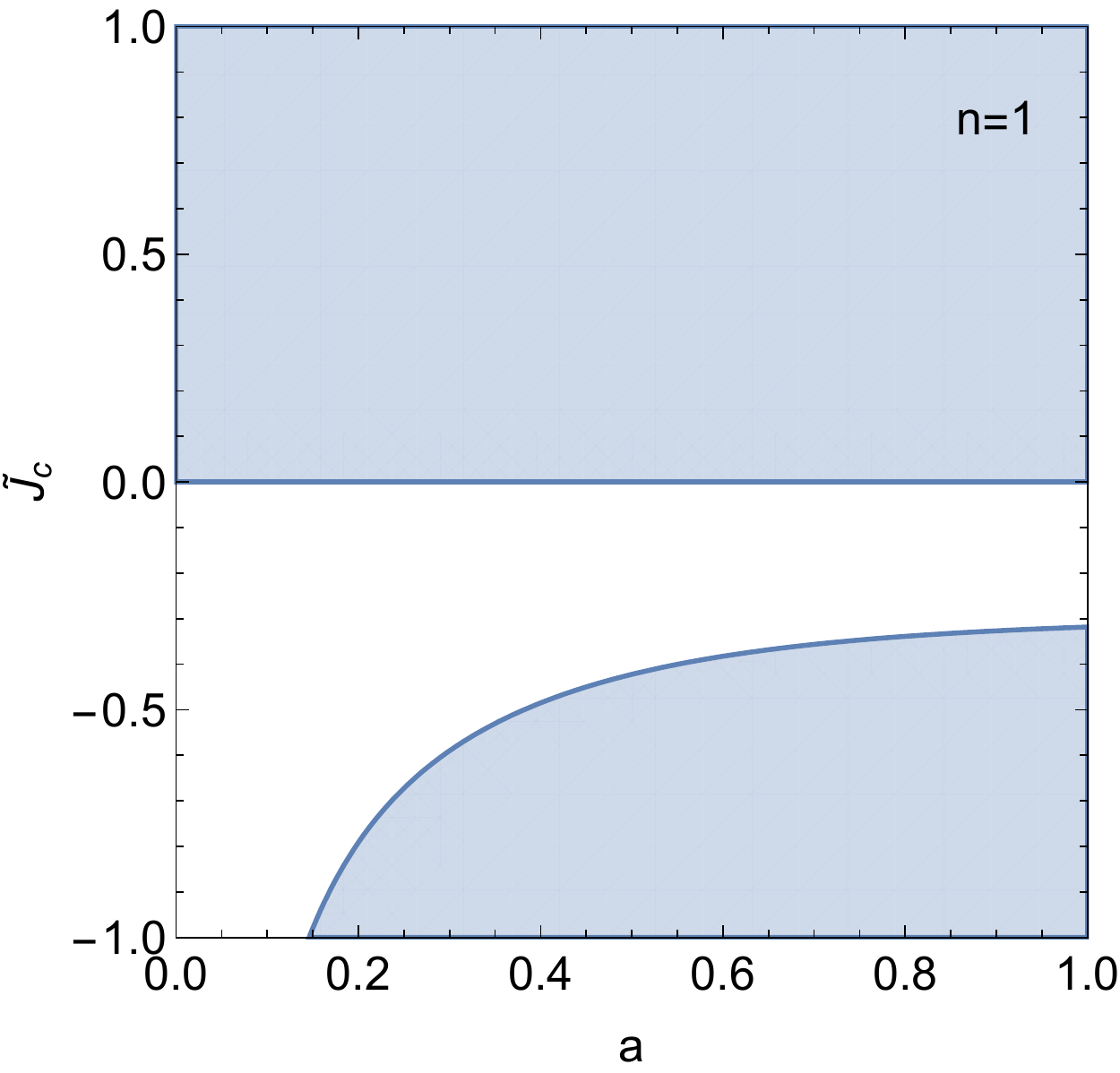}
    \includegraphics[width=0.32\textwidth]{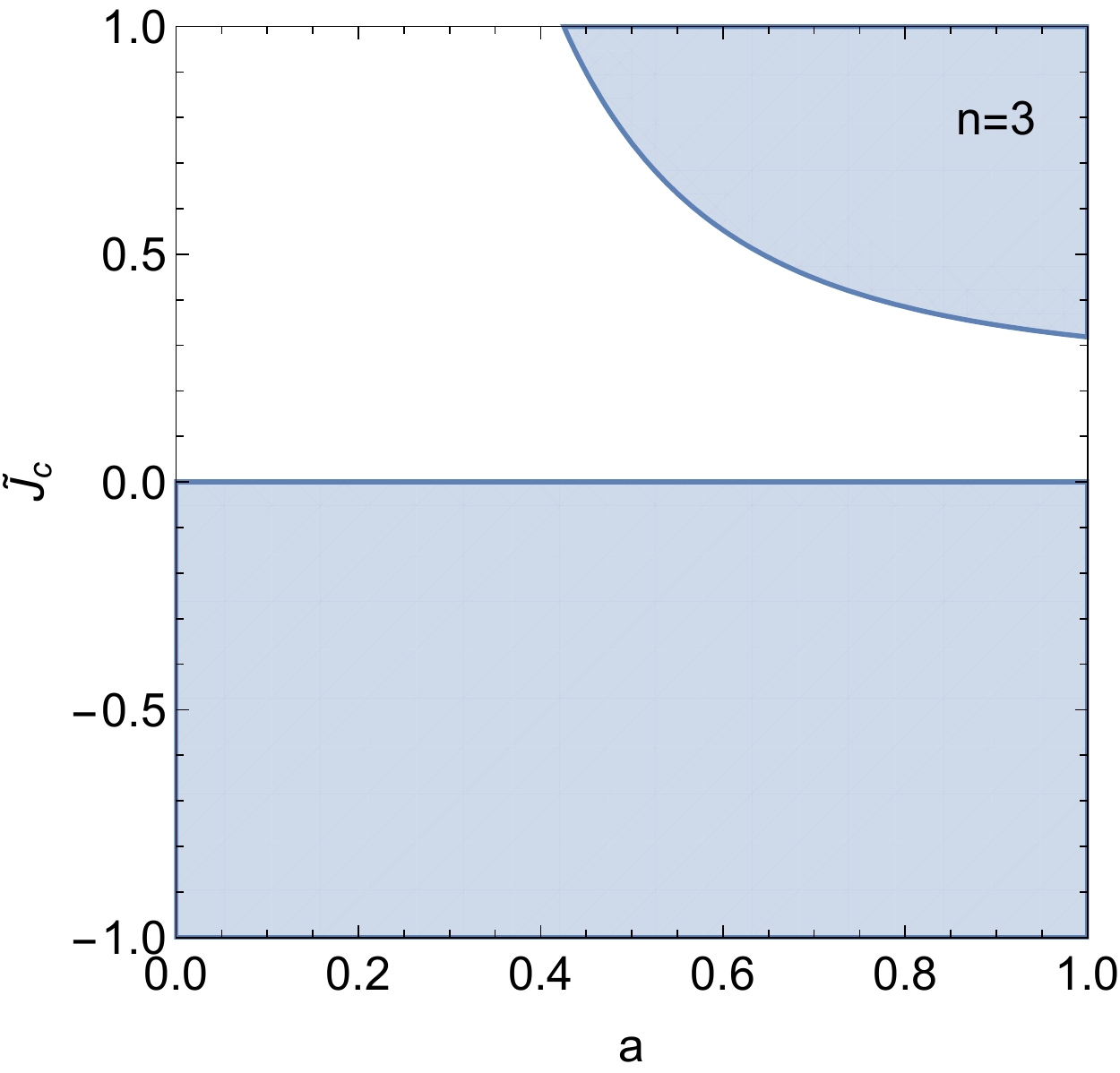}
    \includegraphics[width=0.32\textwidth]{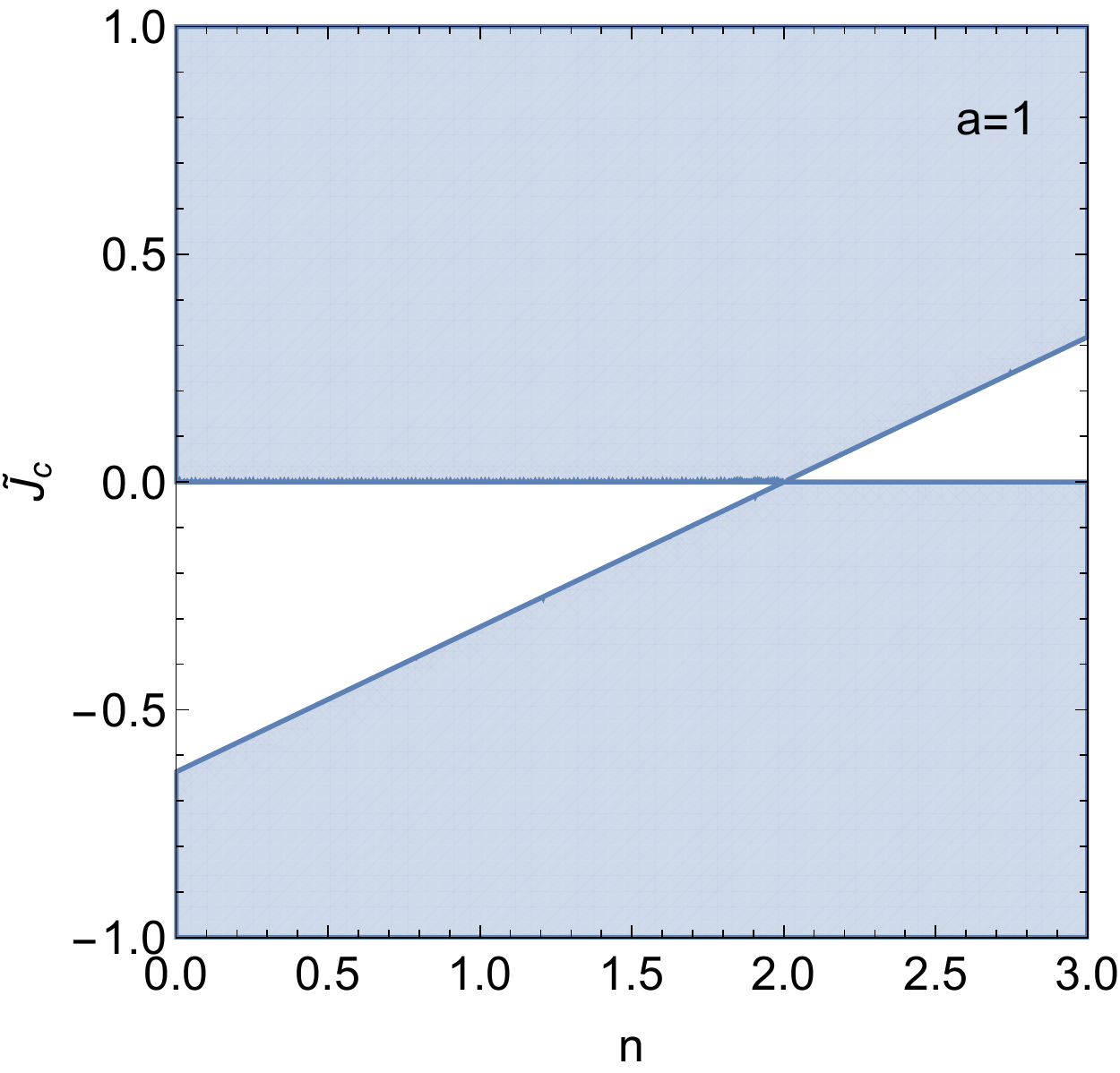}
	\caption{The allowed parameter space (shaded region) so that $D>0$ for $n=1$ (left) and $n=3$ (center) and $a=1$ for $\Omega_{m,0}=0.3$. In the case of $n=2$, all values of $\tilde{J}_c$ are allowed.}
	\label{fig:stability}
\end{figure*}

Then, inequality \eqref{eq:ineqD} implies that in order for the system to be stable we must have either $\tilde{J}_c>0$ for $0<n\le2$ or a complicated set of expressions that can however be easily derived from Eq.~\eqref{eq:ineqD} with algebraic manipulations. For $n=2$ the inequality is automatically satisfied for any value of $\tilde{J}_c$ as $\alpha_K=0$ as can be seen from Eq.~\eqref{eq:alphak}. We show the complicated parameter space that is allowed for $n=1$ and $n=3$ as a function of scale factor $a$ but also as a function of $n$ for $a=1$, in Fig.~\ref{fig:stability}.

\subsection{Analytic solutions for the growth \label{sec:degen}}
Furthermore, in this case we can also find approximate solutions to the growth equation Eq.~\eqref{eq:ODE-growth} in matter domination for $n=2$. To do this, we first do a series expansion around $a=0$ to the $G_{\textrm{eff}}$ of Eq.~\eqref{eq:ODEGeff}, which gives:
\be
G_{\textrm{eff}}/G_N=1+\frac{\sqrt{2} \tilde{J}_c}{3 \Omega_{m,0} H(a)/H_0},
\ee
which we can use to solve Eq.~\eqref{eq:ODE-growth} in matter domination, where $H(a)/H_0\simeq \sqrt{\Omega_{m,0} a^{-3}}$. Then, we get
\be
\delta_m(a)=\frac{3^{5/3} \Omega_{m,0}^{5/4} \Gamma \left(\frac{8}{3}\right) }{2^{5/4} \tilde{J}_c^{5/6}}a^{-1/4}I_{\frac{5}{3}}\left(\frac{2^{7/4} \sqrt{\tilde{J}_c}}{3 \Omega_{m,0}^{3/4}}a^{3/4}\right), \label{eq:growthn2HDES}
\ee
where $I_n(z)$ is the modified Bessel function of the first kind and $\Gamma(n)$ is the usual Gamma function. Using Eq.~\eqref{eq:growthn2HDES} and the definition of the growth rate $f\sigma_8(a)\equiv f(a)\cdot \sigma(a)=\sigma_8 a \delta_m'(a)/\delta_m(a=1)$, we can calculate the latter exactly. However, it is instructive to perform a series expansion around $a=1$, which gives:
\bea
f\sigma_8(a)&\simeq& \sigma_8\Bigg(\frac{1}{2} \left(\frac{5 \alpha_1}{\alpha_2}-3\right)\nn\\
&+&\frac{1}{4} \left(-\frac{5 \alpha_1}{\alpha_2}+\frac{2 \sqrt{2} \tilde{J}_c}{\Omega_{m,0}^{3/2}}+9\right) (a-1)+\cdots\Bigg),~~~~~~~\label{eq:HDESfs8}
\eea
where we have defined the parameters
\bea
\alpha_1 &=& \, _0F_1\left(\frac{5}{3};\frac{2 \sqrt{2} \tilde{J}_c}{9 \Omega_{m,0}^{3/2}}\right),\\
\alpha_2 &=& \, _0F_1\left(\frac{8}{3};\frac{2 \sqrt{2} \tilde{J}_c}{9 \Omega_{m,0}^{3/2}}\right),
\eea
where $\, _0F_1(c_1,z)$ is a hypergeometric function.

As can be seen from Eq.~\eqref{eq:HDESfs8} there is a strong degeneracy between $\tilde{J}_c$ and $\sigma_8$, which can also be demonstrated by doing a series expansion of $f\sigma_8(a=1)$ for small $\tilde{J}_c$, which gives
\be
f\sigma_8(a=1)\simeq \sigma_8\left(1+\frac{\tilde{J}_c}{4 \sqrt{2} \Omega_{m,0}^{3/2}}+\cdots\right).
\ee
which implies that if we keep the growth today given constant, i.e., $f\sigma_8(a=1)=C_0=\textrm{const}.$ then $\sigma_8$ will scale roughly as
\be
\sigma_8 \simeq C_0 \left(1-\frac{\tilde{J}_c}{4 \sqrt{2} \Omega_{m,0}^{3/2}}+\cdots\right).\label{eq:degenHDES}
\ee
Since $\Omega_{m,0}$ is strongly constrained from Planck, we expect that the low redshift $f\sigma_8$ data will exhibit a degeneracy between $\tilde{J}_c$ and $\sigma_8$. More specifically, by inspecting Eq.~\eqref{eq:degenHDES} we expect a strong negative correlation between the two parameters and this is exactly what we see from the actual Markov Chain Monte Carlo (MCMC) that we present in later sections. This degeneracy is interesting as it can potentially alleviate the soft $2\sigma$ tension between the growth rate data ($\sigma_8=0.88$) and Planck ($\sigma_8=0.831)$, which has been extensively discussed in the literature, see Ref.~\cite{Nesseris:2017vor,Sagredo:2018ahx} and references therein.


\section{Numerical solutions \label{Section:Numerical-Solution}}
Here we present the numerical solutions of the two models, the KGB and HDES, that we described in the previous section.

\subsection{The KGB model}
\subsubsection{The attractor}
\begin{figure*}[!t]
\includegraphics[width=0.49\textwidth]{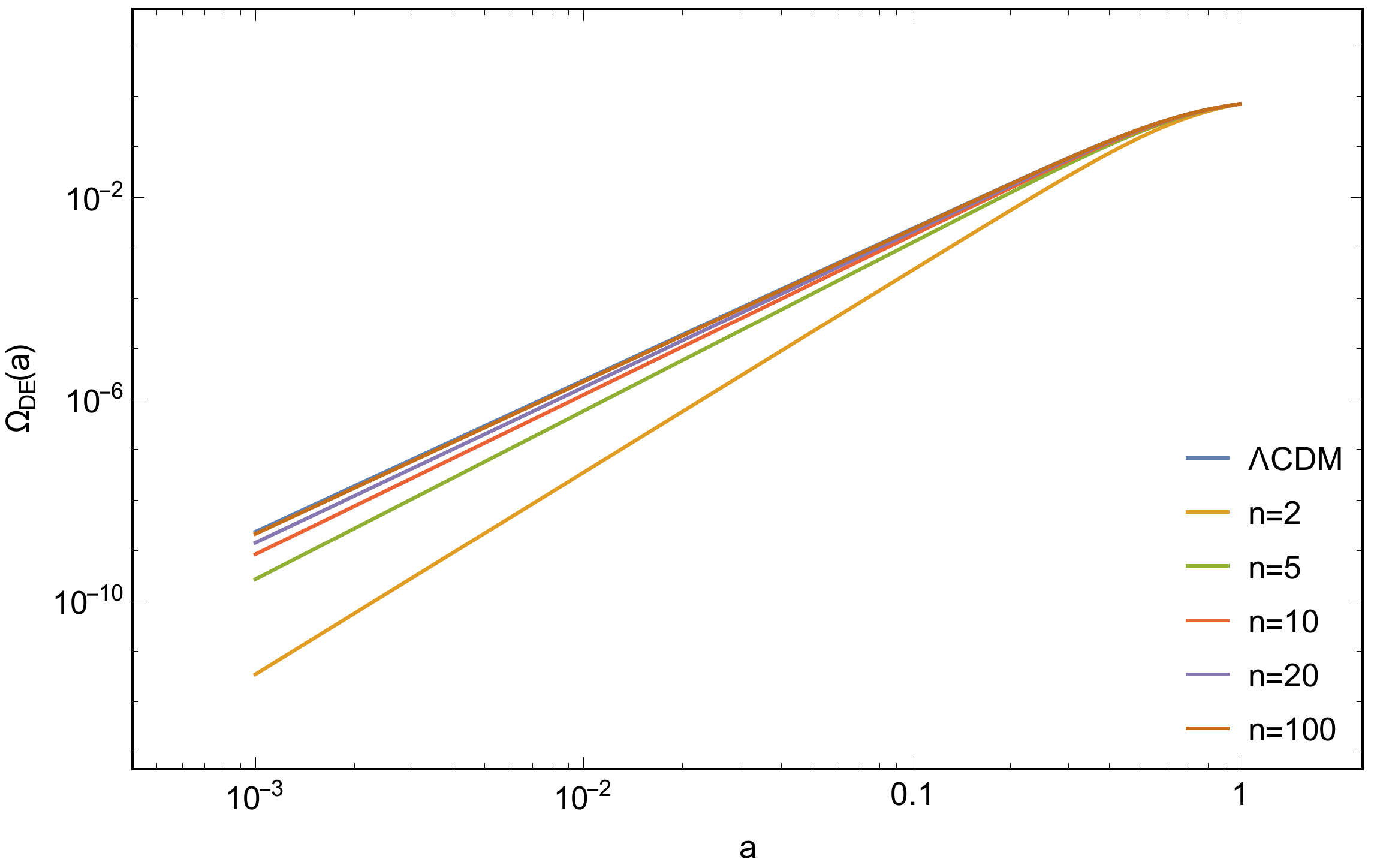}
\includegraphics[width=0.49\textwidth]{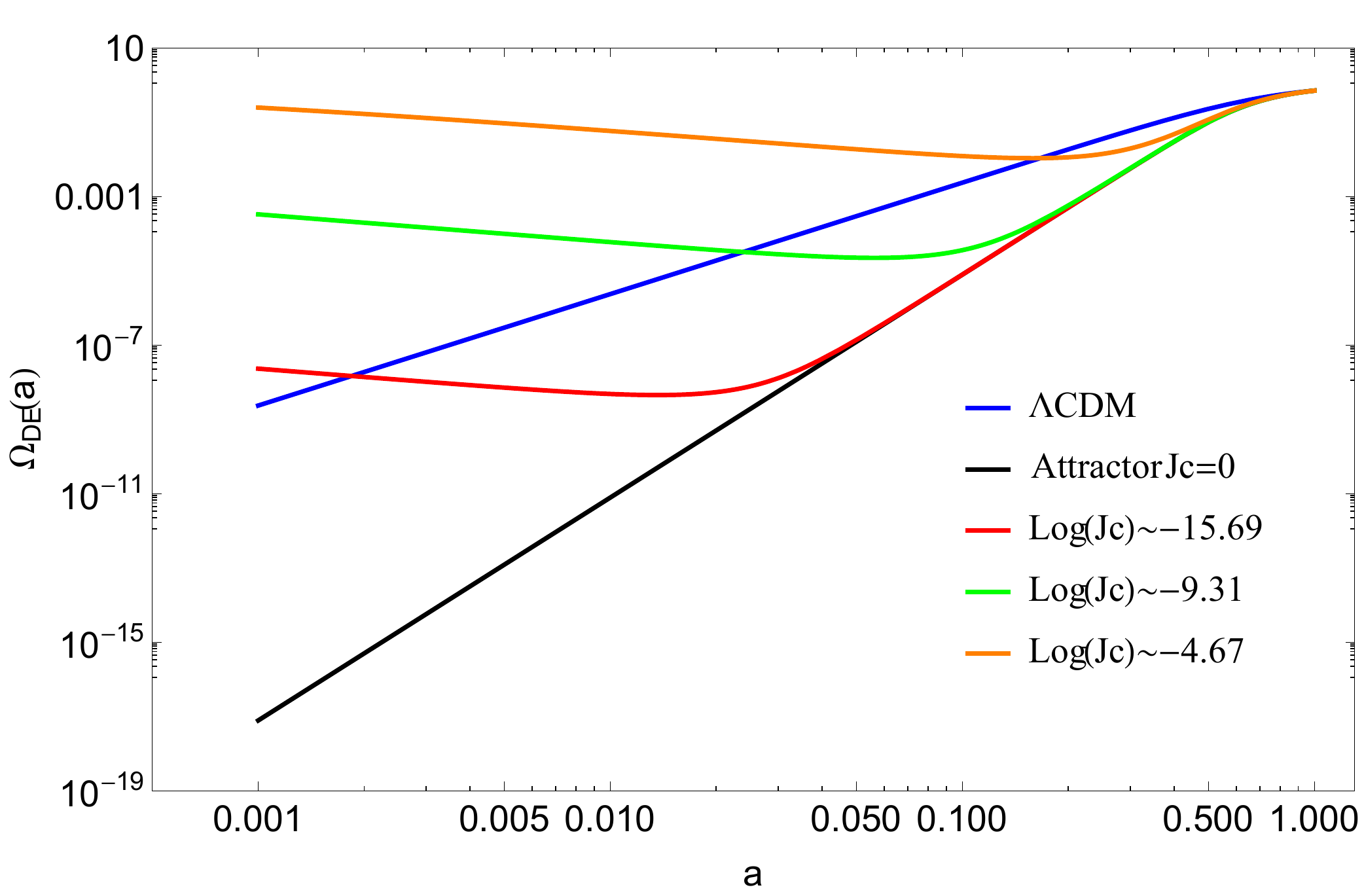}
\caption{The DE density for the KGB model for various values of n (left) and for the KGB model (n=1) for the attractor and three general cases outside the attractor given by different values of $J_c$, chosen so as to highlight the differences of these models with respect to GR. The left panel clearly shows that as $n$ grows the DE density approaches that of the \lcdm model.\label{fig:kgb1}}
\end{figure*}

To explore the possibility of working outside the attractor we only need to use Eqs.~(\ref{eq:327}) and (\ref{eq:jo}), as these constrain $J_c$ and $\alpha$ with $H(a=1)=H_0$. To parameterize the deviation from the attractor we will use the parameter $J_c$. An illustrative example is found in Fig. \ref{fig:kgb1} where we plot the dark energy density $\Omega_{DE}$ with respect to the scale factor for several values of $n$ (left) and $J_c$ (right). The values of values for $J_c$ were chosen so as to highlight the differences of these models with respect to GR.

In the KGB model the DE density can be written via Eq.~\eqref{eq:dde} as
\bea
    \Omega_{DE}&=&\frac{\rho_{DE}}{\rho_c},\\
    \rho_{DE}&=&-K+K_{X}\dot{\phi}^{2}-G_{3\phi}\dot{\phi}^2+3G_{3X}H\dot{\phi}^3.
\eea
From Fig.~\ref{fig:kgb1} we can see that working outside the attractor for the KGB model $(n=1)$ we might find new parts of the parameter space and new phenomenology. In the right panel of Fig.~\ref{fig:kgb1}, we see that the orange line can be ruled out because it predicts a very high value for the DE density at early times. The red and green lines, although outside the attractor solution, are plausible solutions that are interesting to analyze in more depth.

\subsubsection{Numerical solution}
In this section we present the results of the numerical solution of the evolution equations. In all cases we will assume $\Omega_{m,0}=0.3$, $k=300H_0$ and $\sigma_{8,0}=0.8$, unless otherwise specified. The reason we choose the specific value of $k=300H_0\sim 0.1\;h/\textrm{Mpc}$ for the wave-number is that it corresponds to the largest value of $k$ we can choose without entering the non-linear regime. Finally, we set the initial conditions for the DE variables to zero at $a_i=10^{-3}$, when we are well inside the matter dominated regime.

\begin{figure*}[!t]
\centering
\includegraphics[width=0.495\textwidth]{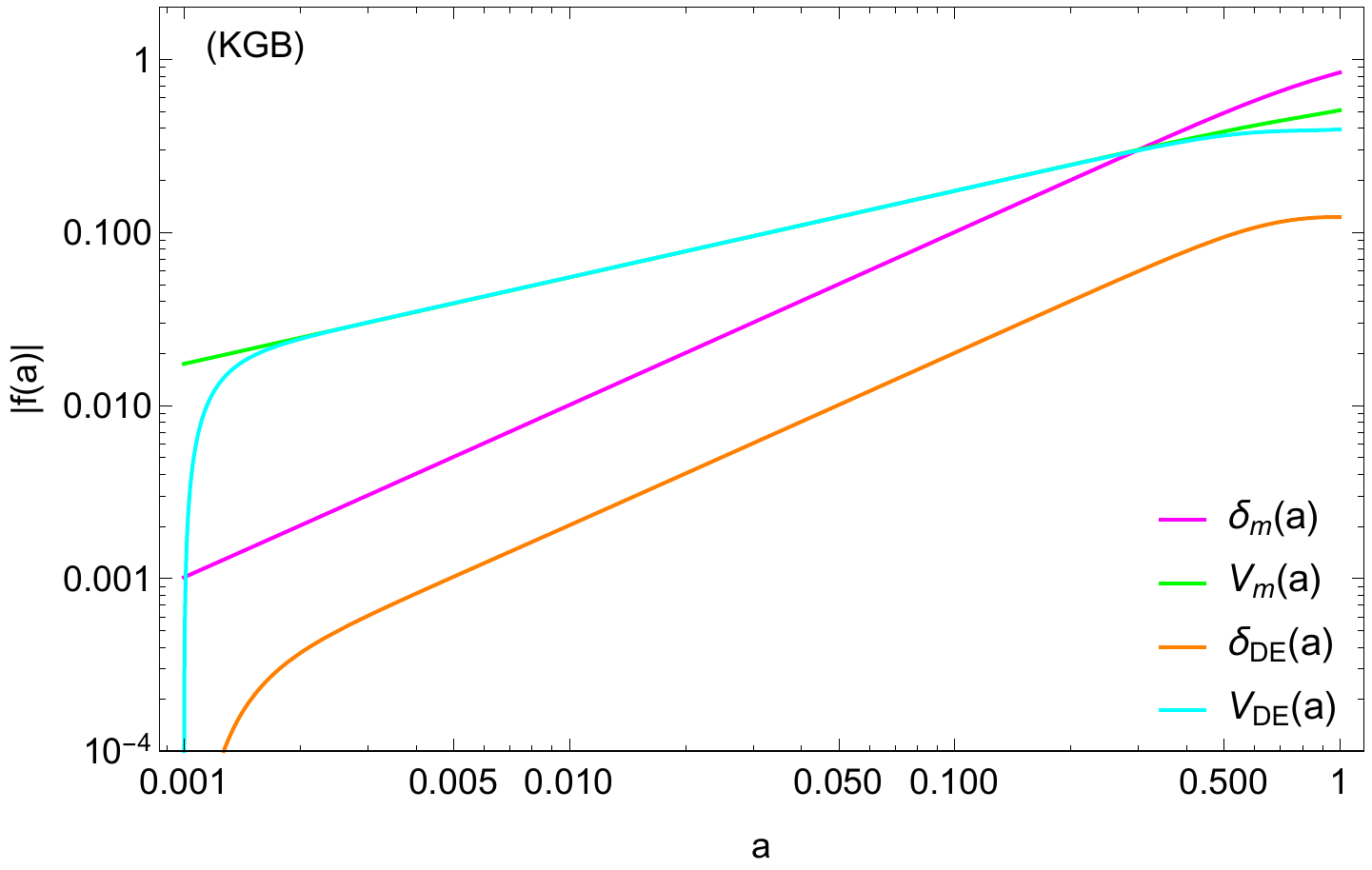}
\includegraphics[width=0.48\textwidth]{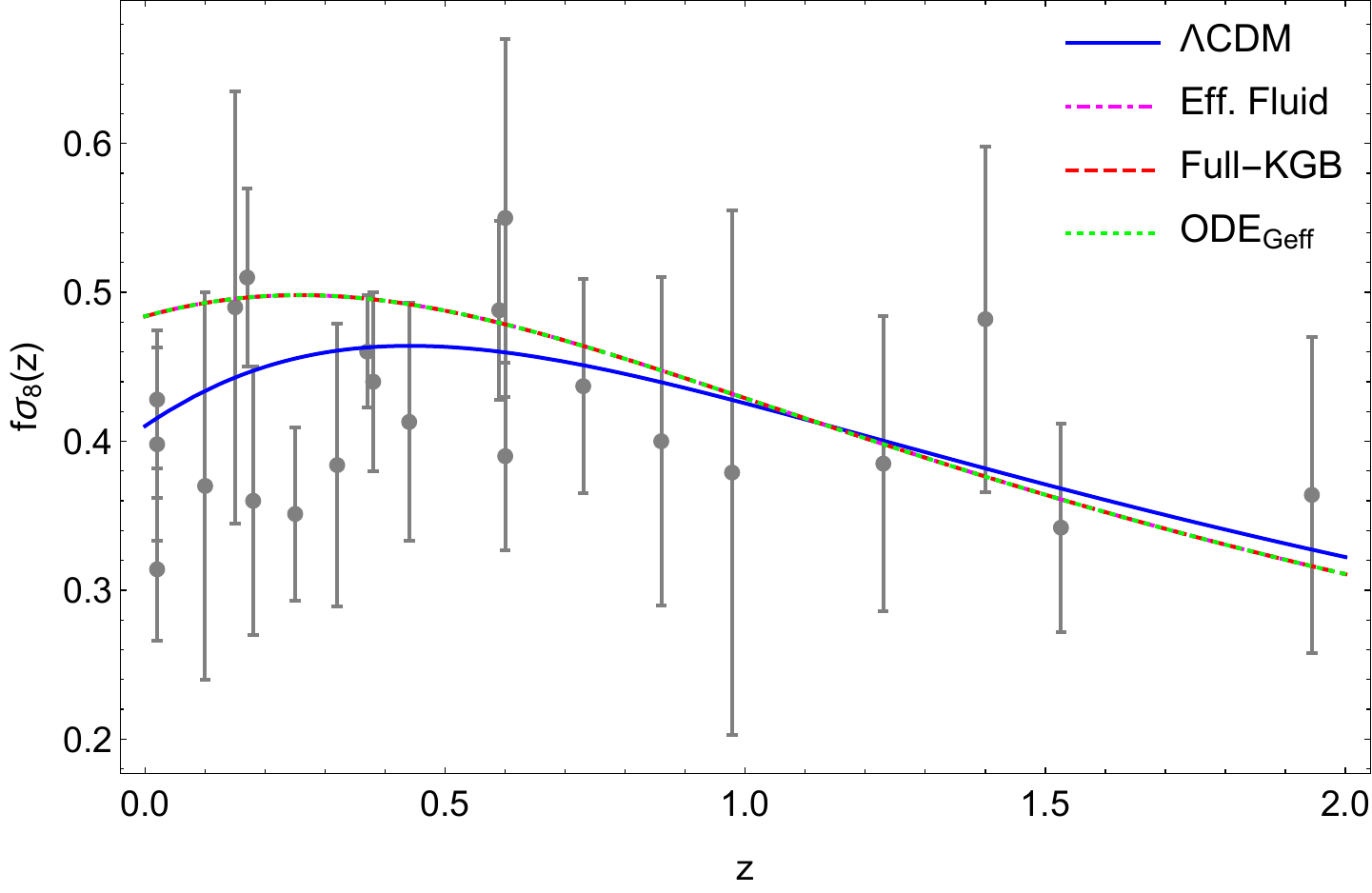}
\caption{Left: The evolution of the matter and effective DE perturbation variables $(\delta_m,V_m,\delta_{DE},V_{DE})$ for the KGB with $n=2$. Right: The evolution of the $f\sigma_8(z)$ parameter for the KGB model with $n=2$ and $\sigma_{8,0}=0.8$ versus the $f\sigma_8$ data compilation from Ref.~\cite{Sagredo:2018ahx}. Here we show the theoretical curves for the ``Full KGB" brute-force solution, the effective fluid approach, the $\Lambda$CDM model and the numerical solution of the $G_{\textrm{eff}}$ equation. As can be seen, the agreement with all approaches is excellent. \label{fig:KGBevo}}
\end{figure*}

Next we will also present our results for the growth rate of matter perturbations parameter $f\sigma_8(a)\equiv f(a)\cdot \sigma(a)$, where $f(a)=\frac{d ln\delta}{d lna}$ is the growth rate and $\sigma(a)=\sigma_{8,0}\frac{\delta(a)}{\delta(1)}$  is  the redshift-dependent rms fluctuations of the linear density field within spheres of radius $R=8 h^{-1} \textrm{\textrm{Mpc}}$, while the parameter $\sigma_{8,0}$ is its value today. The $f\sigma_8(a)$ parameter is important as it can be shown to be not only independent of the bias $b_1$, but also a good discriminator of DE models. The reason for this is that in linear theory the quadrupole contribution to the galaxy power spectrum in redshift space is sensitive only to the combination $f\sigma_8(a)$.

Specifically, here we will compare the numerical solutions for the following cases:
\begin{itemize}
  \item The numerical solution of the full system of equations given by Eqs.~\eqref{eq:1}-\eqref{eq:field-equation-horndeski-4}, which however we rewrite in terms of $\delta X= \dot{\phi} \dot{\delta \phi}-\dot{\phi}^2 \Psi$ as the system is more stable this way. We call this case ``Full KGB".
  \item The numerical solution of the effective fluid approach given by Eqs.~\eqref{Eq:evolution-delta}-\eqref{Eq:evolution-V}. We call this case ``Eff. Fluid".
  \item The numerical solution of the growth factor equation \eqref{eq:ODE-growth}. We call this case ``ODE-Geff".
  \item The \lcdm model.
\end{itemize}

In the left panel of Fig.~\ref{fig:KGBevo} we show the evolution of the matter and effective DE perturbation variables $(\delta_m,V_m,\delta_{DE},V_{DE})$ for the KGB for $n=2$. In the right panel we show the evolution of the $f\sigma_8(z)$ parameter for the KGB model for $n=2$ and $\sigma_{8,0}=0.8$ versus the $f\sigma_8$ data compilation from Ref.~\cite{Sagredo:2018ahx}. We show the theoretical curves for the ``Full KGB" brute-force solution, the effective fluid approach, the $\Lambda$CDM model and the numerical solution of the $G_{\textrm{eff}}$ equation. As can be seen, the agreement with all approaches is excellent.

An interesting thing to note in Fig.~\ref{fig:KGBevo} is that $V_{DE}>\delta_{DE}$ and $V_{DE}\sim V_m$ at intermediate redshift. The reason for this is that in the effective fluid approach the DE velocity perturbations are not always subdominant, as it would be expected in a general DE fluid. This can be seen by remembering that the velocity perturbations are actually a component of the effective energy momentum tensor, namely the $T^0_i$ part, thus they contain some of the main contributions of the Modified Gravity (MoG) theory and can be in some cases rather large. See, for example, Eqs.~\eqref{eq:effectTmnvde} and \eqref{Eq:evolution-V} for the definition of $V_{DE}$ and Eqs.~\eqref{eq:field-equation-horndeski-3} and \eqref{eq:efftheta0} for all of the extra terms that are rewritten as $V_{DE}$.

As an example, also consider the case of quintessence and k-essense, where $V_{DE}$ is proportional to the scalar field perturbations, see Eqs.~\eqref{eq:vdequint} and \eqref{eq:vdekess} respectively. In the case of $f(R)$, $V_{DE}$ is given by \eqref{eq:vdefR} and is proportional to $\dot{F}/F$, which parameterizes the deviations from GR, so it is a proxy for the $f(R)$ modified gravity perturbations.

However, in the case of the KGB model the subhorizon approximation fails when the parameter $n$ is large. This can easily be seen by calculating the large $n$ limit of the $G_{\textrm{eff}}$ parameter via Eq.~\eqref{eq:ODEGeff}:
\be
G_{\textrm{eff}}/G_N\simeq 1+\frac{2 a^3 (1-\Omega_{m,0})}{5 \Omega_{m,0}},
\ee
which at $a=1$ tends to $G_{\textrm{eff}}/G_N\simeq \frac35+\frac{2}{5\Omega_{m,0}}$, which is different from unity as expected at this limit. However, in general deviations of $G_{\textrm{eff}}/G_N$ from unity on such scales are not problematic as screening mechanisms play an important role. In any case, our finding is in agreement with what was previously found in Ref.~\cite{Kimura:2010di}, namely: the quasistatic approximation breaks down for the model due to the rapid oscillations of the scalar field. As a result, in what follows we will only focus on our new designer model, which does not suffer from this issue.

\subsection{Designer Model}
We now focus on our designer model HDES, given by Eq.~\eqref{eq:bestdes}. Again, we will consider the numerical solutions for the following cases:
\begin{itemize}
  \item The numerical solution of the full system of equations given by Eqs.~\eqref{eq:1}-\eqref{eq:field-equation-horndeski-4}, which however we rewrite in terms of $\delta X= \dot{\phi} \dot{\delta \phi}-\dot{\phi}^2 \Psi$ as the system is more stable this way. We call this case ``Full-DES".
  \item The numerical solution of the effective fluid approach given by Eqs.~\eqref{Eq:evolution-delta}-\eqref{Eq:evolution-V}. We call this case ``Eff. Fluid".
  \item The numerical solution of the growth factor equation \eqref{eq:ODE-growth}. We call this case ``ODE-Geff".
  \item The \lcdm model.
\end{itemize}

As mentioned in the previous sections, we can absorb the constant $c_0$ in that of $J_c$, so we will only vary the latter, i.e., we set $\tilde{c}_0=1$. Furthermore, since the model is stable for all values of $J_c$ when $n=2$, we will consider this case when studying cosmological constraints. Again, we use $\Omega_{m,0}=0.3$, $k=300H_0$ and $\sigma_{8,0}=0.8$, unless otherwise specified.

\begin{figure*}[!t]
\centering
\includegraphics[width=0.49\textwidth]{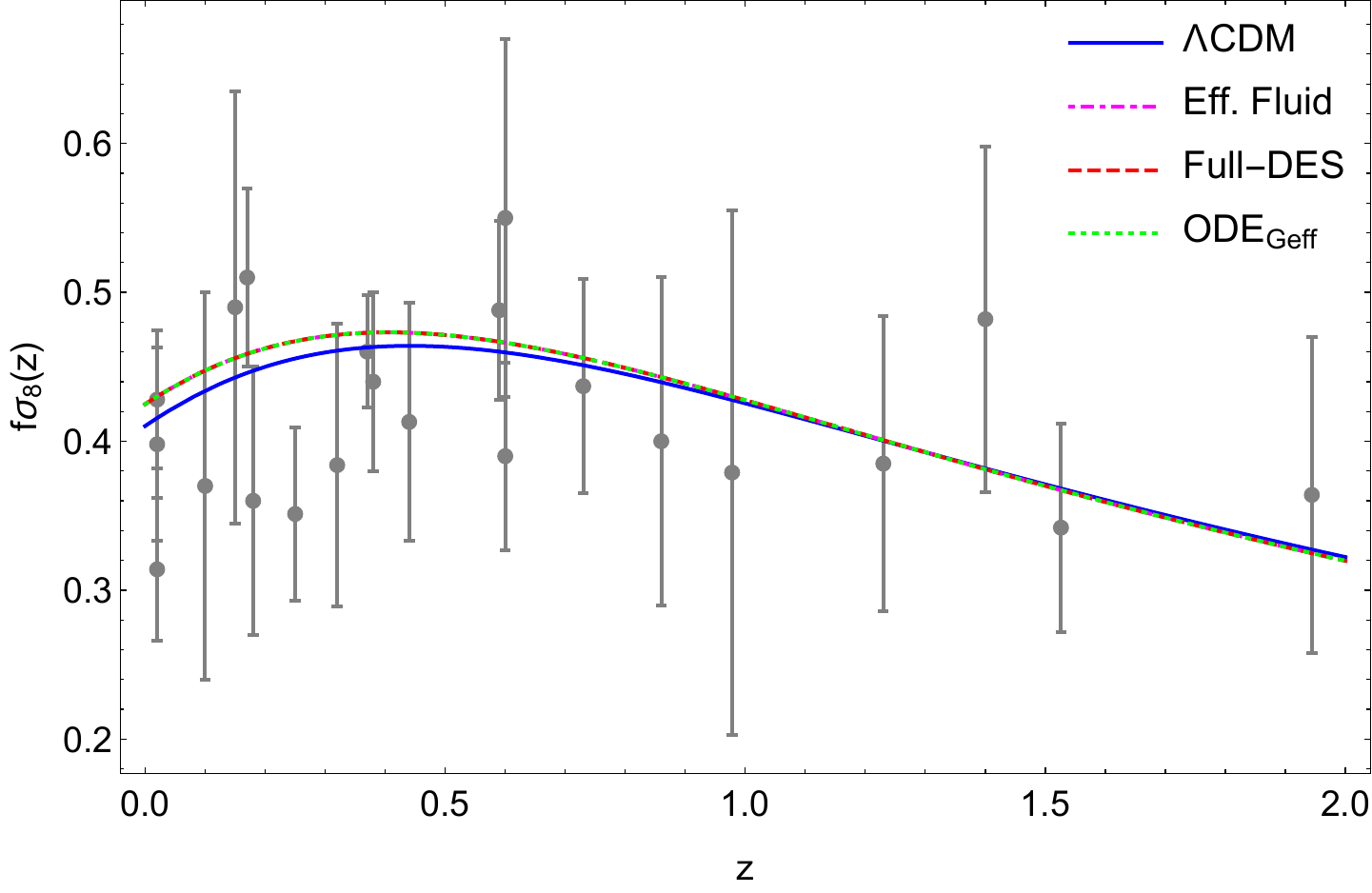}
\includegraphics[width=0.49\textwidth]{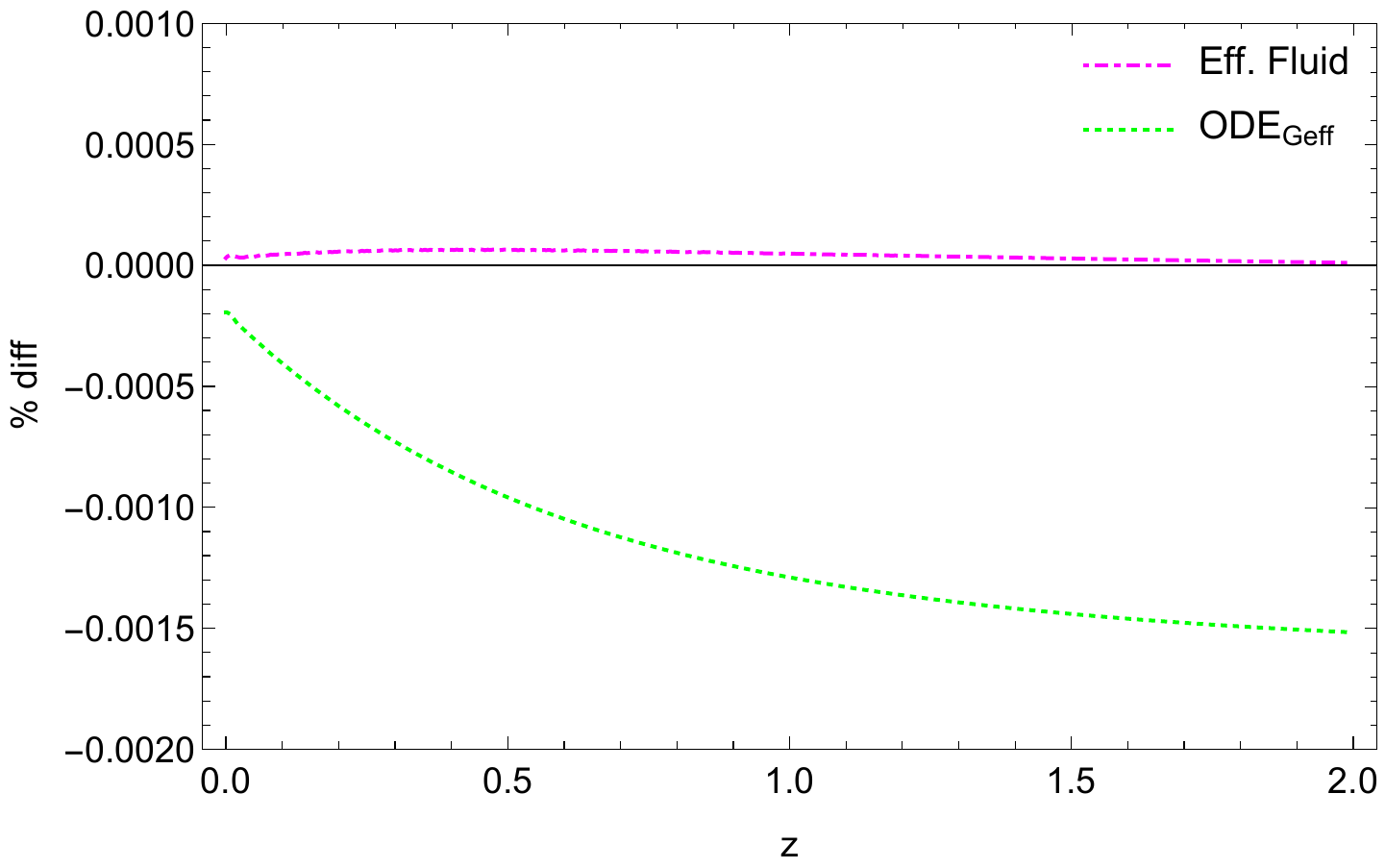}
\caption{Left: We show the evolution of the $f\sigma_8(z)$ parameter for the HDES model with $n=2$, $\tilde{J}_c=5\cdot10^{-2}$ and $\sigma_{8,0}=0.8$ versus the $f\sigma_8$ data compilation from Ref.~\cite{Sagredo:2018ahx}. The values of values for $\tilde{J}_c$ were chosen so as to highlight the differences of these models with respect to GR. Here we show the theoretical curves for the HDES model for the ``Full-DES" brute-force numerical  solution, the effective fluid approach, the $\Lambda$CDM model and the numerical solution of the $G_{\textrm{eff}}$ equation. As can be seen, the agreement with all approaches is excellent. Right: The percent difference between the ``Full-DES" brute-force numerical solution and the effective fluid approach (magenta dot dashed line) and the numerical solution of the growth factor equation \eqref{eq:ODE-growth} (green dotted line). \label{fig:HDESevo}}
\end{figure*}

In the left panel of Fig.~\ref{fig:HDESevo} we show the evolution of the $f\sigma_8(z)$ parameter for the HDES model with $n=2$, $\tilde{J}_c=5\cdot10^{-2}$ and $\sigma_{8,0}=0.8$. The values of values for $\tilde{J}_c$ were chosen so as to highlight the differences of these models with respect to GR. We show the theoretical curves for the HDES model for the ``Full-DES" brute-force numerical  solution, the effective fluid approach, the $\Lambda$CDM model and the numerical solution of the $G_{\textrm{eff}}$ equation. As can be seen, the agreement with all approaches is excellent. In the right panel of the same figure we show the percent difference between the ``Full-DES" brute-force numerical solution and the effective fluid approach (magenta dot dashed line) and the numerical solution of the growth factor equation \eqref{eq:ODE-growth} (green dotted line).

\subsection{Modifications to CLASS and the ISW effect.}
Here we will present our modifications to the CLASS Boltzmann code, which we call EFCLASS. We will compare the outcome with the hi\_CLASS code, which solves the full set of dynamical equations but at the cost of significantly more complicated modifications. At the same time, we will also compare with a brute force calculation of the ISW effect as in our previous paper \cite{Arjona:2018jhh}.

\begin{figure*}[!t]
\centering
\includegraphics[width=0.49\textwidth]{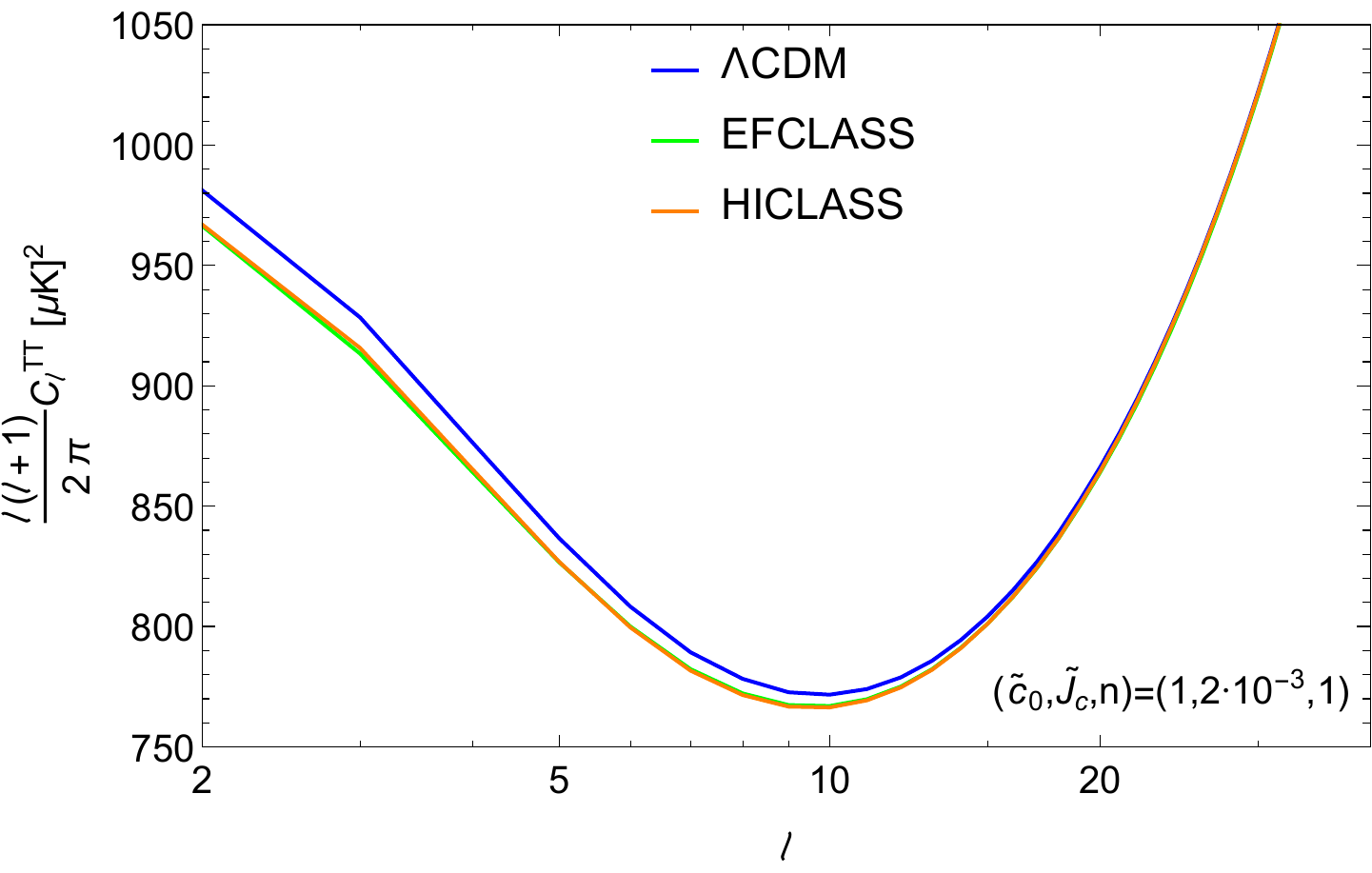}
\includegraphics[width=0.49\textwidth]{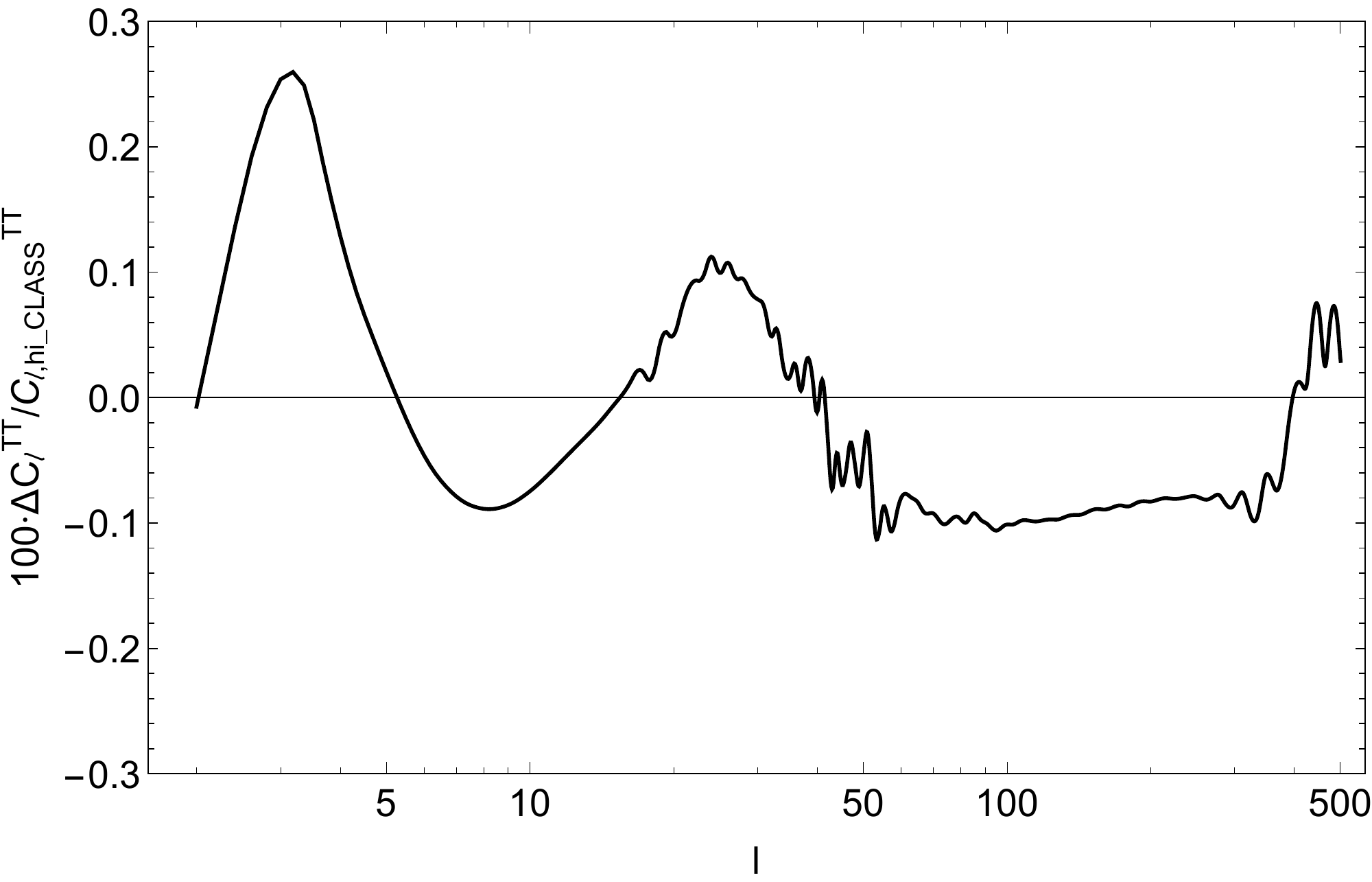}
\caption{Left: The low-$\ell$ multipoles of the TT CMB spectrum for a flat universe with $\Omega_{m,0}=0.3$, $n_s=1$, $A_s=2.3 \cdot 10^{-9}$, $h=0.7$ and $(\tilde{c_0},\tilde{J_c},n)=(1,2\cdot 10^{-3},1)$. The values of values for $\tilde{J}_c$ were chosen so as to highlight the differences of these models with respect to GR.
Our EFCLASS code is denoted by the green line, hi\_CLASS by the orange line and for reference the \lcdm with a blue line. Right: The percent difference of our code with hi\_CLASS as a reference. As can be seen, our simple modification achieves roughly $\sim 0.1\%$ accuracy across all multipoles.  \label{fig:classcls}}
\end{figure*}

In order to modify the CLASS code in our effective fluid approach we only need two functions, the DE velocity and the anisotropic stress \cite{Arjona:2018jhh}. In the case of the HDES model, the anisotropic stress $\pi_{DE}$ is zero, as can be seen from Eq.~(\ref{eq:field-equation-horndeski-4}), since $G_{4\phi}=0$. Therefore, we only need the DE velocity which we can easily be obtained from Eq.~\eqref{eq:effthetakgb}, however we found that this approach is not very stable numerically. Hence, in order to have a consistent solution, we solve Eq.~\eqref{Eq:evolution-V} for $V_{DE}$ and since $w_{DE}=-1$, the only variable we need is the effective pressure $\delta P_{DE}$ given by Eq.~\eqref{eq:effpreskgb}. The expressions are rather cumbersome, but for $n=1$ we have
\be
V_{DE}\simeq \left(-\frac{14 \sqrt{2}}{3} \Omega_{m,0}^{-3/4} \tilde{J_c}~H_0~ a^{1/4}\right)\frac{\bar{\rho}_m}{\bar{\rho}_{DE}} \delta_m.\label{eq:VDEHDES}
\ee
In the left panel of Fig.~\ref{fig:classcls} we show the low-$\ell$ multipoles of the TT CMB spectrum for a flat universe with $\Omega_{m,0}=0.3$, $n_s=1$, $A_s=2.3 \cdot 10^{-9}$, $h=0.7$ and $(\tilde{c_0},\tilde{J_c},n)=(1,2\cdot 10^{-3},1)$. Our EFCLASS code is denoted by the green line, hi\_CLASS by the orange line and for reference the \lcdm with a blue line. On the right panel of Fig.~\ref{fig:classcls} we show the percent difference of our code with hi\_CLASS as a reference\footnote{In this case we did not use $n=2$ as we found that in this case hi\_CLASS crashes and we cannot compare with that code.}. As can be seen, our simple modification achieves roughly $\sim 0.1\%$ accuracy across all multipoles.

We also compare our results with a brute force calculation of the Integrated Sachs-Wolfe (ISW) effect. In this case the power spectrum is given by \cite{Song:2006ej}:
\bea
C_\ell^{\textrm{ISW}}=4\pi \int \frac{dk}{k} I_\ell^{\textrm{ISW}}(k)^2 \frac{9}{25} \frac{k^3 P_{\zeta}}{2\pi^2},\label{eq:clsISWtheory}
\eea
where $I_\ell^{\textrm{ISW}}(k)$ is a kernel that depends on the line of sight integral of the growth and a bessel function and $P_{\zeta}$ is the power spectrum (see Ref. \cite{Song:2006ej} and Appendix A of Ref. \cite{Arjona:2018jhh}), and is given by the primordial power spectrum times a transfer function
\be
\frac{k^3 P_{\zeta}}{2\pi^2}=A_s \left(\frac{k}{k_0}\right)^{n_s-1} T(k)^2,
\ee
where $A_s$ is the primordial amplitude, $k_0$ is the pivot scale and $T(k)$ is the usual matter-radiation Bardeen, Bond, Kaiser and Szalay (BBKS) transfer function (see Eq. (7.71) in Ref.~\cite{Dodelson:2003ft}).

\begin{figure*}[!t]
\centering
\includegraphics[width=0.49\textwidth]{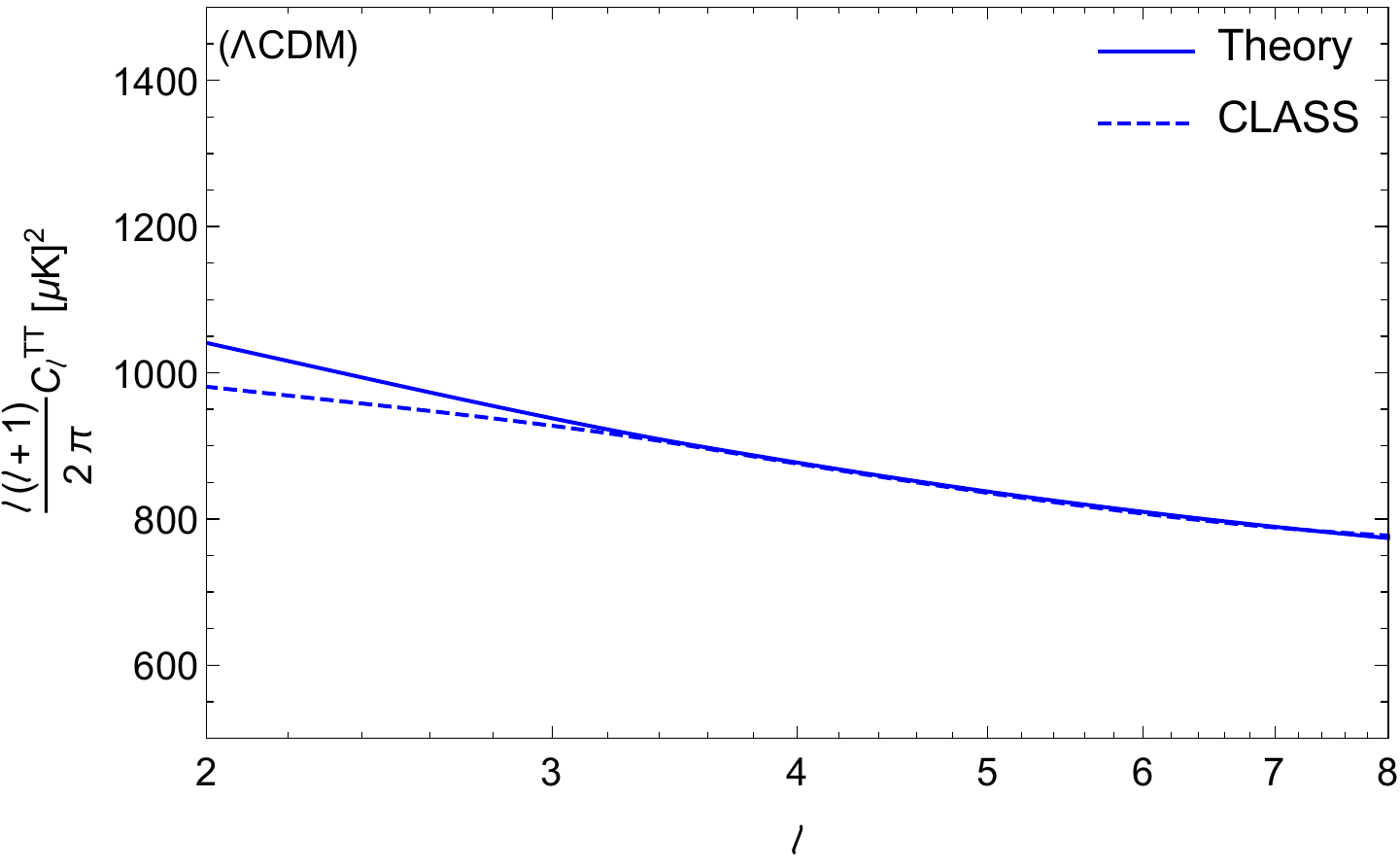}
\includegraphics[width=0.49\textwidth]{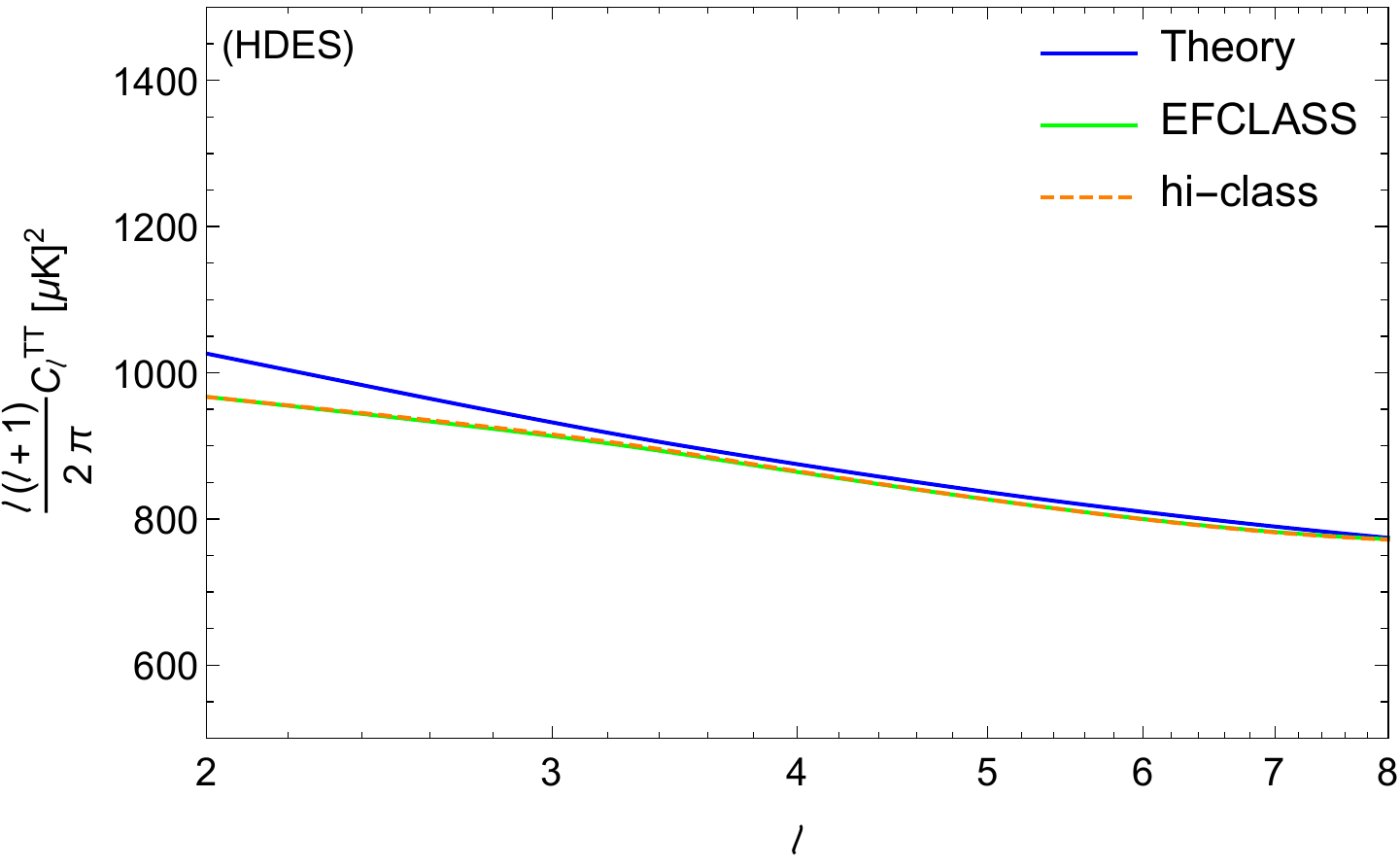}
\caption{The ISW effect and a comparison with CLASS/hi\_CLASS for the \lcdm model (left) and the HDES model (right), for the same parameters as in Fig.~\ref{fig:classcls}. We see that there is excellent agreement for all multipoles, except $\ell=2$ due to the use of the BBKS transfer function which is accurate only up to $10\%$ at large scales. \label{fig:classclsisw}}
\end{figure*}

In Fig.~\ref{fig:classclsisw} we present the results for the calculation of the ISW effect and a comparison with CLASS/hi\_CLASS for the \lcdm model (left) and the HDES model (right), for the same parameters as in Fig.~\ref{fig:classcls}. We see that there is excellent agreement for all multipoles, except $\ell=2$. The reason for this is that we have used the BBKS formula for the transfer function $T(k)$ which is very accurate at small scales, but only at the level of $10\%$ on large scales, i.e., small multipoles.

\section{Cosmological constraints}
\label{Section:mcmc}

Here we present the cosmological constraints for the $n=2$ HDES and \lcdm models discussed in previous sections. We use the latest cosmological observations including the supernovae type Ia (SnIa), Baryon Acoustic Oscillations (BAO), CMB and the Hubble expansion H(z) data. Specifically, we use the Pantheon SnIa compilation of Ref.~\cite{Scolnic:2017caz}, the BAO measurements from 6dFGS \cite{Beutler:2011hx}, SDDS \cite{Anderson:2013zyy}, BOSS CMASS \cite{Xu:2012hg}, WiggleZ \cite{Blake:2012pj}, MGS \cite{Ross:2014qpa}, BOSS DR12 \cite{Gil-Marin:2015nqa} and DES Y1 \cite{Abbott:2017wcz}. For the CMB we use the shift parameters $(R, l_a)$ based on the \textit{Planck 2018} release \cite{Aghanim:2018eyx} and as derived by Ref.~\cite{Zhai:2018vmm}. We assume the existence of three families of neutrinos with $N_{\textrm{eff}}=3.046$.

Furthermore, we also incorporate the direct measurements of the Hubble expansion $H(z)$ data. These can be derived in two ways: by the clustering of galaxies or quasars and by the differential age method. The former provides direct measurements of the Hubble parameter by measuring the BAO peak in the radial direction from the clustering of galaxies or quasars \cite{Gaztanaga:2008xz}. The latter method obtains the Hubble parameter via the redshift drift of distant objects over significant time periods, usually a decade or longer. This is possible as in GR the Hubble parameter can be expressed via the rate of change of the redshift $H(z)=-\frac{1}{1+z}\frac{dz}{dt}$ \cite{Jimenez:2001gg}. These methods result in a compilation of 36 Hubble parameter $H(z)$ data points, which for clarity we show in Table \ref{tab:Hzdata} along with their corresponding references.

The growth-rate data used here are obtained via the redshift-space distortions (RSD). These are sensitive probes of the Large Scale Structure (LSS) and can measure the quantity $f\sigma_8(a)\equiv f(a)\cdot \sigma(a)$, which is a product of the growth rate $f(a)=\frac{d ln\delta}{d lna}$ and the redshift-dependent rms fluctuations $\sigma(a)=\sigma_{8,0}\frac{\delta(a)}{\delta(1)}$ of the linear density field within spheres of radius $R=8 h^{-1} \textrm{\textrm{Mpc}}$. In this notation the parameter $\sigma_{8,0}$ is the value of the rms fluctuations today and is a direct measure of the amplitude of fluctuations in linear scales.

We should mention that $f\sigma_8(a)$ can be estimated via the ratio of the monopole to the quadrupole of the redshift-space power spectrum $P(k)$. The latter is sensitive on the quantity $\beta=f/b_1$, where $f$ is the growth-rate as defined earlier and $b_1$ is the galaxy bias \cite{Percival:2008sh,Song:2008qt,Nesseris:2006er}. In all cases we assume linear theory. The combination $f\sigma_8(a)$ not only is independent of bias, as the latter completely cancels out, but it has also been demonstrated to be an excellent discriminator of DE models as it probes the dynamics of a given gravitational theory and not only the geometric of space-time \cite{Song:2008qt}. The covariances of the data and how to make the necessary corrections for the Alcock-Paczynski effect are given in Refs.~\cite{Sagredo:2018ahx,Nesseris:2017vor,Kazantzidis:2018rnb}, while other related analyses with these data can be found in Refs.~\cite{Basilakos:2018arq,Basilakos:2017rgc,Basilakos:2016nyg,Arjona:2018jhh}.

In this paper we use the growth-rate data compilation of Ref.~\cite{Sagredo:2018ahx}, which we show in Table~\ref{tab:fs8tab} for completeness, along with the corresponding references for each point. This dataset was analyzed in Ref.~\cite{Sagredo:2018ahx} with the ``Internal Robustness method" of Ref.~\cite{Amendola:2012wc}, by examining combinations of subsets and it was shown that this specific dataset is indeed internally robust.

With these in mind, our total likelihood function $L_{\rm tot}$ can be given as the product of the separate likelihoods of the data (we assume they are statistically independent) as follows:
$$
L_{\rm tot}=L_{\rm SnIa} \times L_{\rm BAO} \times L_{\rm H(z)} \times L_{\rm CMB}\times L_{\rm growth},
$$
which is related to the total $\chi^2$ via $\chi^{2}_{\rm tot}=-2\log{L_{\rm tot}}$ or
\be
\chi^{2}_{\rm tot}=\chi^{2}_{\rm SnIa}+\chi^{2}_{\rm BAO}+\chi^{2}_{\rm H(z)}+
\chi^{2}_{\rm cmb}+\chi^{2}_{\rm growth}.\label{eq:chi2eq}
\ee

Calculating the best-fit is not enough, but we also need to study the statistical significance of our constraints. To achieve this we make use of the well known Akaike Information Criterion (AIC)~\cite{Akaike1974}. The AIC estimator is given (assuming Gaussian errors) by
\begin{eqnarray}
{\rm AIC} = -2 \ln {\cal L}_{\rm max}+2k_p+\frac{2k_p(k_p+1)}{N_{\rm dat}-k_p-1} \label{eq:AIC}\;,
\end{eqnarray}
where $k_p$ and $N_{\rm dat}$ stand for the number of free parameters and the total number of data points respectively. For other similar statistical tools see also Ref.~\cite{Liddle:2007fy}. In this analysis we have 1048 data points from the Pantheon set, 3 from the CMB shift parameters, 10 from the BAO measurements, 22 from the growth measurements and finally 36 $H(z)$ points, for a total of $N_{\rm dat}=1118$.

The AIC can be interpreted similarly to the $\chi^2$, i.e. a smaller relative value signifies a better fit to the data. To apply this statistic to model selection we take the pair difference between models $\Delta {\rm  AIC}={\rm AIC}_{\rm model}-{\rm AIC}_{\rm min}$. This can in principle be interpreted with the Jeffreys' scale in the following manner: when $4<\Delta {\rm AIC} <7$ this indicates positive evidence against the model with higher value of ${\rm AIC}_{\rm model}$, while in the case when $\Delta {\rm AIC} \ge 10$ it can be interpreted as strong evidence. On the other hand, when $\Delta {\rm AIC} \le 2$, then this means that the two models are statistically equivalent. However, in Ref.~\cite{Nesseris:2012cq} it has been shown that in general the Jeffreys' scale can sometimes lead to misleading conclusions, and thus it should be interpreted with care.

Finally, our total $\chi^2$ is given by Eq.~(\ref{eq:chi2eq}) while the parameter vectors (assuming a spatially flat Universe) are given by: $p_{\Lambda \textrm{CDM}}=\left(\Omega_{m,0}, 100\Omega_b h^2, h,\sigma_8\right)$ for the \lcdm and  $p_{\textrm{HDES}}=\left(\Omega_{m,0}, 100\Omega_b h^2, h, \tilde{J_c},\sigma_8 \right)$ for the HDES model. Using the aforementioned cosmological data and methodology, we can obtain the best-fit parameters and their uncertainties via the MCMC method based on a Metropolis-Hastings algorithm. The codes used in the analysis were written by one of the authors.\footnote{The MCMC code for Mathematica used in the analysis is freely available at \url{http://members.ift.uam-csic.es/savvas.nesseris/}.} The priors we assumed for the parameters are given by $\Omega_{m,0} \in[0.1, 0.5]$, $\Omega_b h^2 \in[0.001, 0.08]$, $\tilde{J_c} \in[-1, 12]$, $h \in[0.4, 1]$, $\sigma_8\in[0,2]$ and we sample $\sim10^5$ MCMC points for each of the two models.

\begin{table}[!t]
\caption{The $H(z)$ data used in the current analysis (in units of $\textrm{km}~\textrm{s}^{-1} \textrm{Mpc}^{-1}$). This compilation is partly based on those of Refs.~\cite{Moresco:2016mzx} and \cite{Guo:2015gpa}.\label{tab:Hzdata}}
\small
\centering
\begin{tabular}{cccccccccc}
\\
\hline\hline
$z$  & $H(z)$ & $\sigma_{H}$ & Ref.   \\
\hline
$0.07$    & $69.0$   & $19.6$  & \cite{Zhang:2012mp}  \\
$0.09$    & $69.0$   & $12.0$  & \cite{STERN:2009EP} \\
$0.12$    & $68.6$   & $26.2$  & \cite{Zhang:2012mp}  \\
$0.17$    & $83.0$   & $8.0$   & \cite{STERN:2009EP}    \\
$0.179$   & $75.0$   & $4.0$   & \cite{MORESCO:2012JH}   \\
$0.199$   & $75.0$   & $5.0$   & \cite{MORESCO:2012JH}   \\
$0.2$     & $72.9$   & $29.6$  & \cite{Zhang:2012mp}   \\
$0.27$    & $77.0$   & $14.0$  & \cite{STERN:2009EP}   \\
$0.28$    & $88.8$   & $36.6$  & \cite{Zhang:2012mp}  \\
$0.35$    & $82.7$   & $8.4$   & \cite{Chuang:2012qt}   \\
$0.352$   & $83.0$   & $14.0$  & \cite{MORESCO:2012JH}   \\
$0.3802$  & $83.0$   & $13.5$  & \cite{Moresco:2016mzx}   \\
$0.4$     & $95.0$   & $17.0$  & \cite{STERN:2009EP}    \\
$0.4004$  & $77.0$   & $10.2$  & \cite{Moresco:2016mzx}   \\
$0.4247$  & $87.1$   & $11.2$  & \cite{Moresco:2016mzx}   \\
$0.44$    & $82.6$   & $7.8$   & \cite{Blake:2012pj}   \\
$0.44497$ & $92.8$   & $12.9$  & \cite{Moresco:2016mzx}   \\
$0.4783$  & $80.9$   & $9.0$   & \cite{Moresco:2016mzx}   \\
\hline\hline
\end{tabular}
\begin{tabular}{cccccccccc}
\\
\hline\hline
$z$  & $H(z)$ & $\sigma_{H}$ & Ref.   \\
\hline
$0.48$    & $97.0$   & $62.0$  & \cite{STERN:2009EP}   \\
$0.57$    & $96.8$   & $3.4$   & \cite{Anderson:2013zyy}   \\
$0.593$   & $104.0$  & $13.0$  & \cite{MORESCO:2012JH}  \\
$0.60$    & $87.9$   & $6.1$   & \cite{Blake:2012pj}   \\
$0.68$    & $92.0$   & $8.0$   & \cite{MORESCO:2012JH}    \\
$0.73$    & $97.3$   & $7.0$   & \cite{Blake:2012pj}   \\
$0.781$   & $105.0$  & $12.0$  & \cite{MORESCO:2012JH} \\
$0.875$   & $125.0$  & $17.0$  & \cite{MORESCO:2012JH} \\
$0.88$    & $90.0$   & $40.0$  & \cite{STERN:2009EP}   \\
$0.9$     & $117.0$  & $23.0$  & \cite{STERN:2009EP}   \\
$1.037$   & $154.0$  & $20.0$  & \cite{MORESCO:2012JH} \\
$1.3$     & $168.0$  & $17.0$  & \cite{STERN:2009EP}   \\
$1.363$   & $160.0$  & $33.6$  & \cite{Moresco:2015cya}  \\
$1.43$    & $177.0$  & $18.0$  & \cite{STERN:2009EP}   \\
$1.53$    & $140.0$  & $14.0$  & \cite{STERN:2009EP}  \\
$1.75$    & $202.0$  & $40.0$  & \cite{STERN:2009EP}  \\
$1.965$   & $186.5$  & $50.4$  & \cite{Moresco:2015cya}  \\
$2.34$    & $222.0$  & $7.0$   & \cite{Delubac:2014aqe}   \\
\hline\hline
\end{tabular}
\end{table}

\begin{table}[!t]
\caption[]{Compilation of the $f\sigma_8(z)$ measurements used in this analysis along with the reference matter density parameter $\Omega_{m_0}$ (needed for the growth correction) and related references.	 \label{tab:fs8tab}}
\begin{center}
\begin{tabular}{ccccccccc}
\hline
\hline
$z$     & $f\sigma_8(z)$ & $\sigma_{f\sigma_8}(z)$  & $\Omega_{m,0}^\text{ref}$ & Ref. \\ \hline
0.02    & 0.428 & 0.0465  & 0.3 & \cite{Huterer:2016uyq}   \\
0.02    & 0.398 & 0.065   & 0.3 & \cite{Turnbull:2011ty},\cite{Hudson:2012gt} \\
0.02    & 0.314 & 0.048   & 0.266 & \cite{Davis:2010sw},\cite{Hudson:2012gt}  \\
0.10    & 0.370 & 0.130   & 0.3 & \cite{Feix:2015dla}  \\
0.15    & 0.490 & 0.145   & 0.31 & \cite{Howlett:2014opa}  \\
0.17    & 0.510 & 0.060   & 0.3 & \cite{Song:2008qt}  \\
0.18    & 0.360 & 0.090   & 0.27 & \cite{Blake:2013nif} \\
0.38    & 0.440 & 0.060   & 0.27 & \cite{Blake:2013nif} \\
0.25    & 0.3512 & 0.0583 & 0.25 & \cite{Samushia:2011cs} \\
0.37    & 0.4602 & 0.0378 & 0.25 & \cite{Samushia:2011cs} \\
0.32    & 0.384 & 0.095  & 0.274 & \cite{Sanchez:2013tga}   \\
0.59    & 0.488  & 0.060 & 0.307115 & \cite{Chuang:2013wga} \\
0.44    & 0.413  & 0.080 & 0.27 & \cite{Blake:2012pj} \\
0.60    & 0.390  & 0.063 & 0.27 & \cite{Blake:2012pj} \\
0.73    & 0.437  & 0.072 & 0.27 & \cite{Blake:2012pj} \\
0.60    & 0.550  & 0.120 & 0.3 & \cite{Pezzotta:2016gbo} \\
0.86    & 0.400  & 0.110 & 0.3 & \cite{Pezzotta:2016gbo} \\
1.40    & 0.482  & 0.116 & 0.27 & \cite{Okumura:2015lvp} \\
0.978   & 0.379  & 0.176 & 0.31 & \cite{Zhao:2018jxv} \\
1.23    & 0.385  & 0.099 & 0.31 & \cite{Zhao:2018jxv} \\
1.526   & 0.342  & 0.070 & 0.31 & \cite{Zhao:2018jxv} \\
1.944   & 0.364  & 0.106 & 0.31 & \cite{Zhao:2018jxv} \\
\hline
\hline
\end{tabular}
\end{center}
\end{table}

\begin{table}[!t]
\begin{center}
\caption{$\Lambda$CDM parameters with $68\%$ limits based on TT,TE,EE+lowP and a flat $\Lambda$CDM model (middle column) or a $w$CDM model (right column); see Ref.~\cite{Aghanim:2018eyx} and the Planck chains archive.\label{tab:planck}}
\begin{tabular}{ccc}\hline \hline
Parameter & Value ($\Lambda$CDM) & Value ($w$CDM) \\
\hline
$\Omega_b h^2$ & $0.02225\pm0.00016$ & $0.02229\pm0.00016$ \\
$\Omega_c h^2$ & $0.1198\pm0.0015$ & $0.1196\pm0.0015$\\
$n_s$ & $0.9645\pm0.0049$ & $0.9649\pm0.0048$\\
$H_0$ & $67.27\pm0.66$ & $>81.3$\\
$\Omega_m$ & $0.3156\pm0.0091$ & $0.203^{+0.022}_{-0.065}$\\
$w$ & $-1$ & $-1.55^{+0.19}_{-0.38}$\\
$\sigma_8$ & $0.831\pm0.013$ & $0.983^{+0.100}_{-0.055}$\\
\hline \hline
\end{tabular}
\end{center}
\end{table}

\begin{table*}[!t]
\caption{The best-fit parameters for the $\Lambda$CDM and the HDES $(n=2)$ models respectively. \label{tab:bestfits}}
\begin{centering}
\begin{tabular}{cccccc}
Model & $\Omega_{m,0}$ & $100\Omega_b h^2$ & $\tilde{J_c}$ & $h$ & $\sigma_8$  \\\hline
Best-fit values &  &  &  & \\\hline
$\Lambda$CDM  & $ 0.311 \pm 0.006 $ & $ 2.243 \pm 0.014 $ & $0$ & $ 0.680 \pm 0.004 $ & $ 0.758 \pm 0.025 $\\\hline
HDES          & $ 0.313 \pm 0.006 $ & $ 2.240 \pm 0.014 $ & $-0.309 \pm 0.244$ & $ 0.678 \pm 0.004 $ & $ 0.911 \pm 0.068 $ \\\hline
\end{tabular}
\par
\end{centering}
\end{table*}

\begin{table}[!t]
\caption{The $\chi^2$ and AIC parameters for the $\Lambda$CDM and the HDES models respectively. \label{tab:chi2AIC}}
\begin{centering}
\begin{tabular}{cccc}
Model & $\chi^2$ & AIC & $\Delta$AIC \\\hline
$\Lambda$CDM & $ 1087.64 $ & $ 1095.68 $ &$ 0 $ \\\hline
HDES         & $ 1086.30 $ & $ 1096.35 $ &$ 0.678 $ \\\hline
\end{tabular}
\par
\end{centering}
\end{table}

\subsection{Results}

\begin{figure*}[!t]
\centering
\includegraphics[width=0.95\textwidth]{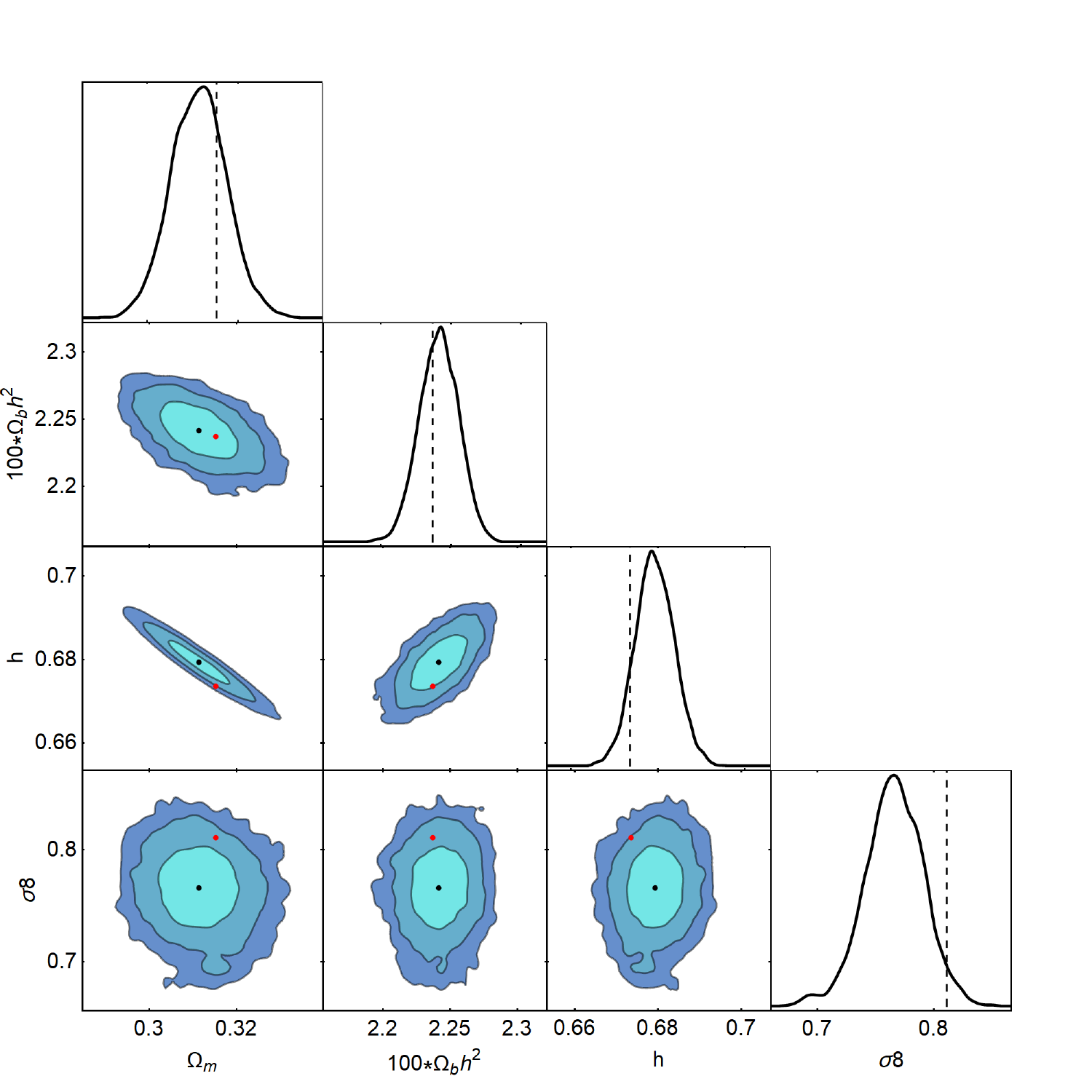}
\caption{The 68.3$\%$, 95.4$\%$ and 99.7$\%$ confidence contours for the \lcdm model, along with the 1D marginalized likelihoods for all parameter combinations. We also highlight with a black point the mean MCMC values and with a red point or dashed vertical line the Planck 2018 concordance cosmology. The latter is based on the TT,TE,EE+lowP spectra, a flat $\Lambda$CDM model and the values are shown in Table \ref{tab:planck}. \label{fig:mcmclcdm}}
\end{figure*}

\begin{figure*}[!t]
\centering
\includegraphics[width=0.95\textwidth]{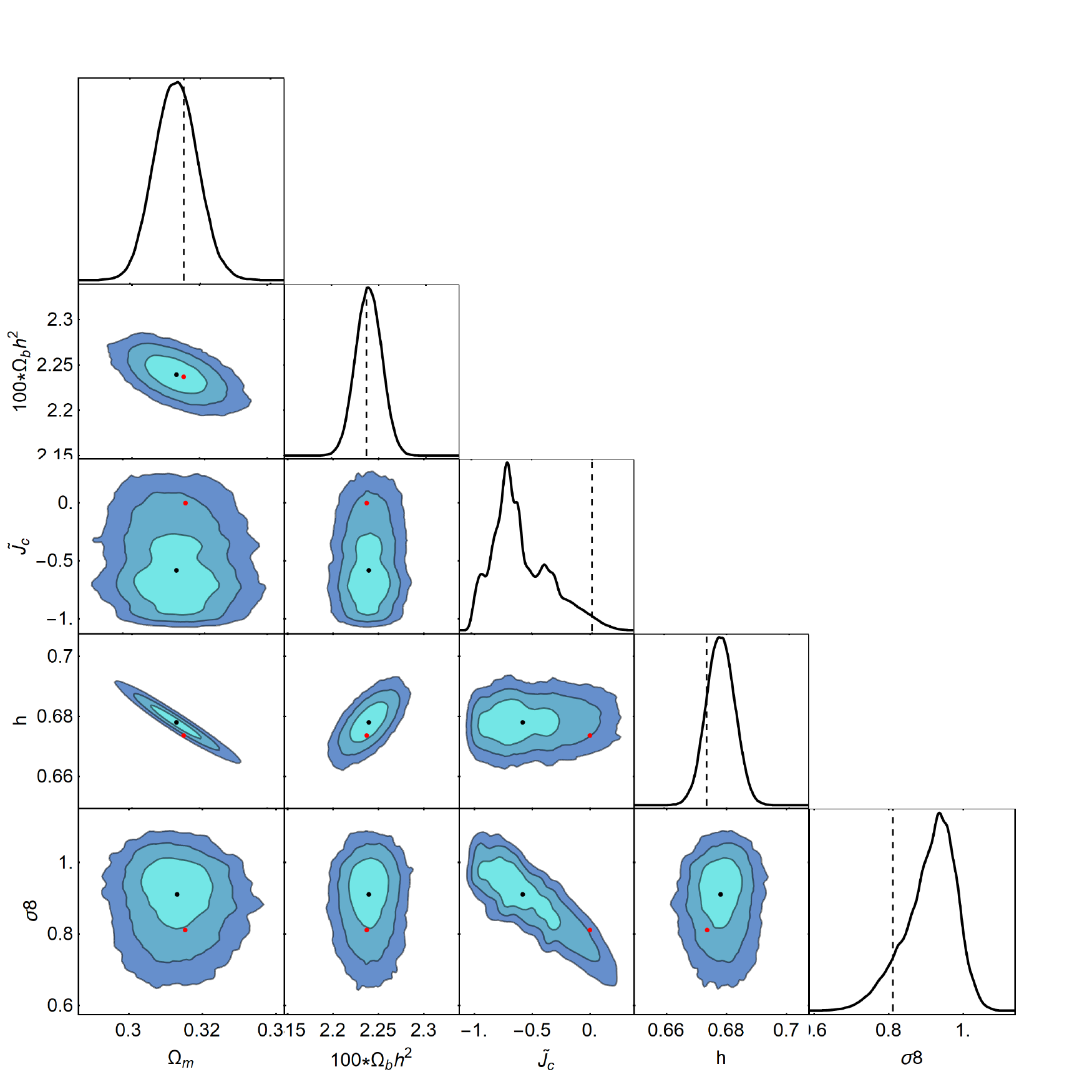}
\caption{The 68.3$\%$, 95.4$\%$ and 99.7$\%$ confidence contours for the HDES $(n=2)$ model, along with the 1D marginalized likelihoods for all parameter combinations. We also highlight with a black point the mean MCMC values and with a red point or dashed vertical line the Planck 2018 concordance cosmology. The latter is based on the TT,TE,EE+lowP spectra, a flat $\Lambda$CDM model and the values are shown in Table \ref{tab:planck}. \label{fig:mcmchdes}}
\end{figure*}

\begin{figure*}[!t]
\centering
\includegraphics[width=0.49\textwidth]{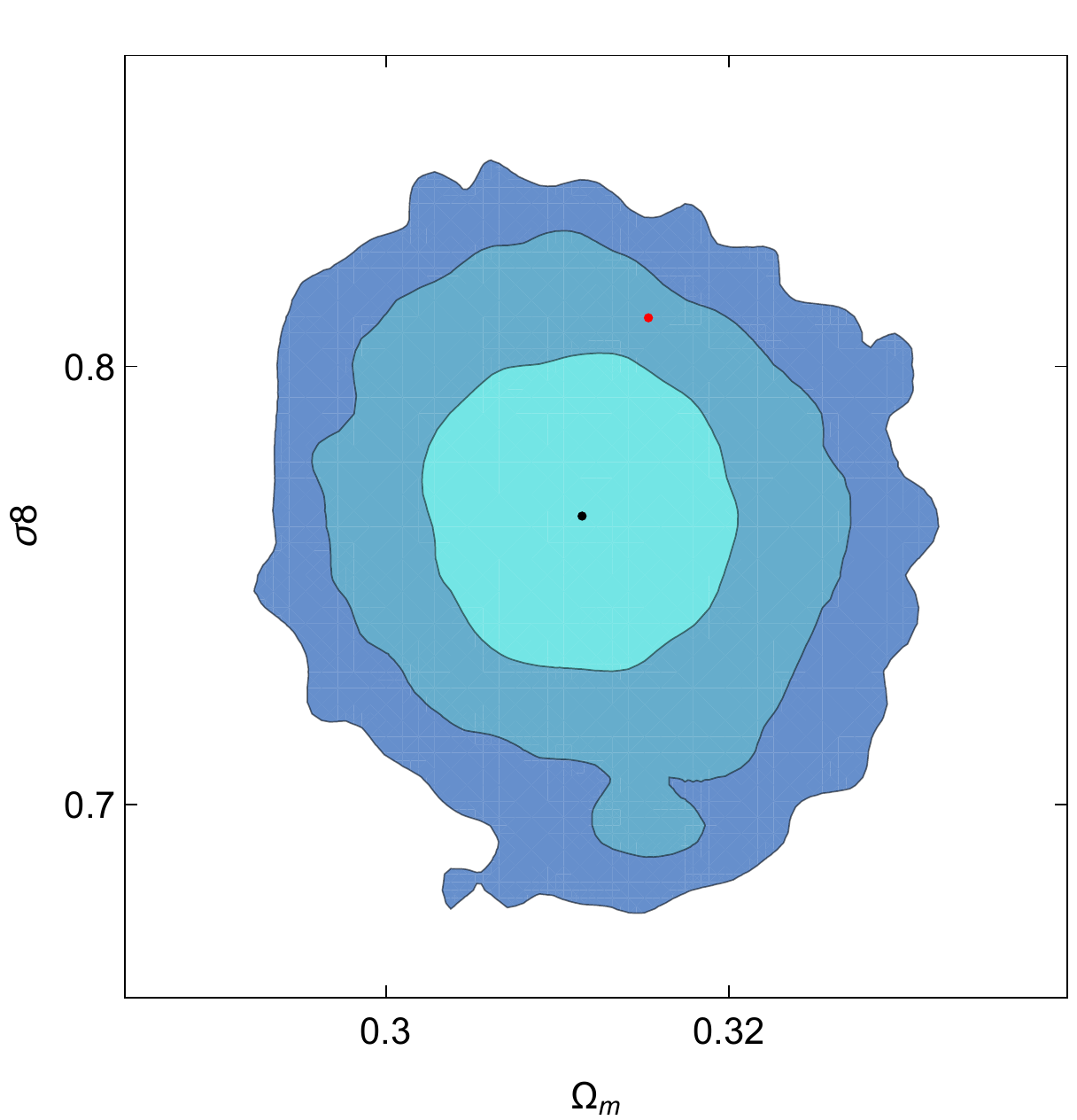}
\includegraphics[width=0.49\textwidth]{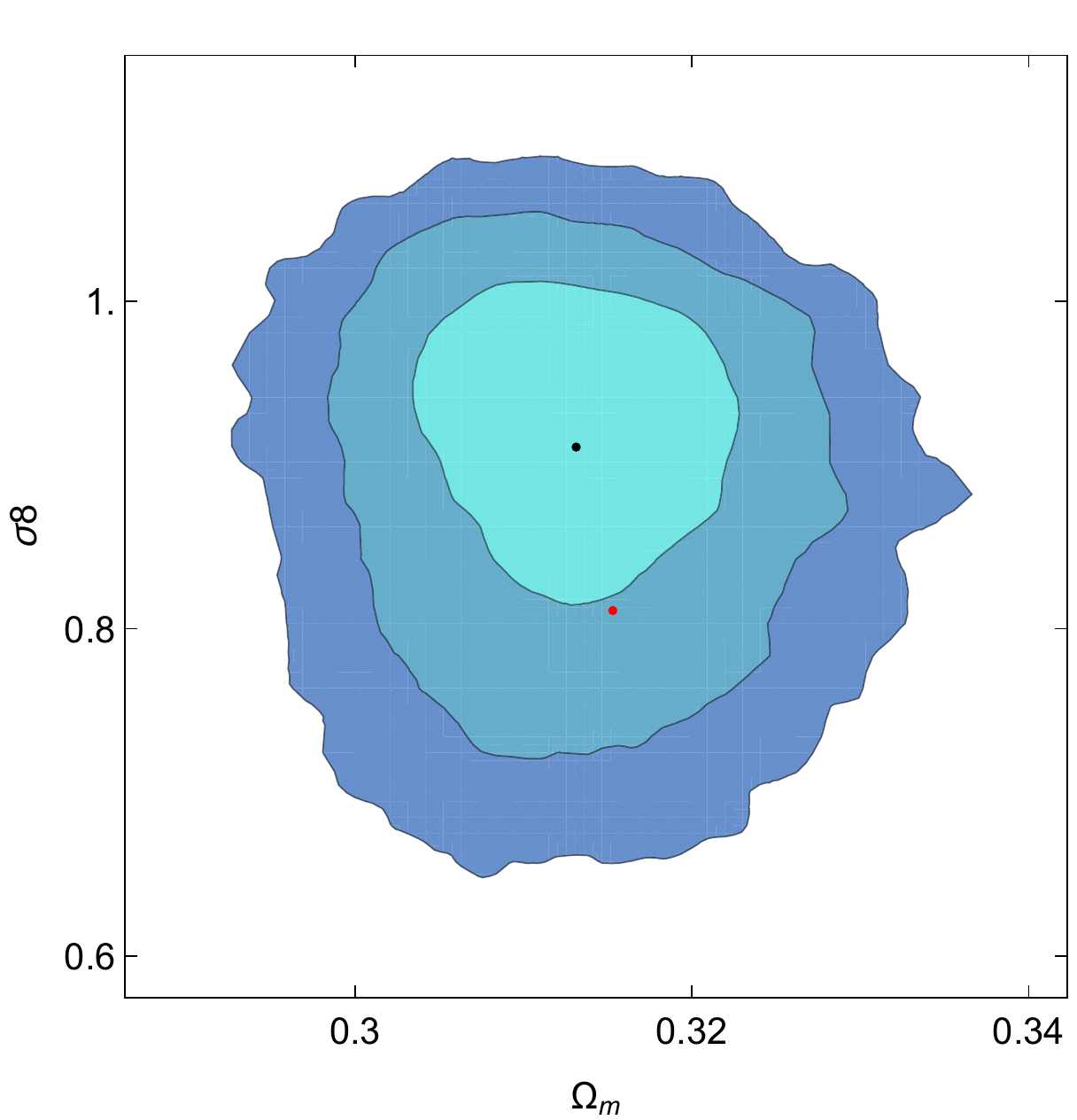}
\caption{The 68.3$\%$, 95.4$\%$ and 99.7$\%$ confidence contours for the \lcdm (left) and the HDES $(n=2)$ (right) models respectively in the $(\Omega_{m,0},\sigma_8)$ plane. We also highlight with a black point the mean MCMC values and with a red point or dashed vertical line the Planck 2018 concordance cosmology. The latter is based on the TT,TE,EE+lowP spectra, a flat $\Lambda$CDM model and the values are shown in Table \ref{tab:planck}. \label{fig:tension}}
\end{figure*}

In Figs.~\ref{fig:mcmclcdm} and \ref{fig:mcmchdes} we show the 68.3$\%$, 95.4$\%$ and 99.7$\%$ confidence contours for the \lcdm and the HDES models, respectively, along with the one-dimensional (1D) marginalized likelihoods for all parameter combinations in the familiar triangle plot. We also highlight with a black point the mean MCMC values and with a red point the Planck 2018 concordance cosmology. The latter is based on the TT,TE,EE+lowP spectra, a flat $\Lambda$CDM model and the values are shown in Table \ref{tab:planck}.

In Tables~\ref{tab:bestfits} and \ref{tab:chi2AIC} we show the best-fit values of the model parameters and the values for the $\chi^2$ and AIC parameters for the $\Lambda$CDM and the HDES model respectively. As can be seen from Tables \ref{tab:bestfits} and \ref{tab:chi2AIC}, we find that as the difference in the AIC parameters is roughly $\sim 0.68$, then both models seem to be statistically equivalent with each other. Furthermore, as seen in Fig.~\ref{fig:mcmchdes}, there is a clear negative correlation between $\tilde{J_c}$ and $\sigma_8$ as we saw in Sec. \ref{sec:degen} and Eq.~\eqref{eq:degenHDES} due to the strong degeneracy between the parameters. This degeneracy is useful as it can potentially alleviate and relax the tension that has been recently observed, see Refs.~\cite{Nesseris:2017vor,Sagredo:2018ahx}. In particular, in Fig~\ref{fig:tension} we show the 68.3$\%$, 95.4$\%$ and 99.7$\%$ confidence contours for the \lcdm (left) and the HDES $(n=2)$ (right) models respectively in the $(\Omega_{m,0},\sigma_8)$ plane. As can be seen, for the HDES model, the best-fit in the $(\Omega_{m,0},\sigma_8)$ plane moves toward higher values of $\sigma_8$, closer to those of Planck.


\section{Conclusions}
\label{Section:conclusions}

The recent discovery of gravitational waves emission from a binary neutron star merger with an optical counterpart, signified a major breakthrough in astrophysics and cosmology as it provided a direct measurement of the speed of propagation of gravitational waves. This observation not only represented an important advance for astronomy, but it also served to greatly reduce the number of alternative models aiming at explaining the current accelerating phase of the Universe. In particular, since the constraint on the speed of propagation of gravitational waves is extremely close to the speed of light, the Horndeski Lagrangian simplified to only three functions. Although this means a notable progress in constraining cosmological models, degeneracies with the \lcdm model remain and must be further investigated.

In this paper we used an effective fluid approach to study the remaining Horndeski Lagrangian. This formalism makes it possible to compare models with different underlying physics (e.g., DE and MG models) in a relatively easy way: each model is mapped to three functions describing the effective fluid, namely, the equation of state $w$, the sound speed $c_s^2$, and the anisotropic stress $\pi$. Even though the remaining Horndeski Lagrangian is now simpler than its original version, finding exact analytical solutions can be quite laborious. Nevertheless, the subhorizon and quasistatic approximations are pretty helpful at overcoming this difficulty.

One of our main results is the set of Eqs.~\eqref{eq:effpres}-\eqref{eq:eff-DE-sound-speed}. These equations along with the equation of state Eq. \eqref{eq:ww} describe the remaining Horndeski Lagrangian in an effective fluid approach under the subhorizon and quasistatic approximations. In this paper, we provide explicit expressions for the effective fluid description of several DE and MG models.

In order to exemplify our results and since we focused on explanations to the late-time accelerating universe, we carried out an analysis where only DM and an effective DE fluid are taken into consideration. A particularly interesting model also included in our formalism is the KGB model. In Sec. \ref{Section:Numerical-Solution} we show our analytical solutions agree pretty well with a full numerical solution of the system of differential equations describing the DM and effective DE perturbations. We also confirm that the subhorizon approximation breaks down for the KGB model due to the rapid oscillations of the scalar field in the large $n$ limit, in agreement with Ref.~\cite{Kimura:2010di}. Also, for the KGB model the background equation for the expansion history $H(a)$ can only be found numerically for $n>1$, thus slowing down the codes significantly.

Due to these problems, we propose a completely new class of Horndeski models based on the designing principle, i.e., fixing the background to a specific model, usually that of the \lcdm and then determining the Lagrangian. Given the freedom in specifying the remaining functions of the Horndeski Lagrangian, we propose a way to find families of models which match a particular background expansion, i.e., the equation of state $w_{DE}$. Since current observations are in good agreement with the standard \lcdm at the background level, we provide equations specifying a $w_{DE}=-1$ designer Horndeski model (see Eqs.~\eqref{eq:bestdes}), which we call HDES. Furthermore, for this model we are able to find exact solutions for the growth $\delta_m(a)$ in the matter domination epoch by solving Eq.~\eqref{eq:ODE-growth}. The solutions we found are given by Eq.~\eqref{eq:HDESfs8} and they imply a degeneracy between $\sigma_8$ and the parameter of the HDES model $\tilde{J_c}$, which can approximately be described via Eq.~\eqref{eq:degenHDES}.

Although fixing the background to \lcdm is a common practice, the treatment of the perturbations might not be rigorous enough in current studies. Public codes solving the perturbation equations for the Horndeski Lagrangian (e.g., hi$\_$CLASS) use ad hoc parametrizations for the $\alpha_i$ functions which differ significantly from our findings that approximate a realistic model (see Eqs. \eqref{eq:bestdesalpha}-\eqref{eq:bestdesalphaT}), see for example Refs.~\cite{Perenon:2019dpc,Noller:2018wyv,SpurioMancini:2019rxy}.

We implemented the parametrized version for the DE effective fluid of our $w_{DE}=-1$ designer Horndeski HDES model in the public code CLASS, which we call EFCLASS, by following the straightforward implementation explained in our previous paper \cite{Arjona:2018jhh}. For the sake of comparison and in order to check the validity of our effective fluid approach, we compared results from our code EFCLASS with the public code hi\_CLASS, which solves numerically the full perturbation equations.

In Fig.~\ref{fig:classcls} we show the CMB angular power spectrum computed with both codes and as can be seen in the right panel of Fig.~\ref{fig:classcls}, the agreement is remarkable and on average on the order of $\sim0.1\%$. Since the hi$\_$CLASS code does not utilize either the subhorizon or the quasistatic approximation, but our EFCLASS does it, we conclude our effective fluid approach is quite accurate and powerful. Furthermore, the main advantage of our method is that while hi\_CLASS requires significant and non-trivial modifications, our EFCLASS code practically only requires the implementation of Eq.~\eqref{eq:VDEHDES}, which is trivial.

We further investigated our $w_{DE}=-1$ designer Horndeski HDES model by computing cosmological constraints with recent data sets using an MCMC analysis. The results of our MCMC analysis are shown in Tables~\ref{tab:bestfits} and \ref{tab:chi2AIC}, where we present the best-fit values of the model parameters and the values for the $\chi^2$ and AIC parameters for the $\Lambda$CDM and the HDES model respectively. We find that as the difference in the AIC parameters is roughly $\sim 0.68$, then both models seem to be statistically equivalent with each other. Furthermore, as seen in Fig.~\ref{fig:mcmchdes}, there is a clear negative correlation between $\tilde{J_c}$ and $\sigma_8$. This can be understood, as we saw in Sec. \ref{sec:degen}, due to the strong degeneracy between the parameters described by Eq.~\eqref{eq:degenHDES}. This degeneracy is useful as it can potentially alleviate the $\sigma_8$ tension that has been recently observed, see Ref.~\cite{Nesseris:2017vor,Sagredo:2018ahx}.


\textbf{Numerical Analysis Files}: The numerical codes used by the authors in the analysis of the paper and our modifications to the CLASS code, which we call EFCLASS, will be released upon publication of the paper on the websites of the EFCLASS \href{https://members.ift.uam-csic.es/savvas.nesseris/efclass.html}{here} and \href{https://github.com/wilmarcardonac/EFCLASS}{here}.

\section*{Acknowledgements}
The authors would like to thank Hector Gil Mar\'{i}n and Tom\'{a}s Ort\'{i}n for useful discussions. They also acknowledge support from the Research Projects FPA2015-68048-03-3P [MINECO-FEDER],  PGC2018-094773-B-C32 and the Centro de Excelencia Severo Ochoa Program SEV-2016-0597. S.N. also acknowledges support from the Ram\'{o}n y Cajal program through Grant No. RYC-2014-15843.

\appendix

\section{Scalar and Gravitational field equations}
\label{Section:appendix-A}

For completeness, in this Appendix we show how to compute both the gravitational and the scalar-field equations derived from the Horndeski action (\ref{eq:action1}).

\subsection{Scalar field equation}

For a function of a single variable with higher derivatives, the stationary values of the functional \cite{courant1953methods}
\bea
& & I[f]=\int_{x_0}^{x_1}\mathcal{L}\left(x,f,f',f'',\cdots,f^{(k)}\right)dx;\hspace{3mm} f' \equiv \frac{df}{dx},\nn \\
& & f'' \equiv \frac{d^2f}{dx^2},\hspace{1mm} f^{(k)} \equiv \frac{d^kf}{dx^k},
\eea
can be obtained from the Euler-Lagrange equation
\begin{equation}
\label{eq:euler}
    \frac{\partial \mathcal{L}}{\partial f}-\frac{d}{dx}\left(\frac{\partial \mathcal{L}}{\partial f'}\right)+\frac{d^2}{dx^2}\left(\frac{\partial \mathcal{L}}{\partial f''}\right)- \dots (-1)^k\frac{d^k}{dx^k}\left(\frac{\partial \mathcal{L}}{\partial f^{k}}\right)=0.
\end{equation}
Since our Lagrangian $\mathcal{L}_i$ functions defined in the Horndeski action (\ref{eq:action1}) depend on the scalar field $\phi$ and its first and second derivatives, we can use the Euler-Lagrange equation (\ref{eq:euler}) to compute the scalar field equation for $\mathcal{L}_2$, $\mathcal{L}_3$ and $\mathcal{L}_4$. For $\mathcal{L}_2$ we have
\bea
\mathcal{L}_2\left(\phi,\partial_{\mu}\phi\right) & = & \frac{\partial \mathcal{L}_2}{\partial \phi}\delta \phi+\frac{\partial \mathcal{L}_2}{\partial_{\mu} \phi}\delta\left(\partial_{\mu}\phi\right) \nn \\
& = & \frac{\partial \mathcal{L}_2}{\partial \phi}\delta \phi-\partial_{\mu}\frac{\partial \mathcal{L}_2}{\partial_{\mu} \phi}\delta\phi,
\eea

\bea
\label{eq:l2b}
\frac{\partial \mathcal{L}_2}{\partial \phi}-\partial_{\mu}\frac{\partial \mathcal{L}_2}{\partial_{\mu} \phi} & = & P^2_{\phi}-\nabla^{\mu}J^2_{\mu} \nn \\
& = & 0.
\eea

Since $\mathcal{L}_2=K\left(\phi,X\right)$, applying Eq.~(\ref{eq:l2b}) leads to

\bea
P^2_{\phi} & = & \frac{\partial \mathcal{L}_2}{\partial \phi} \nn \\
& = & K_{\phi},
\eea

\bea
\nabla^{\mu}J^2_{\mu} & = & \partial_{\mu}\frac{\partial \mathcal{L}_2}{\partial_{\mu} \phi} \nn \\
& = & \nabla^{\mu}\left(\frac{\partial K}{\partial^{\mu}\phi}\right) \nn \\
& = & \nabla^{\mu}\left(\frac{\partial K}{\partial X}\frac{\partial X}{\partial^{\mu}\phi}\right) \nn \\
& = & -\nabla^{\mu}\left(K_{X}\nabla_{\mu}\phi\right),
\eea
where we have replaced the partial derivatives by covariant derivatives and we are using the fact that $X=-\frac{1}{2}\partial_{\mu}\phi \partial^{\mu}\phi$. Hence, for $\mathcal{L}_2$ the scalar field equation reads

\begin{equation}
K_{\phi}+\nabla^{\mu}\left(K_{X}\nabla_{\mu}\phi\right)=0.
\end{equation}

For the term $\mathcal{L}_3$ we follow the same approach

\bea
\mathcal{L}_3\left(\phi,\partial_{\mu}\phi,\partial_{\mu}\partial_{\nu}\phi\right) & = & \frac{\partial \mathcal{L}_3}{\partial \phi}\delta \phi+\frac{\partial \mathcal{L}_3}{\partial_{\mu} \phi}\delta\left(\partial_{\mu}\phi\right) \nn \\
& + & \frac{\partial \mathcal{L}_3}{\partial_{\mu}\partial_{\nu} \phi}\delta\left(\partial_{\mu}\partial_{\nu}\phi\right) \nn \\
& = & \frac{\partial \mathcal{L}_3}{\partial \phi}\delta \phi-\partial_{\mu}\frac{\partial \mathcal{L}_3}{\partial_{\mu} \phi}\delta\phi \nn \\
& + & \partial_{\mu}\partial_{\nu}\frac{\partial \mathcal{L}_3}{\partial_{\mu}\partial_{\nu} \phi}\delta\phi,
\eea

\begin{equation}
\label{eq:l3b}
\frac{\partial \mathcal{L}_3}{\partial \phi}-\partial_{\mu}\frac{\partial \mathcal{L}_3}{\partial_{\mu} \phi}+\partial_{\mu}\partial_{\nu}\frac{\partial \mathcal{L}_3}{\partial_{\mu}\partial_{\nu} \phi}=0.
\end{equation}

Knowing that $\mathcal{L}_3=-G_3\left(\phi,X\right)\left[\Box \phi=g^{\mu \nu}\nabla_{\mu}\nabla_{\nu}\phi\right]$, applying Eq.~(\ref{eq:l3b}) gives

\begin{equation}
\frac{\partial \mathcal{L}_3}{\partial \phi}=-G_{3\phi}\Box \phi,
\end{equation}

\bea
\partial_{\mu}\frac{\partial \mathcal{L}_3}{\partial_{\mu} \phi} & = & \nabla^{\mu}\left(\frac{\partial G_3}{\partial^{\mu}\phi}\Box \phi\right)\nn \\
& = & \nabla^{\mu}\left(\frac{\partial G_3}{\partial X}\frac{\partial X}{\partial^{\mu}\phi}\Box \phi\right) \nn \\
& = & -\nabla^{\mu}\left(G_{3X}\nabla_{\mu}\phi\Box \phi\right),
\eea

\bea
\partial_{\mu}\partial_{\nu}\frac{\partial \mathcal{L}_3}{\partial_{\mu}\partial_{\nu} \phi} & = & -\nabla_{\mu}\left(\nabla_{\nu}g^{\mu \nu}G_3\right) \nn \\
& = & -\nabla^{\mu}\left(G_{3\phi}\nabla_{\mu}\phi+G_{3X}\nabla_{\mu}X\right),
\eea
where we have replaced again the partial derivatives by covariant derivatives. We can then conclude that, for $\mathcal{L}_3$ the scalar field equation reads
\bea
& & -G_{3\phi}\Box \phi-\nabla^{\mu}\left(G_{3X}\nabla_{\mu}\phi\Box \phi\right) \nn \\
& & -\nabla^{\mu}\left(G_{3\phi}\nabla_{\mu}\phi\right)-\nabla^{\mu}\left(G_{3X}\nabla_{\mu}X\right)=0.
\eea
and we make the following assignment
\begin{align}
    P^3_{\phi}&=\nabla_{\mu}G_{3\phi}\nabla^{\mu}\phi,\\
    \nabla^{\mu}J^3_{\mu}&=\nabla^{\mu}\left(-G_{3X}\nabla_{\mu}\phi+G_{3X}\nabla_{\mu}X+2G_{3\phi}\nabla_{\mu}\phi\right).
\end{align}

For $\mathcal{L}_4$ we have
\begin{equation}
    \mathcal{L}_4\left(\phi\right)=\frac{\partial \mathcal{L}_4}{\partial \phi}\delta\phi,
\end{equation}

\begin{equation}
\label{eq:l4}
\frac{\partial \mathcal{L}_4}{\partial \phi}=P^4_{\phi}=0.
\end{equation}

Since $\mathcal{L}_4=G_4\left(\phi\right)R$, applying Eq.~(\ref{eq:l4}) leads to
\begin{equation}
    P^4_{\phi}=G_{4\phi}R.
\end{equation}

Our result for the scalar field equation considering $G_{4X}=0$ and $G_5=0$ is in full agreement with Ref.~\cite{Kobayashi:2011nu}. Hence, the scalar-field equation can be written as
\begin{equation}
    \nabla^{\mu}\left(\sum_{i=2}^{4}J^{i}_{\mu}\right)=\sum^{4}_{i=2}P^{i}_{\phi}.
\end{equation}

\subsection{Gravitational field equations}

Defining the arbitrary functions $\mathcal{L}_i$ from the action (\ref{eq:action1}) as

\bea
\mathcal{L}_2&=&K\left(\phi,X\right),\\
\mathcal{L}_3&=&-G_3\left(\phi,X\right)\Box \phi,\\
\mathcal{L}_4&=&G_4(\phi) R,
\eea
we can then vary the action with respect to the metric tensor; using the principle of least action, this leads to

\begin{equation}
    \delta S=\delta S_2+\delta S_3+\delta S_4+ \delta \left(\sqrt{-g}\mathcal{L}_m\right)=0.
\end{equation}

For $\delta S_2$ we have
\begin{align}
    \delta S_2&=\int d^4x \left[\delta \sqrt{-g}K+\sqrt{-g}\delta K\right],
\end{align}
and using the fact that
\begin{equation}
    \delta\sqrt{-g}=-\frac{1}{2}\sqrt{-g}g_{\mu \nu}\delta g^{\mu \nu},
\end{equation}
and that the variation of $K$ with respect to the metric can be written as
\begin{equation}
    \delta K\left(\phi, X\right)=K_{X}\delta g^{\mu \nu}\left(-\frac{1}{2}\nabla_{\mu}\phi\nabla_{\nu}\phi\right),
\end{equation}
we get
\begin{align}
    \delta S_2&=\int d^4x \sqrt{-g}\delta g^{\mu \nu} \left[-\frac{1}{2}Kg_{\mu \nu}-\frac{1}{2}K_{X}\nabla_{\mu}\phi\nabla_{\nu}\phi \right].
\end{align}

For $\delta S_3$ we have

\begin{align}
   \delta S_3&=\int d^4x \left[-\delta \sqrt{-g}G_3\Box \phi-\sqrt{-g}\delta\left(G_3\Box \phi\right)\right].
\end{align}
The variations of $G_3$ with respect to the metric can be written as
\bea
\delta \left(G_3\left(\phi, X\right)\Box \phi\right)&=&\delta G_3\Box \phi + G_3\delta \left(\Box \phi\right)\nn \\ 
& = & G_{3X}\delta g^{\mu \nu}\left(-\frac{1}{2}\nabla_{\mu}\phi\nabla_{\nu}\phi\right)\Box \phi \nn \\
& + & G_3\delta \left(\Box \phi\right),
\eea
hence
\bea
& & \delta S_3 = \int d^4x\sqrt{-g} \left[\frac{1}{2}g_{\mu \nu}\delta g^{\mu \nu}G_3\Box \phi \right. \nn \\
& & \left. +\frac{1}{2}\delta g^{\mu \nu}G_{3X}\Box \phi \nabla_{\mu}\phi\nabla_{\nu}\phi+G_3\delta \left(\Box \phi\right)\right].\label{eq:A30}
\eea
The last term of the above equation can be expanded in the following way
\begin{align}
    \delta \Box\phi &=\delta g^{ab}\nabla_a\nabla_b\phi+g^{ab}\delta \left(\nabla_a\nabla_b\phi\right)\nn \\
    &=\delta g^{ab}\nabla_a\nabla_b\phi+\Box \left(\delta \phi\right)-g^{ab}\delta \Gamma^{\gamma}_{ab}\partial_{\gamma}\phi,
\end{align}
since
\begin{equation}
    \nabla_a\nabla_b\phi=\partial_a\partial_b\phi-\Gamma^{\gamma}_{ab}\partial_{\gamma}\phi,
\end{equation}
and
\begin{equation}
    \delta\left(\nabla_a\nabla_b\phi\right)=\nabla_a\nabla_b\left(\delta \phi\right)-\delta\Gamma^{\gamma}_{ab}\partial_{\gamma}\phi.
\end{equation}
Also we have that $g^{ab}\Gamma^{\gamma}_{ab}=\ldots=-\nabla_a\delta g^{\gamma a}+\frac{1}{2}g_{ab}g^{\gamma \lambda}\nabla_{\lambda}\delta g^{ab}$, so we get for the last term in Eq.~\eqref{eq:A30}:
\begin{widetext}
\begin{align}
    \delta S_{\textrm{last-term}}&=\int d^4x\sqrt{-g}\left(-G_3\right)\left(\delta g^{ab}\nabla_a\nabla_b\phi+\Box \delta \phi+\left(\nabla_a\delta g^{\gamma a}-\frac{1}{2}g_{ab}g^{\gamma \lambda}\nabla_{\lambda}\delta g^{ab}\right)\partial_{\gamma}\phi\right)\nn \\
    &=\int d^4x\sqrt{-g}\left[-\delta g^{\mu \nu}\left(\nabla_{\mu}\nabla_{\nu}\right)G_3+\delta g^{\gamma a}\nabla_a\left(G_3\nabla_{\gamma}\phi\right)-\frac{1}{2}\delta g^{ab}g_{ab}g^{\gamma \lambda}\nabla_{\lambda}\left(G_3\nabla_{\gamma}\phi\right)\right]\nn \\
    &=\int d^4x\sqrt{-g}\left[-\delta g^{\mu \nu}\left(\nabla_{\mu}\nabla_{\nu}\right)G_3+\delta g^{\mu \nu}\nabla_{\nu}\left(G_3\nabla_{\mu}\phi\right)-\frac{1}{2}\delta g^{\mu\nu}g_{\mu\nu}\nabla^{\gamma}\left(G_3\nabla_{\gamma}\phi\right)\right]\nn \\
    &=\int d^4x\sqrt{-g}\delta g^{\mu \nu}\left[\left(\nabla_{(\mu}\phi\right)\left(\nabla_{\nu)}G_3\right)-\frac{1}{2}g_{\mu\nu}\nabla^{\gamma}\left(G_3\nabla_{\gamma}\phi\right)\right].
\end{align}
\end{widetext}
Combining all terms we have
\bea
& & \delta S_3=\int d^4x\sqrt{-g}\delta g^{\mu \nu} \left[\frac{1}{2}G_{3X}\Box \phi \nabla_{\mu}\phi\nabla_{\nu}\phi \right. \nn \\
& & \left. +\nabla_{(\mu}G_3\nabla_{\nu)}\phi-\frac{1}{2}g_{\mu\nu}\nabla_{\lambda}G_3\nabla^{\lambda}\phi\right].
\eea

For $\delta S_4$ we have
\begin{align}
   \delta S_4&=\int d^4x \left[\delta \sqrt{-g}G_4 R+\sqrt{-g}G_4\delta R\right],
\end{align}
where
\bea
\delta R & = & \delta\left(g^{\mu \nu}R_{\mu \nu}\right) \nn \\
& = & R_{\mu \nu}\delta g^{\mu \nu}+g^{\mu \nu}\delta R_{\mu \nu} \nn \\
& = & R_{\mu \nu}\delta g^{\mu \nu}+g^{\mu \nu}\left(\nabla_{\rho}\delta \Gamma^{\rho}_{\nu \mu}-\nabla_{\nu}\delta \Gamma^{\rho}_{\rho \mu}\right).\label{eq:A37}
\eea
Since $\delta \Gamma^{\lambda}_{\mu \nu}$ is the difference of two connections, it should transform as a tensor. Therefore, it can be written as
\begin{equation}
    \delta \Gamma^{\lambda}_{\mu \nu}=\frac{1}{2}g^{\lambda \alpha}\left(\nabla_{\mu}\delta g_{\alpha \nu}+\nabla_{\nu}\delta g_{\alpha \mu}-\nabla_{\alpha}\delta g_{\mu \nu}\right).\label{eq:A38}
\end{equation}
Then, substituting Eq.~\eqref{eq:A38} into \eqref{eq:A37}, we get
\begin{equation}
    \delta R=R_{\mu\nu}\delta g^{\mu \nu}+g_{\mu \nu}\Box\left(\delta g^{\mu \nu}\right)-\nabla_{\mu}\nabla_{\nu}\left(\delta g^{\mu \nu}\right),
\end{equation}
hence
\begin{widetext}
\begin{align}
\delta S_4&=\int d^4x \sqrt{-g}\left[-\frac{1}{2}g_{\mu \nu}\delta g^{\mu \nu}G_4R+G_4R_{\mu \nu}\delta g^{\mu \nu}+G_4\left(g_{\mu \nu}\Box\left(\delta g^{\mu \nu}\right)-\nabla_{\mu}\nabla_{\nu}\left(\delta g^{\mu \nu}\right)\right)\right]\nn \\
&=\int d^4x \sqrt{-g}\delta g^{\mu \nu}\left[G_{\mu \nu}G_4+g_{\mu \nu}\Box G_4-\nabla_{\mu}\nabla_{\nu}G_4 +\text{total derivatives}\right]\nn \\
&=\int d^4x \sqrt{-g}\delta g^{\mu \nu}\left[G_{\mu \nu}G_4+g_{\mu \nu}\left(G_{4\phi}\Box\phi-2X G_{4\phi\phi}\right)-G_{4\phi}\nabla_{\mu}\nabla_{\nu}\phi-G_{4\phi\phi}\nabla_{\mu}\phi\nabla_{\nu}\phi +\text{total derivatives}\right]
\end{align}
\end{widetext}
where
\bea
-\nabla_{\mu}\left(\nabla_{\nu}G_4\right) & = & -\nabla_{\mu}\left(\nabla_{\nu}\phi G_{4\phi}\right) \nn \\
& = & -\nabla_{\mu}\nabla_{\nu}\phi G_{4\phi}-\nabla_{\mu}\phi\nabla_{\nu}\phi G_{4\phi\phi},~~~~~~~~~
\eea
\bea
g_{\mu \nu}\Box G_4 & = & g_{\mu \nu}\left(g^{ab}\nabla_a\nabla_bG_4\right) \nn \\
& = & g_{\mu \nu}\left(g^{ab}\nabla_a\left(\nabla_b\phi G_{4\phi}\right)\right) \nn \\
& = & g_{\mu \nu}\left(g^{ab}\nabla_a\nabla_b\phi G_{4\phi}+g^{ab}\nabla_b\phi\nabla_a G_{4\phi}\right) \nn \\
& = & g_{\mu \nu}\left(\Box \phi G_{4\phi}-2XG_{4\phi\phi}\right).
\eea
Since the energy-momentum tensor is defined as
\begin{equation}
    T^{(m)}_{\mu \nu}=-\frac{2}{\sqrt{-g}}\frac{\delta \left(\sqrt{-g}\mathcal{L}_m\right)}{\delta g^{\mu \nu}},
\end{equation}
the gravitational field equation can be written
\bea
& &  T^{(m)}_{\mu \nu} = -K_{X}\nabla_{\mu}\phi\nabla_{\nu}\phi -Kg_{\mu \nu} + G_{3X}\Box \phi \nabla_{\mu}\phi \nabla_{\nu}\phi \nn \\
& & + 2\nabla_{(\mu}G_3\nabla_{\nu)}\phi - g_{\mu \nu}\nabla_{\lambda}G_3\nabla^{\lambda}\phi + 2G_4G_{\mu \nu} \nn \\ & & + 2g_{\mu \nu}\left(G_{4\phi}\Box \phi-2XG_{4\phi \phi}\right) - 2G_{4\phi}\nabla_{\mu}\nabla_{\nu}\phi \nn \\
& & - 2G_{4\phi\phi}\nabla_{\mu}\phi\nabla_{\nu}\phi.
\eea

\section{Coefficients}
\label{Section:appendix-B}

Here we show the coefficients for the perturbations in the Horndeski theory in Eq. \eqref{eq:action1}. They are given by:
\bea
A_1 & = & -3\dot{\phi}^3G_{3X}+12HG_4+6\dot{\phi}G_{4\phi},
\eea
\bea
A_2 & = & -\dot{\phi}\left(K_{X}+\dot{\phi}^2K_{XX}\right) + 2\dot{\phi}G_{3\phi} \nn \\
& & -  3H\dot{\phi}^2\left(3G_{3X}+\dot{\phi}^2G_{3XX}\right)+\dot{\phi}^3G_{3\phi X} \nn \\
& & +  6HG_{4\phi},
\eea
\bea
A_3 & = & 4G_4,
\eea
\bea
A_4 & = & \dot{\phi}^2\left(K_{X}+\dot{\phi}^2K_{XX}\right)-2\dot{\phi}^2G_{3\phi}-\dot{\phi}^4G_{3\phi X} \nn \\
& & +  3H\dot{\phi}^3\left(4G_{3X}+\dot{\phi}^2G_{3XX}\right)\nn \\
& & -  12H\left(HG_4+\dot{\phi}G_{4\phi}\right),
\eea
\bea
A_6 & = & -\dot{\phi}^2G_{3X}+2G_{4\phi},
\eea
\bea
\mu & = & -K_{\phi}+\dot{\phi}^2K_{\phi X}-\dot{\phi}^2G_{3\phi \phi}+3H\dot{\phi}^3G_{3\phi X} \nn \\
& & -  6H^2G_{4\phi}-6H\dot{\phi}G_{4\phi\phi},
\eea
\bea
B_1 & = & 12G_4,
\eea
\bea
B_2 & = & -3\dot{\phi}^2G_{3X}+6G_{4\phi},
\eea
\bea
B_3 & = & 12\left(\dot{\phi}G_{4\phi}+3HG_4\right),
\eea
\bea
B_4 & = & 3\left[\dot{\phi}K_{X}-2\dot{\phi}G_{3\phi}-2\dot{\phi}\ddot{\phi}G_{3X} \right. \nn \\
& - & \left. \dot{\phi}^3\left(G_{3\phi X}+\ddot{\phi}G_{3XX}\right)+4HG_{4\phi}+4\dot{\phi}G_{4\phi\phi}\right],~~~~~~~~
\eea
\bea
B_5 & = & 3\left(\dot{\phi}^3G_{3X}-4HG_4-2\dot{\phi}G_{4\phi}\right),
\eea
\bea
B_6 & = & 4G_4, \hspace{2mm} B_7=4G_{4\phi}, \hspace{2mm} B_8=4G_4,
\eea

\bea
B_9 & = & -3K_X\dot{\phi}^{2}+6G_{3\phi}\dot{\phi}^{2}+3G_{3\phi X}\dot{\phi}^{4} \nn \\
& + & 12G_{3X}\dot{\phi}^{2}\ddot{\phi}+3G_{3XX}\dot{\phi}^{4}\ddot{\phi}-36G_4H^2-24G_4\dot{H} \nn \\
& - & 24G_{4\phi}H\dot{\phi}-12G_{4\phi\phi}\dot{\phi}^{2}-12G_{4\phi}\ddot{\phi},
\eea
and using Eq.(\ref{eq:friedmann2-horndeski}) to eliminate $G_4$ in favor of $K$ we can express $B_9$ as
\bea
B_9 & = & 3\left(2K-\dot{\phi}^2K_{X}+2\dot{\phi}^2\ddot{\phi}G_{3X}+\dot{\phi}^4G_{3\phi X} \right. \nn \\
& + & \left. \dot{\phi}^4\ddot{\phi}G_{3XX}\right),
\eea
\bea
\nu & = & K_{\phi}-\dot{\phi}^2\left(G_{3\phi \phi}+\ddot{\phi}G_{3\phi X}\right)\nn \\
& + & 2\left(3H^2+2\dot{H}\right)G_{4\phi} + 2\left(\ddot{\phi} + 2H\dot{\phi}\right)G_{4\phi\phi} \nn \\
& + & 2\dot{\phi}^2G_{4\phi\phi\phi},
\eea
\bea
C_1 & = & 4G_4,
\eea
\bea
C_2 & = & -\dot{\phi}^2G_{3X}+2G_{4\phi},
\eea
\bea
C_3 & = & \dot{\phi}^3G_{3X}-4HG_4-2\dot{\phi}G_{4\phi},
\eea
\bea
C_4 & = & \dot{\phi}\left(K_{X}-2G_{3\phi}+2G_{4\phi\phi}\right)\nn \\
& + & H\left(3\dot{\phi}^2G_{3X}-2G_{4\phi}\right),
\eea
\bea
D_1 & = & -3\left(\dot{\phi}^2G_{3X}-2G_{4\phi}\right),
\eea
\bea
D_2 & = & -K_{X}-\dot{\phi}^2K_{XX}+2G_{3\phi}-6H\dot{\phi}G_{3X}+\dot{\phi}^2G_{3\phi X} \nn \\
& - & 3H\dot{\phi}^3G_{3XX},
\eea
\bea
D_3 & = & -3\left(\dot{\phi}K_{X}-2\dot{\phi}G_{3\phi}+6H\dot{\phi}^2G_{3X}+2\dot{\phi}\ddot{\phi}G_{3X} \right. \nn \\
& + & \left. \dot{\phi}^3G_{3\phi X}+\dot{\phi}^3\ddot{\phi}G_{3XX}-8HG_{4\phi}\right),
\eea
\bea
D_4 & = & \frac{d}{dt}D_2+3HD_2  \nn \\
& = & - 3HK_X-K_{\phi X}\dot{\phi}-K_{\phi XX}\dot{\phi}^3-K_{XXX}\dot{\phi}^3\ddot{\phi} \nn \\
& - & 3K_{XX}\left(H\dot{\phi}^2+\dot{\phi}\ddot{\phi}\right)+6HG_{3\phi} \nn \\
& + & 2G_{3\phi \phi}\dot{\phi}-6G_{3X}\left(3H^2\dot{\phi}+\dot{H}\dot{\phi}+H\ddot{\phi}\right) \nn \\
& + & G_{3\phi X}\left(-3H\dot{\phi}^2+4\dot{\phi}\ddot{\phi}\right)+G_{3\phi \phi X}\dot{\phi}^3 \nn \\
& - & 3G_{3XX}\dot{\phi}^2\left(3H^2\dot{\phi}+\dot{H}\dot{\phi}+5H\ddot{\phi}\right) \nn \\
&  + & G_{\phi XX}\left(H\dot{\phi}^3\ddot{\phi}-3H\dot{\phi}^4\right)-3G_{3XXX}H\dot{\phi}^4\ddot{\phi},~~~~~~~~~~~
\eea
\bea
D_5 & = & \dot{\phi}\left(K_{X}+\dot{\phi}^2_{XX}-2G_{3\phi}-\dot{\phi}^2G_{3\phi X}\right) \nn \\
& + & 3H\left(3\dot{\phi}^2G_{3X}+\dot{\phi}^4G_{3XX}-2G_{4\phi}\right),
\eea

\bea
D_7 & = &4G_{4\phi},
\eea

\bea
D_8 & = & 9H\dot{\phi}^{-1}K+3K_{\phi} \nn \\
& - & 3\left(\ddot{\phi}+3H\dot{\phi}\right)K_X-3\dot{\phi}^{2}\left(K_{\phi X}+\ddot{\phi}K_{XX}\right) \nn \\
& + & 3\left(2\ddot{\phi}+3H\dot{\phi}\right)G_{3\phi} \nn \\
& - & 9\dot{\phi}\left(3H\ddot{\phi}+3H^2\dot{\phi}+\dot{H}\dot{\phi}\right)G_{3X} \nn \\
& + & 3\dot{\phi}^{2}G_{3\phi \phi}+3\dot{\phi}^{2}\left(\ddot{\phi}-3H\dot{\phi}\right)G_{3\phi X} \nn \\
& - & 9H\dot{\phi}^{3}\ddot{\phi}G_{3XX}+18H\dot{\phi}^{-1}\left(3H^2+2\dot{H}\right)G_4 \nn \\
& + & 18\dot{\phi}^{-1}\left(H\ddot{\phi}+4H^2\dot{\phi}+\dot{H}\dot{\phi}\right)G_{4\phi} \nn \\
& + & 18H\dot{\phi}G_{4\phi \phi},
\eea
and using Eqs. (\ref{eq:friedmann2-horndeski})  and (\ref{eq:scalar-field-equation-zeroorder}) we find that
\bea
D_8 & = & 0,
\eea
\bea
D_9 & = & -K_{X}+2G_{3\phi}-4H\dot{\phi}G_{3X} \nn \\
& - & \ddot{\phi}\left(2G_{3X}+\dot{\phi}^2G_{3XX}\right)-\dot{\phi}^2G_{3\phi X},\\
D_{10} & = & -\dot{\phi}^2G_{3X}+2G_{4\phi},\\
D_{11} & = & K_{\phi}+\left(\ddot{\phi}+3H\dot{\phi}\right)K_{X}+\dot{\phi}^2\left(4\ddot{\phi}+3H\dot{\phi}\right)K_{XX} \nn \\
& + & \dot{\phi}^4\left(K_{\phi XX}+\ddot{\phi}K_{XXX}\right) - 2\left(\ddot{\phi} +  3H\dot{\phi}\right)G_{3\phi} \nn \\
& + & 9\dot{\phi}\left(2H\ddot{\phi}+3H^2\dot{\phi}+\dot{H}\dot{\phi}\right)G_{3X}-\dot{\phi}^2G_{3\phi \phi} \nn \\
&-&\dot{\phi}^2\left(5\ddot{\phi} - 3H\dot{\phi}\right)G_{3\phi X} - \dot{\phi}^4G_{3\phi \phi X} \nn \\
&+& 3\dot{\phi}^3\left(7H\ddot{\phi} + 3H^2\dot{\phi} + \dot{H}\dot{\phi}\right)G_{3XX}  \nn \\
&-&\dot{\phi}^4\left(\ddot{\phi}-3H\dot{\phi}\right)G_{3\phi XX}+3H\dot{\phi}^5\ddot{\phi}G_{3XXX} \nn \\
&-& 6\left(2H^2+\dot{H}\right)G_{4\phi},\\
M^2 & = & -K_{\phi \phi}+\left(\ddot{\phi}+3H\dot{\phi}\right)K_{\phi X}+\dot{\phi}^2K_{\phi \phi X}+\dot{\phi}^2\ddot{\phi}K_{\phi X X}  \nn \\
&-&\ddot{\phi}\left[2G_{3\phi\phi}+\dot{\phi}^2G_{3\phi\phi X} \right. \nn \\
&-& \left. 3H\dot{\phi}\left(2G_{3\phi X}+\dot{\phi}^2G_{3\phi XX}\right)\right] \nn \\
&-& 6H\dot{\phi}G_{3\phi\phi} + 3\dot{\phi}^2\left(3H^2+\dot{H}\right)G_{3\phi X}-\dot{\phi}^2G_{3\phi\phi\phi}\nn \\ &+& 3H\dot{\phi}^3G_{3\phi\phi X}-6\left(2H^2+\dot{H}\right)G_{4\phi\phi}.
\eea
For the DE effective perturbation equations we found the following coefficients
\bea
\mathcal{F}_1 & = &\left(A_6-B_7\right)B_7G_4G^2_{4\phi}\Bigl(B_7G_4 \nn \\
&-&\left(B_6-2\right)G_{4\phi}\Bigr),
\eea
\bea
\mathcal{F}_2 & = &  \left(A_6-B_7\right)B_7G_4\Bigl(3\nu G^2_{4\phi}+2B_2G^2_{4\phi \phi}\dot{\phi}^2 \nn \\
&-&G_{4\phi}\Bigl(B_4G_{4\phi\phi}\dot{\phi}+B_2G_{4\phi\phi\phi}\dot{\phi}^2+B_2G_{4\phi\phi}\ddot{\phi}\Bigr)\Bigr)\nn \\
&+&G^2_{4\phi}\Bigl(B_2B_7\left(B_7-A_6\right)G_{4\phi\phi}\dot{\phi}^2+G_{4\phi}\Bigl(B_9\left(B^2_7-2D_9\right)\nn \\
&+&\left(A_6-B_7\right)B_7\left(B_4\dot{\phi}+B_2\ddot{\phi}\right)\Bigr)\Bigr),
\eea
\bea
\mathcal{F}_3 & = & B_6B_9M^2G^3_{4\phi},
\eea
\bea
\mathcal{F}_4 & = & G_{4\phi},
\eea
\bea
\mathcal{F}_5 & = & B_6\left(A^2_6-2A_6B_7+B_6D_9\right)G^2_{4\phi},
\eea
\bea
\mathcal{F}_6 & = & -B^2_6M^2G^2_{4\phi},
\eea
\bea
\mathcal{F}_7 & = & G_{4\phi}\Bigl(A_6\left(A_6-B_7\right)B_7G_4\nn \\
&+&\left(B_6-2\right)\left(B_6D_9-A_6B_7\right)G_{4\phi}\Bigr),
\eea
\bea
\mathcal{F}_8 & =&\Bigl(G^2_{4\phi}\Bigl(A_4\left(B^2_7-B_6D_9\right)-\left(B_6-2\right)B_6M^2\nn \\
&+&6\left(B_7^2-B_6D_9\right)H^2+A_2\left(A_6-B_7\right)B_7\dot{\phi}\Bigr)\nn \\
&-&\left(A_6-B_7\right)B_7G_4\left(\mu G_{4\phi}+A_2G_{4\phi\phi}\dot{\phi}\right)\Bigr),
\eea
\bea
\mathcal{F}_9 & = &B_6M^2G^2_{4\phi}\left(A_4+6H^2\right)
\eea
\bea
\mathcal{F}_{10} & = &G_{4\phi}\Bigl(\left(A_6-B_7\right)B_7C_4G_4 \nn \\
&+&\left(B^2_7-B_6D_9\right)G_{4\phi}\left(C_3+2H\right)\Bigr)\nn \\
&+&\left(A_6-B_7\right)B_7C_2\left(G^2_{4\phi}-G_4G_{4\phi\phi}\right)\dot{\phi},
\eea
\bea
\mathcal{F}_{11}&=B_6M^2G^2_{4\phi}\left(C_3+2H\right).
\eea

The coefficients for the KGB DE effective perturbation equations are

\bea
\hat{\mathcal{F}}_2 & = &-B_9D_9+3A_6\nu-6D_9\left(3H^2+2\dot{H}\right),
\eea
\bea
\hat{\mathcal{F}}_3 & = &M^2\left(B_9+18H^2+12\dot{H}\right)
\eea
\bea
\hat{\mathcal{F}}_5 & = & A^2_6+B_6D_9 ,
\eea
\bea
\hat{\mathcal{F}}_6 & = & -B_6M^2,
\eea
\bea
\hat{\mathcal{F}}_7 & = & -A^2_6-\left(B_6-2\right)D_9,
\eea
\bea
\hat{\mathcal{F}}_8 & = &A_4D_9+M^2\left(B_6-2\right)+A_6\mu+6D_9H^2,
\eea
\bea
\hat{\mathcal{F}}_9 & = &- M^2\left(A_4+6H^2\right),
\eea
\bea
\hat{\mathcal{F}}_{10} & = &M^2\left(C_3+2H\right),
\eea
\bea
\hat{\mathcal{F}}_{11} & = &A_6C_4-C_3D-9-2D_9H.
\eea

\bibliography{efa-horndeski}

\end{document}